\documentclass[11pt,a4paper]{amsart}
\usepackage[dvips,colorlinks=true,linkcolor=blue]{hyperref}
\usepackage{subfigure,caption2}
\usepackage{amssymb}
\usepackage{graphicx}
\usepackage[rflt]{floatflt}
\numberwithin{equation}{section}
\newcommand{\smallpagebreak}{{\par\vspace{2 mm}\noindent}}

\newcommand{\demo}{\par\noindent{\it Proof.\/} \ }

\newcommand{\dsize}{\textstyle}
\newcommand{\D}{\displaystyle}

\newcommand{\R}{{\mathbb R}}

\newcommand{\Z}{{\mathbb Z}}
\newcommand{\N}{{\mathbb N}}
\newcommand{\C}{{\mathbb C}}

\providecommand{\babs}[1]{\left\bracevert#1\right\bracevert}

\newcommand{\re}{{\rm Re}\,}
\newcommand{\im}{{\rm Im}\,}
\newcommand{\ind}{\text{ind}\,}
\newcommand{\res}{{\rm res}\, }

\newcommand{\dist}{{\rm dist}\,}
\newcommand{\mes}{{\rm mes}\,}
\newcommand{\vers}{\operatornamewithlimits{\to}}
\theoremstyle{plain}
\newtheorem{Th}{Theorem}[section]
\newtheorem{Le}{Lemma}[section]
\newtheorem{Pro}{Proposition}[section]
\newtheorem{Cor}{Corollary}[section]
\theoremstyle{definition}
\newtheorem{Rem}{Remark}[section]
\newtheorem{Def}{Definition}[section]
\title{Weakly resonant tunneling interactions for adiabatic
  quasi-periodic Schr{\"o}dinger operators}
\author{Alexander Fedotov} \author{Fr{\'e}d{\'e}ric Klopp}
\address[Alexander Fedotov]{Department of Mathematical Physics, St
  Petersburg State University, 1, Ulia\-novskaja, 198904 St
  Petersburg-Petrodvorets, Russia}
\email{\href{mailto:fedotov@mph.phys.spbu.ru}{fedotov@mph.phys.spbu.ru}}
\address[Fr{\'e}d{\'e}ric Klopp]{LAGA, Institut Galil{\'e}e, U.R.A 7539 C.N.R.S,
  Universit{\'e} Paris-Nord, Avenue J.-B.  Cl{\'e}ment, F-93430 Villetaneuse,
  France}
\email{\href{mailto:klopp@math.univ-paris13.fr}{klopp@math.univ-paris13.fr}}
\keywords{quasi periodic Schr{\"o}dinger equation, two resonating
wells, pure point spectrum, absolutely continuous spectrum,
complex WKB method, monodromy matrix}
\subjclass{34E05, 34E20, 34L05, 34L40}
\thanks{F.K.'s research was partially supported by the program RIAC
  160 at Universit{\'e} Paris 13 and by the FNS 2000 ``Programme Jeunes
  Chercheurs''. A.F. thanks the LAGA, Universit{\'e} Paris 13 for its kind
  hospitality. Both authors thank the Mittag-Leffler Institute where
  part of this work was done.}
\begin{document}
\begin{abstract}
  In this paper, we study spectral properties of the one dimensional
  periodic Schr{\"o}dinger operator with an adiabatic quasi-periodic
  perturbation. We show that in certain energy regions the
  perturbation leads to resonance effects related to the ones observed
  in the problem of two resonating quantum wells. These effects affect
  both the geometry and the nature of the spectrum. In particular,
  they can lead to the intertwining of sequences of intervals
  containing absolutely continuous spectrum and intervals containing
  singular spectrum. Moreover, in regions where all of the spectrum is
  expected to be singular, these effects typically give rise to
  exponentially small "islands" of absolutely continuous spectrum.
  \vskip.5cm
  \par\noindent   \textsc{R{\'e}sum{\'e}.}
  Cet article est consacr{\'e} {\`a} l'{\'e}tude du spectre d'une famille
  d'op{\'e}rateurs quasi-p{\'e}riodiques obtenus comme perturbations
  adiabatiques d'un op{\'e}rateur p{\'e}riodique fix{\'e}. Nous montrons que, dans
  certaines r{\'e}gions d'{\'e}nergies, la perturbation entra{\^\i}ne des
  ph{\'e}nom{\`e}nes de r{\'e}sonance similaires {\`a} ceux observ{\'e}s dans le cas de
  deux puits. Ces effets s'observent autant sur la g{\'e}om{\'e}trie du
  spectre que sur sa nature. En particulier, on peut observer un
  entrelacement de type spectraux i.e. une alternance entre du spectre
  singulier et du spectre absolument continu. Un autre ph{\'e}nom{\`e}ne
  observ{\'e} est l'apparition d'{\^\i}lots de spectre absolument continu dans
  du spectre singulier d{\^u}s aux r{\'e}sonances.
\end{abstract}
\setcounter{section}{-1}
\maketitle
\section{Introduction}
\label{sec:intro}
The present paper is devoted to the analysis of the family of
one-dimensional quasi-periodic Schr{\"o}dinger operators acting on
$L^2(\R)$ defined by
\begin{equation}
  \label{family}
  H_{z,\varepsilon}=-\frac{d^2}{dx^2}+V(x-z)+\alpha
  \cos(\varepsilon x).
\end{equation}
We assume that
\begin{description}
\item[(H1)] $V:\ \R\to\R$ is a non constant, locally square
  integrable, $1$-periodic function;
\item[(H2)] $\varepsilon$ is a small positive number chosen such that
  $2\pi/\varepsilon$ be irrational;
\item[(H3)] $z$ is a real parameter;
\item[(H4)] $\alpha$ is a strictly positive parameter that we will
  keep fixed in most of the paper.
\end{description}
\noindent
As $\varepsilon$ is small, the operator~\eqref{family} is a slow
perturbation of the periodic Schr{\"o}dinger operator
\begin{equation}
  \label{Ho}
  H_0=-\frac{d^2}{dx^2}+V\,(x)
\end{equation}
acting on $L^2(\R)$. To study~\eqref{family}, we use the asymptotic
method for slow perturbations of one-dimensional periodic
equations developed in~\cite{MR2002h:81069} and~\cite{Fe-Kl:03e}.\\
The results of the present paper are follow-ups on those obtained
in~\cite{MR2003f:82043,Fe-Kl:01b,Fe-Kl:03f} for the
family~\eqref{family}. In these papers, we have seen that the spectral
properties of $H_{z,\varepsilon}$ at energy $E$ depend crucially on
the position of the {\it spectral window}
$\mathcal{F}(E):=[E-\alpha,E+\alpha]$ with respect to the spectrum of
the unperturbed operator $H_0$. Note that the size of the window is
equal to the amplitude of the adiabatic perturbation. In the present
paper, the relative position is described in
figure~\ref{interactingfigure} i.e., we assume that there exists $J$,
an interval of energies, such that, for all $E\in J$, the spectral
window $\mathcal{F}(E)$ covers the edges of two neighboring spectral
bands of $H_0$ (see assumption (TIBM)). In this case, one can say that
the spectrum in $J$ is determined by the interaction of the
neighboring spectral bands induced by the adiabatic perturbation.\\
The central object of our study is the monodromy equation, a
finite difference equation determined by the monodromy matrix for the
family~\eqref{family} of almost periodic operators.  The monodromy
matrix for almost periodic equations with two frequencies was
introduced in~\cite{MR2003f:82043}. The passage from~\eqref{family} to
the monodromy equation is a non trivial generalization of the
monodromization idea from the study of difference equations with
periodic coefficients on the real line, see~\cite{MR2003i:39025}.\\
%
\begin{floatingfigure}[r]{3.5cm}
  \centering
  \includegraphics[bbllx=71,bblly=662,bburx=176,bbury=721,width=3.5cm]{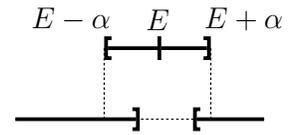}
  \caption{``Interacting'' bands}\label{interactingfigure}
\end{floatingfigure}
%
\noindent Let us now briefly describe our results and the heuristics
underlying them. Let $\mathbf{E}(\kappa)$ be the dispersion relation
associated to $H_0$ (see section~\ref{sec:le-quasi-moment}) ; consider
the {\it real} and {\it complex iso-energy curves}, respectively
$\Gamma_\R$ and $\Gamma$, defined by
\begin{gather}
  \label{isoenr}\hskip-3cm
  \Gamma_\R:=\{(\zeta,\kappa)\in\R^2;\
  \mathbf{E}(\kappa)+\alpha\cdot \cos(\zeta)=E\},\\
  \label{isoen}\hskip-3cm
  \Gamma:=\{(\zeta,\kappa)\in\C^2;\ 
  \mathbf{E}(\kappa)+\alpha\cdot \cos(\zeta)=E\}.
\end{gather}
The dispersion relation $\kappa\mapsto\mathbf{E}(\kappa)$ being
multi-valued, in~\eqref{isoen}, we ask that the equation be satisfied
at least for one of the possible values of $\mathbf{E}(\kappa)$.\\
The curves $\Gamma$ and $\Gamma_\R$ are both $2\pi$-periodic in the
$\kappa$- and $\zeta$-directions; they are described in details in
section~\ref{sec:iso-energy-curve-1}. The connected components of
$\Gamma_\R$ are called {\it real branches} of $\Gamma$.
%
\begin{floatingfigure}[r]{7.5cm}
  \centering
  \includegraphics[bbllx=71,bblly=566,bburx=247,bbury=721,width=7cm]{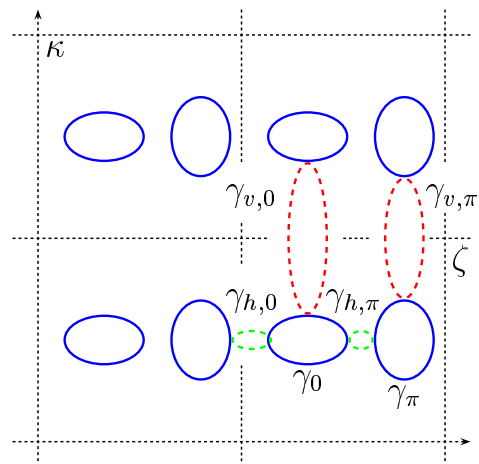}
  \caption{The adiabatic phase space}\label{TIBMfig:actions}
\end{floatingfigure}
%
\vskip.1cm Consider an interval $J$ such that, for $E\in J$, the
assumption on the relative position of the spectral window and the
spectrum of $H_0$ described above is satisfied (see
figure~\ref{interactingfigure}). Then, the curve $\Gamma_\R$ consists
of a infinite union of connected components, each of which is
homeomorphic to a torus ; there are exactly two such components in
each periodicity cell, see figure~\ref{TIBMfig:actions}. In this
figure, each square represents a periodicity cell. The connected
components of $\Gamma_\R$ are represented by full lines;
we denote two of them by $\gamma_0$ and $\gamma_\pi$.\\
The dashed lines represent loops on  $\Gamma$ that connect certain
connected components of $\Gamma_\R$; one can distinguish between the
``horizontal'' loops and the ``vertical'' loops. There are two
special horizontal loops denoted by $\gamma_{h,0}$ and
$\gamma_{h,\pi}$; the loop $\gamma_{h,0}$ (resp.
$\gamma_{h,\pi}$) connects $\gamma_0$ to $\gamma_\pi-(2\pi,0)$ (resp.
$\gamma_0$ to $\gamma_\pi$). In the same way, there are two special
vertical loops denoted by $\gamma_{v,0}$ and $\gamma_{v,\pi}$ ; the
loop $\gamma_{v,0}$ (resp.  $\gamma_{v,\pi}$) connects $\gamma_0$ to
$\gamma_0+(0,2\pi)$ (resp. $\gamma_\pi$ to $\gamma_\pi+(0,2\pi)$).
\par The standard semi-classical heuristic suggests the following
spectral behavior. To each of the loops $\gamma_0$ and $\gamma_\pi$,
one associates a phase obtained by integrating the fundamental
$1$-form on $\Gamma$ along the given loop; let $\Phi_0=\Phi_0(E)$
(resp.  $\Phi_{\pi}=\Phi_{\pi}(E)$) be one half of the phase
corresponding to $\gamma_0$ (resp. $\gamma_\pi$). Each of these
phases defines a quantization condition
\begin{equation}
  \label{eq:3}
  \frac{1}{\varepsilon}\Phi_0(E)=\frac\pi2+n\pi\quad\quad\text{ and
  }\quad\quad
  \frac{1}{\varepsilon}\Phi_\pi(E)=\frac\pi2+n\pi,\quad\quad n\in\N.
\end{equation}
Each of these conditions defines a sequence of energies in $J$, say
$(E_0^{(l)})_l$ and $(E_\pi^{(l')})_{l'}$. For $\varepsilon$
sufficiently small, the spectrum of $H_{z,\varepsilon}$ in $J$
should then be located in a neighborhood of these energies.
\par Moreover, to each of the complex loops $\gamma_{h,0}$,
$\gamma_{h,\pi}$, $\gamma_{v,0}$ and $\gamma_{v,\pi}$, one naturally
associates an action obtained by integrating the fundamental $1$-form
on $\Gamma$ along the loop. For $b\in\{0,\pi\}$ and $a\in\{v,h\}$, we
denote by $S_{a,b}$ the action associated to $\gamma_{a,b}$ multiplied
by $i/2$. For $E\in\R$, all these actions are real. One orients the
loops so that they all be positive. Finally, we define tunneling
coefficients as
\begin{equation*}
 t_{a,b}=e^{-S_{a,b}/\varepsilon},\quad b\in\{0,\pi\},\ a\in\{v,h\}.
\end{equation*}
When the real iso-energy curve consists in a single torus per
periodicity cell (see~\cite{MR2003f:82043}), the spectrum of
$H_{z,\varepsilon}$ is contained in a sequence of intervals
described as follows:
\begin{itemize}
\item each interval is neighboring a solution of the quantization
  condition;
\item the length of the interval is of order the largest tunneling
  coefficient associated to the loop;
\item the nature of the spectrum is determined by the ratio of the
  vertical tunneling coefficient to the horizontal one:
  \begin{itemize}
  \item if this ratio is large, the spectrum is singular;
  \item if the ratio is small, the spectrum is absolutely continuous.
  \end{itemize}
\end{itemize}
In the present case, one must moreover take into account the possible
interactions between the tori living in the same periodicity cell.
Similarly to what happens in the standard ``double well'' case
(see~\cite{MR81j:81010,MR85d:35085,He-Sj:84}), this effect only plays
an important role when the two energies, generated each by one of the
tori, are sufficiently close to each other. In this paper, we do not
consider the case when these energies are ``resonant'', i.e. coincide
or are ``too close'' to one another, but we can ``go'' up to the case
of exponentially close energies.
\smallpagebreak Let $E_0$ be an energy satisfying the quantization
condition~\eqref{eq:3} defined by $\Phi_0$; let $\delta$ be the
distance from $E_0$ to the sequence of energies satisfying the
quantization condition~\eqref{eq:3} defined by $\Phi_\pi$. We now
discuss the possible cases depending on this distance. Let us just add
that, as the sequences of energies satisfying the quantization
equation given by $\Phi_0$ or $\Phi_\pi$ play symmetric roles, in this
discussion, the indexes $0$ and $\pi$ can be interchanged freely.
\par First, we assume that, for some fixed $n>1$, this distance is of
order at least $\varepsilon^n$. In this case, near $E_0$, the states
of the system don't ``see'' the other lattice of tori, those obtained
by translation of the torus $\gamma_\pi$; nor do they ``feel'' the
associated tunneling coefficient $t_{v,\pi}$. Near $E_0$, everything
is as if there was a single torus, namely a translate of $\gamma_0$,
per periodicity cell.  Near $E_0$, the spectrum of $H_{z,\varepsilon}$
is located in a interval of length of order of the largest of the
tunneling coefficients $t_{v,0}$ and $t_h=t_{h,0}t_{h,\pi}$ (see
section~\ref{sec:acti-integr-tunn-1}).  And, the nature of the
spectrum is determined by quotient $t_{v,0}/t_h$.\\
So, in the energy region not too close to solutions to both
quantization conditions in~\eqref{eq:3}, we see that the spectrum is
contained in two sequences of exponentially small intervals. For each
sequence, the nature of the spectrum is obtained from comparing the
vertical to the horizontal tunneling coefficient for the torus
generating the sequence. As the tunneling coefficients for both tori
are roughly ``independent'' (see section~\ref{sec:remark}), it may
happen that the spectrum for one of the interval sequences be singular
while it be absolutely continuous for the other sequence. If this is
the case, one obtains numerous Anderson transitions i.e., thresholds
separating a.c. spectrum from singular spectrum (see
figure~\ref{fig:alt_spec}).
\par Let us now assume that $\delta$ is exponentially small, i.e. of
order $e^{-\eta/\varepsilon}$ for some fixed positive $\eta$ (not too
large, see section~\ref{sec:la-descr-prec}). This means that we
approach the case of resonant energies. Note that, this implies that
there is exactly one energy $E_\pi$ satisfying~\eqref{eq:3} for
$\Phi_\pi$ that is exponentially close to $E_0$; all other energies
satisfying~\eqref{eq:3} for $\Phi_\pi$ are at least at a distance of
order $\epsilon$ away from $E_0$.\\
Then, one can observe two new phenomena. First, there is a repulsion
of $I_0$ and $I_\pi$, the intervals corresponding to $E_0$ and $E_\pi$
respectively containing spectrum. This phenomenon is similar to the
splitting phenomenon observed in the double well problem
(see~\cite{MR81j:81010,MR85d:35085,He-Sj:84}). Second, the interaction
can change the nature of the spectrum: the spectrum that would be
singular for intervals sufficiently distant from each other can become
absolutely continuous when they are close to each other, see
Fig.~\ref{fig:res_tun}.  To explain this phenomenon, assume, for
simplicity, that $t_{v,0}$ and $t_{v,\pi}$, the ``vertical'' tunneling
coefficients associated to the tori $\gamma_0$ and $\gamma_\pi$, are
of the same order (in $\varepsilon$), i.e. $t_{v,0}\sim t_{v,\pi}\sim
t_v$. Then, if $|E_0-E_\pi|\sim \varepsilon^n$, on each of the
intervals $I_0$ and $I_\pi$, the nature of the spectrum is determined
by the same ratio $t_v/t_h$. If $|E_0-E_\pi|\sim
e^{-\eta/\varepsilon}$, the two arrays of tori begin to ``feel'' one
another: they form an array for which the tori from both arrays play
equivalent roles. In result, the ``horizontal'' tunneling becomes
stronger: it appears that $t_h$ has to be replaced by the effective
``horizontal'' tunneling coefficient $t_{h,{\rm eff}}=t_h/
\text{dist}(E_0, E_\pi)$, and the ratio $t_v/t_h$ has to be replaced
by $t_v/t_{h,{\rm eff}}$. So, the singular spectrum on the intervals
$I_0$ and $I_\pi$ ``tends to turn'' into absolutely continuous one.
\par There is one more case that will not be discussed in the present
paper: it is the case when $\delta\sim e^{-\eta/\varepsilon}$ with no
restriction on $\eta$ positive or, even, when $\delta$ vanishes. This
is the case of strong resonances; it reveals interesting new spectral
phenomena and is studied in detail in~\cite{Fe-Kl:04b}.


%
\section{The results}
\label{sec:MainResults}
\noindent We now state our assumptions and results in a precise way.
\subsection{The periodic operator}
\label{sec:periodic-operator}
This section is devoted to the description of elements of the spectral
theory of one-dimensional periodic Schr{\"o}dinger operator $H_0$ that we
need to present our results. For more details and proofs we refer to
section~\ref{S3} and to~\cite{Eas:73,MR2002f:81151}.
\subsubsection{The spectrum of $H_0$}
\label{sec:son-spectre}
The spectrum of the operator $H_0$ defined in~\eqref{Ho} is a union of
countably many intervals of the real axis, say $[E_{2n+1},\,E_{2n+2}]$
for $n\in\N$ , such that
\begin{gather*}
  E_1<E_2\le E_3<E_4\dots E_{2n}\le E_{2n+1}<E_{2n+2}\le \dots\,,\\
  E_n\to+\infty,\quad n\to+\infty.
\end{gather*}
This spectrum is purely absolutely continuous. The points
$(E_{j})_{j\in\N}$ are the eigenvalues of the self-adjoint operator
obtained by considering the differential polynomial~\eqref{Ho} acting
in $L^2([0,2])$ with periodic boundary conditions (see~\cite{Eas:73}).
The intervals $[E_{2n+1},\,E_{2n+2}]$, $n\in\N$, are the {\it spectral
  bands}, and the intervals $(E_{2n},\,E_{2n+1})$, $n\in\N^*$, the
{\it spectral gaps}. When $E_{2n}<E_{2n+1}$, one says that the $n$-th
gap is {\it open}; when $[E_{2n-1},E_{2n}]$ is separated from the rest
of the spectrum by open gaps, the $n$-th band is said to be {\it
  isolated}.
\smallpagebreak From now on, to simplify the exposition, we suppose
that 
\begin{description}
\item[(O)] all the gaps of the spectrum of $H_0$ are open.
\end{description}
\subsubsection{The Bloch quasi-momentum}
\label{sec:le-quasi-moment}
Let $x\mapsto\psi(x,E)$ be a non trivial solution to the periodic
Schr{\"o}din\-ger equation $H_0\psi=E\psi$ such that, for some $\mu\in\C$,
$\psi\,(x+1,E)=\mu \,\psi\,(x,E)$, $\forall x\in\R$. This solution is
called a {\it Bloch solution} to the equation, and $\mu$ is the {\it
  Floquet multiplier} associated to $\psi$. One may write
$\mu=\exp(ik)$; then, $k$ is the {\it Bloch quasi-momentum} of the
Bloch solution $\psi$.
\smallpagebreak It appears that the mapping $E\mapsto k(E)$ is an
analytic multi-valued function; its branch points are the points
$E_1$, $E_2$, $E_3$, $\dots$, $E_n$, $\dots$. They are all of ``square
root'' type.
\smallpagebreak The dispersion relation $k\mapsto{\mathbf E}(k)$ is
the inverse of the Bloch quasi-momentum.  We refer to
section~\ref{SS3.2} for more details on $k$.
\subsection{A ``geometric'' assumption on the energy region under study}
\label{sec:main-assumption-w}
Let us now describe the energy region where our study will be valid.
\smallpagebreak The spectral window centered at $E$, $\mathcal{F}(E)$,
is the range of the mapping $\zeta\in\R\mapsto E-\alpha\cos(\zeta)$.
\smallpagebreak In the sequel, $J$ always denotes a compact interval
such that, for some $n\in\N^*$ and for all $E\in J$, one has
\begin{description}
\item[(TIBM)]
  $[E_{2n},E_{2n+1}]\subset\dot{\mathcal F}(E)$ and  ${\mathcal
    F}(E)\subset]E_{2n-1},E_{2n+2}[$.
\end{description}
where $\dot{\mathcal F}(E)$ is the interior of $\mathcal{F}(E)$ (see
figure~\ref{interactingfigure}).\\
Actually, in the analysis, one has to distinguish between the cases
$n$ odd and $n$ even. From now on, we assume that, in the assumption
(TIBM), $n$ is even. The case $n$ odd is dealt with in the same way.
The spectral results are independent of whether $n$ is even or odd.
\begin{Rem}
  \label{rem:5}
  As all the spectral gaps of $H_0$ are assumed to be open, as their
  length tends to $0$ at infinity, and, as the length of the spectral
  bands goes to infinity at infinity, it is clear that, for any non
  vanishing $\alpha$, assumption (TIBM) is satisfied in any gap at a
  sufficiently high energy; it suffices that this gap be of length
  smaller than $2\alpha$.
\end{Rem}
\subsection{The definitions of the phase integrals and the tunneling
  coefficients}
\label{sec:iso-energy-curve}
We now give precise definitions of the phase integrals and the
tunneling coefficients appearing in the introduction.
\subsubsection{The complex momentum and its branch points}
\label{sec:complex-momentum-its}
The phase integrals and the tunneling coeffi-
%
\begin{floatingfigure}[r]{7cm}
  \centering
  \includegraphics[bbllx=71,bblly=606,bburx=275,bbury=721,width=7cm]{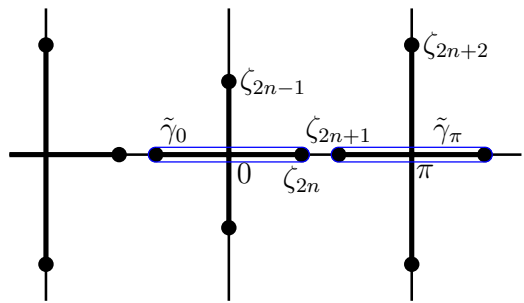}
  \caption{The branch points}\label{bp-tibm}
\end{floatingfigure}
%
\noindent cients are expressed in terms of integrals of the {\it
  complex momentum}. Fix $E$ in $J$. The complex momentum
$\zeta\mapsto\kappa(\zeta)$ is defined by
\begin{equation}
  \label{complex-mom}
  \hskip-5cm\kappa(\zeta)=k(E-\alpha\cos(\zeta)).
\end{equation}
As $k$, $\kappa$ is analytic and multi-valued. The set $\Gamma$
defined in~\eqref{isoen} is the graph of the function $\kappa$.  As
the branch points of $k$ are the points $(E_i)_{i\in\N}$, the branch
points of $\kappa$ satisfy
\begin{equation}
  \label{BPCM}
  \hskip-5cm E-\alpha\cos(\zeta)=E_j,\ j\in\N^*.
\end{equation}
As $E$ is real, the set of these points is symmetric with res-\\pect
to the real axis, to the imaginary axis; it is $2\pi$-periodic in
$\zeta$. All the branch points of $\kappa$ lie in the set
$\arccos(\R)$ which consists of the real axis and all the translates
of the imaginary axis by a multiple of $\pi$.\\
As the branch points of the Bloch quasi-momentum, the branch points of
$\kappa$ are of ``square root'' type.
\vskip.1cm Due to the symmetries, it suffices to describe the branch
points in the half-strip $\{\zeta;\ \im\zeta\geq0,\ 
0\leq\re\zeta\leq\pi\}$. These branch points are described in detail
in section~\ref{sec:kappa}. In figure~\ref{bp-tibm}, we show some of
them.  The points $(\zeta_j)_j$ satisfy~\eqref{BPCM}; one has
\begin{equation*}
 0<\zeta_{2n}<\zeta_{2n+1}<\pi,\quad
 0<\im\zeta_{2n+2}<\im\zeta_{2n+3}<\cdots,\quad
 0<\im\zeta_{2n-1}<\cdots<\im\zeta_1.
\end{equation*}
\subsubsection{The contours}
\label{sec:contours}
To define the phases and the tunneling coefficients, we introduce some
integration contours in the complex $\zeta$-plane.\\
These loops are shown in figure~\ref{bp-tibm} and~\ref{TIBMfig:2}. The
loops $\tilde\gamma_{0}$, $\tilde\gamma_{\pi}$, $\tilde\gamma_{h,0}$,
$\tilde\gamma_{h,\pi}$, $\tilde\gamma_{v,0}$ and
$\tilde\gamma_{v,\pi}$ are simple loops going once around respectively
the intervals $[-\zeta_{2n},\zeta_{2n}]$,
$[\zeta_{2n+1},2\pi-\zeta_{2n+1}]$, $[-\zeta_{2n+1},-\zeta_{2n}]$,
$[\zeta_{2n},\zeta_{2n+1}]$, $[\zeta_{2n-1},\overline{\zeta_{2n-1}}]$
and $[\zeta_{2n+2},\overline{\zeta_{2n+2}}]$.\\
In section~\ref{sec:clos-curv-compl}, we show that, on each of the
above loops, one can fix a continuous branch of the complex momentum.\\
Consider $\Gamma$, the complex iso-energy curve defined
by~\eqref{isoen}. Define the projection $\Pi:
(\zeta,\kappa)\in\Gamma\mapsto\zeta\in\C$. The fact that, on each of
the loops $\tilde\gamma_{0}$, $\tilde\gamma_{\pi}$,
$\tilde\gamma_{h,0}$, $\tilde\gamma_{h,\pi}$, $\tilde\gamma_{v,0}$ and
$\tilde\gamma_{v,\pi}$, one can fix a continuous branch of the complex
momentum implies that each of these loops is the projection on the
complex plane of some loop in $\Gamma$ i.e., for
$\tilde\gamma\in\{\tilde\gamma_{0},
\tilde\gamma_{\pi},\tilde\gamma_{h,0},
\tilde\gamma_{h,\pi},\tilde\gamma_{v,0},\tilde\gamma_{v,\pi}\}$, there
exists $\gamma\subset\Gamma$ such that $\tilde\gamma=\Pi(\gamma)$. In
sections~\ref{sec:real-branches} and~\ref{sec:complex-loops}, we give
the precise definitions of the curves $\gamma_{0}$, $\gamma_{\pi}$,
$\gamma_{h,0}$, $\gamma_{h,\pi}$, $\gamma_{v,0}$ and $\gamma_{v,\pi}$
represented in figures~\ref{bp-tibm} and~\ref{TIBMfig:actions} and
show that they project onto the curves $\tilde\gamma_{0}$,
$\tilde\gamma_{\pi}$, $\tilde\gamma_{h,0}$, $\tilde\gamma_{h,\pi}$,
$\tilde\gamma_{v,0}$ and $\tilde\gamma_{v,\pi}$ respectively.
\subsubsection{The phase integrals, the action integrals and the tunneling
  coefficients}
\label{sec:acti-integr-tunn-1}
The results described below are proved in
section~\ref{sec:phase-integr-tunn}.\\
%
%
\begin{floatingfigure}[r]{7cm}
  \centering
  \includegraphics[bbllx=71,bblly=606,bburx=275,bbury=721,width=7cm]{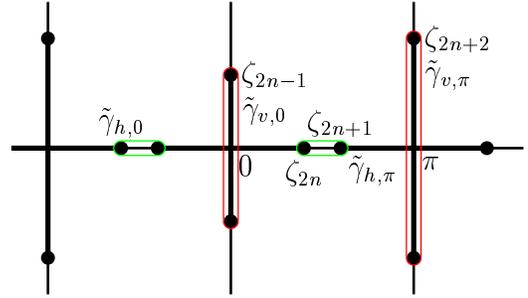}
  \caption{The loops for the phases}\label{TIBMfig:2}
\end{floatingfigure}
%
\noindent Let $\nu\in\{0,\pi\}$. To the loop $\gamma_\nu$, we associate
the {\it phase integral} $\Phi_\nu$ defined as
\begin{equation}
  \label{Phi:Gamma}
  \hskip-5cm\Phi_\nu(E)=\frac12\oint_{\tilde\gamma_\nu}\kappa\,d\zeta,
\end{equation}
where $\kappa$ is a branch of the complex momentum that is continuous
on $\tilde \gamma_\nu$. The function $E\mapsto\Phi_\nu(E)$ is real analytic
and does not vanish on $J$. The loop $\tilde \gamma_\nu$ is oriented so that
$\Phi_\nu(E)$ be positive. One shows that, for all $E\in J$,
\begin{equation}
  \label{eq:21}
  \hskip-5cm \quad\Phi_0'(E)<0\quad\text{ and
  }\quad\Phi_\pi'(E)>0.
\end{equation}
To the loop $\gamma_{v,\nu}$, we associate the {\it vertical action
  integral} $S_{v,\nu}$ defined as
\begin{equation}
  \label{Sv:Gamma}
  \hskip-5cm S_{v,\nu}(E)=-\frac i2\oint_{\tilde \gamma_{v,\nu}}\kappa d\zeta,
\end{equation}
\vskip.1cm\noindent where $\kappa$ is a branch of the complex momentum
that is continuous on $\tilde \gamma_{v,\nu}$. The {\it vertical tunneling
  coefficient} is defined to be
\begin{equation}
  \label{tv}
  t_{v,\nu}(E)=\exp\left(-\frac1\varepsilon S_{v,\nu}(E)\right).
\end{equation}
The function $E\mapsto S_{v,\nu}(E)$ is real analytic and does not
vanish on $J$.  The loop $\tilde \gamma_{v,\nu}$ is oriented so that
$S_{v,\nu}(E)$ be positive.
\smallpagebreak The index $\nu$ being chosen as above, we define
{\it horizontal action integral} $S_{h,\nu}$ by
\begin{equation}
  \label{Sh:Gamma}
  S_{h,\nu}(E)=-\frac i2\oint_{\tilde \gamma_{h,\nu}} \kappa(\zeta)\,d\zeta,
\end{equation}
where $\kappa$ is a branch of the complex momentum that is continuous
on $\tilde \gamma_{h,\nu}$. The function $E\mapsto S_{h,\nu}(E)$ is real
analytic and does not vanish on $J$.  The loop $\tilde \gamma_{h,\nu}$ is
oriented so that $S_{h,\nu}(E)$ be positive.  The {\it horizontal
  tunneling coefficient} is defined as
\begin{equation}
  \label{th}
  t_{h,\nu}(E)=\exp\left(-\frac1\varepsilon S_{h,\nu}(E)\right).
\end{equation}
In section~\ref{sec:t-h-0-pi}, we check that 
\begin{equation}
  \label{parity-h}
  S_{h,0}(E)=S_{h,\pi}(E)\quad\text{ and
  }\quad t_{h,0}(E)=t_{h,\pi}(E).
\end{equation}
One defines
\begin{equation}
  \label{eq:15}
  S_h(E)=S_{h,0}(E)+S_{h,\pi}(E)\quad\text{ and }\quad
  t_h(E)=t_{h,0}(E)\cdot t_{h,\pi}(E).
\end{equation}
In~\eqref{Phi:Gamma},~\eqref{Sv:Gamma}, and~\eqref{Sh:Gamma}, only the
sign of the integral depends on the choice of the branch of $\kappa$;
this sign was fixed by orienting the integration contour; for more
details, see sections~\ref{sec:clos-curv-compl} and~\ref{t,Phi:kappa}.
\subsection{Ergodic family}
\label{sec:une-famille-ergod}
Before discussing the spectral properties of $H_{z,\varepsilon}$, we
recall some general well known results from the spectral theory of
ergodic operators.
\smallpagebreak As $2\pi/\varepsilon$ is supposed to be irrational,
the function $x\mapsto V(x-z)+\alpha \cos(\varepsilon x)$ is
quasi-periodic in $x$, and the operators defined by~\eqref{family}
form an ergodic family (see~\cite{Pa-Fi:92}).
\smallpagebreak The ergodicity implies the following consequences:
\begin{enumerate}
\item the spectrum of $H_{z,\varepsilon}$ is almost surely independent
  of $z$ (\cite{Av-Si:83,MR94h:47068});
\item the absolutely continuous spectrum and the singular spectrum are
  almost surely independent of $z$ (\cite{MR94h:47068,MR2000f:47060});
\item the discrete spectrum is empty (\cite{MR94h:47068});
\item the Lyapunov exponent exists for almost all $z$ and is
  independent of $z$ (\cite{MR94h:47068}); it is defined in the
  following way: let $x\mapsto\psi(x)$ be the solution to the Cauchy
  problem
  \begin{equation*}
    H_{z,\varepsilon}\psi=E\psi,\quad\psi_{|x=0}=0,\quad
    \psi'_{|x=0}=1,
  \end{equation*}
  the following limit (when it exists) defines the Lyapunov exponent:
  \begin{equation*}
      \Theta(E)=\Theta(E,\varepsilon):=\lim_{x\to+\infty}\frac{\log
    \left(\sqrt{|\psi(x,E,z)|^2+|\psi'(x,E,z)|^2}\right)}{|x|}.
  \end{equation*}
\item the absolutely continuous spectrum is the essential closure of
  the set of $E$ where $\Theta(E)=0$ (the Ishii-Pastur-Kotani Theorem,
  see~\cite{MR94h:47068});
\item the density of states exists for almost all $z$ and is
  independent of $z$ (\cite{MR94h:47068}); it is defined in the
  following way: for $L>0$, let $H_{z,\varepsilon;L}$ be the operator
  $H_{z,\varepsilon}$ restricted to the interval $[-L,L]$ with the
  Dirichlet boundary conditions; for $E\in \R$; the following limit
  (when it exists) defines the density of states:
  \begin{equation*}
    N(E)=N(E,\varepsilon):=\lim_{L\to+\infty}\frac{\#\{\text{
    eigenvalues of }H_{z,\varepsilon;L}\text{less then or equal
    to}E\}}{2L};
  \end{equation*}
\item the density of states is non decreasing; the spectrum of
  $H_{z,\varepsilon}$ is the set of points of increase of the density
  of states.
\end{enumerate}
\subsection{A coarse description of the location of the spectrum in $J$}
\label{sec:une-descr-gross}
Henceforth, we assume that the assumptions (H) and (O) are satisfied
and that $J$ is a compact interval satisfying (TIBM). Moreover, we
suppose that
\begin{description}
\item[(T)]\quad $\D 2\pi\cdot\min_{E\in J}\min(\im\zeta_{2n-2}(E),
  \,\im\zeta_{2n+3}(E))>\max_{E\in J}\max(S_h(E),\,S_{v,0}(E),
  \,S_{v,\pi}(E))$.
\end{description}
Note that (T) is verified if the spectrum of $H_0$ has two successive
bands that are sufficiently close to each other and sufficiently far
away from the remainder of the spectrum (this can be checked
numerically on simple examples, see section~\ref{sec:numer-comp}). In
section~\ref{sec:comm-gener-remarks}, we will discuss this assumption
further.
\smallpagebreak Define
  \begin{equation}
    \label{eq:6}
    \delta_0:=\frac12\inf_{E\in J}\min(S_h(E),S_{v,0}(E),S_{v,\pi}(E))>0.
  \end{equation}
\smallpagebreak We prove
\begin{Th}
  \label{thr:2} 
  Fix $E_*\in J$. For $\varepsilon$ sufficiently small, there exists
  $V_*\subset \C$, a neighborhood of $E_*$, and two real analytic
  functions $E\mapsto\check \Phi_0(E,\varepsilon)$ and
  $E\mapsto\check\Phi_\pi(E,\varepsilon)$, defined on $V_*$ satisfying
  the uniform asymptotics
  \begin{equation}
    \label{eq:17}
    \check\Phi_0(E,\varepsilon)=\Phi_0(E)+o(\varepsilon),\quad
    \check\Phi_\pi(E,\varepsilon)=\Phi_\pi(E)+o(\varepsilon)\quad\text{where
    }\sup_{E\in V_*}|\varepsilon^{-1}o(\varepsilon)|\vers_{\varepsilon\to0}0,
  \end{equation}
  such that, if one defines two finite sequences of points in $J\cap
  V_*$, say $(E_{0}^{(l)})_l:=(E_{0}^{(l)}(\varepsilon))_l$ and
  $(E_{\pi}^{(l')})_{l'}:=(E_{\pi}^{(l')}(\varepsilon))_{l'}$, by
  \begin{equation}
    \label{eq:11}
    \frac1{\varepsilon}\check\Phi_0(E_0^{(l)},\varepsilon)=\frac\pi2+\pi
    l\quad\text{ and }\quad\frac1{\varepsilon}\check
    \Phi_\pi(E_\pi^{(l')},\varepsilon)=\frac\pi2+\pi l',\quad
    (l,\,l')\in\N^2,
  \end{equation}
  then, for all $z$, the spectrum of $H_{z,\varepsilon}$ in $J\cap
  V_*$ is contained in the union of the intervals
  \begin{equation*}
    \label{eq:16}
    I_0^{(l)}:=
    E_{0}^{(l)}+[-e^{-\delta_0/\varepsilon},e^{-\delta_0/\varepsilon}]
    \quad\text{ and }\quad
    I_{\pi}^{(l')}:=E_{\pi}^{(l')}+[-e^{-\delta_0/\varepsilon},
    e^{-\delta_0/\varepsilon}]
  \end{equation*}
  that is
  \begin{equation*}
    \sigma(H_{z,\varepsilon})\cap J\cap V_*\subset\left(\bigcup_{l}
    I_{0}^{(l)}\right)\bigcup\left(\bigcup_{l'}
    I_{\pi}^{(l')}\right).
  \end{equation*}
\end{Th}
\noindent In the sequel, to alleviate the notations, we omit the
reference to $\varepsilon$ in the functions $\check\Phi_0$ and
$\check\Phi_\pi$.
\smallpagebreak By~\eqref{eq:21} and~\eqref{eq:17},  there exists $C>0$
such that, for $\varepsilon$ sufficiently small, the points defined
in~\eqref{eq:11}  satisfy
\begin{gather}
  \label{eq:7}
  \frac{1}{C}\varepsilon\le E_0^{(l)}-E_0^{(l-1)}\le
  C\varepsilon,\\
  \label{eq:8}
  \frac{1}{C}\varepsilon\le E_{\pi}^{(l)}-E_{\pi}^{(l-1)}\le
  C\varepsilon.
\end{gather}
Moreover, for $\nu\in\{0,\pi\}$, in the interval $J\cap V_*$, the
number of points $E_{\nu}^{(l)}$ is of order $1/\varepsilon$.
\smallpagebreak In the sequel, we refer to the points $E_{0}^{(l)}$
(resp. $E_{\pi}^{(l)}$), and, by extension, to the intervals
$I_{0}^{(l)}$ (resp. $I_{\pi}^{(l)}$) attached to them, as of type $0$
(resp. type $\pi$).
\smallpagebreak By~\eqref{eq:7} and~\eqref{eq:8}, the intervals of
type $0$ (resp. $\pi$) are two by two disjoints; any interval of type
$0$ (resp. $\pi$) intersects at most a single interval of type $\pi$
(resp. $0$).
\subsection{A precise description of the location of the spectrum in $J$}
\label{sec:la-descr-prec}
We now describe the spectrum of $H_{z,\varepsilon}$ in the intervals
defined in Theorem~\ref{thr:2}.  Let us assume the interval under
consideration is of type $\pi$. One needs to distinguish two cases
whether this interval intersects or not an interval of type $0$.  The
intervals of one of the families that do not intersect any interval of
the other family are called {\it non-resonant}, the others being the
{\it resonant} intervals.
\smallpagebreak In the present paper, we only study the non-resonant
intervals; the resonant one are studied in~\cite{Fe-Kl:04b}. The
non-resonant is the simplest of the two cases; nevertheless, one
already sees that new spectral phenomena occur.
\begin{Rem}
  \label{rem:2}
  One may wonder whether resonances occur. They do occur.  Recall that
  the derivatives of $\Phi_\pi$ and $\Phi_0$ are of opposite signs on
  $J$, see~\eqref{eq:21}. Hence, as $\varepsilon$ decreases, on $J$,
  the points of type $0$ and $\pi$ move toward each other (at least,
  in the leading order in $\varepsilon$).  The motion being
  continuous, they meet.
  \smallpagebreak Nevertheless, for a generic $V$, there are only a
  few resonant intervals in $J$. On the other hand, for symmetric $V$,
  there may be numerous resonant energies; e.g., if $V$ is even, then
  the sequences $(E_0^{(l)})_l$ and $(E_\pi^{(l')})_{l'}$ coincide and
  all the intervals are resonant! This is due to the fact that the
  cosine is even; it is not true if $\alpha\cos(\cdot)$ is replaced by
  a generic potential.
\end{Rem}
\noindent We will describe our results for the intervals of type
$\pi$; {\it mutandi mutandis}, the results for the intervals of type
$0$ are the same. One has
\begin{Th}
  \label{th:tib:sp:1}
  Assume the conditions of Theorem~\ref{thr:2} are satisfied. For
  $\varepsilon$ sufficiently small, let $(I_0^{(l')})_{l'}$ and
  $(I_\pi^{(l)})_{l}$ be the finite sequences of intervals defined in
  Theorem~\ref{thr:2}. Consider $l$ such that, for any $l'$,
  $I_\pi^{(l)}\cap I_0^{(l')}=\emptyset$. Then, the spectrum of
  $H_{z,\varepsilon}$ in $I_\pi^{(l)}$ is contained $\check
  I_\pi^{(l)}$, the interval centered at the point
  \begin{equation}
    \label{eq:24}
    \check E_\pi^{(l)}=E_\pi^{(l)}+\varepsilon\,\frac{\Lambda_n(V)}{2
      \check\Phi_\pi'(E_\pi^{(l)})}\,t_h(E_\pi^{(l)})\,
    \tan\left(\frac{\check\Phi_0(E_\pi^{(l)})}{\varepsilon}\right),
  \end{equation}
  and of length
   \begin{equation}  
   \label{eq:19} \left|\check I_\pi^{(l)}\right| =
    \frac{2\varepsilon}{\check\Phi_\pi'(E_\pi^{(l)})}
    \left(\frac{t_h(E_\pi^{(l)})}
      {2\left|\cos\left(\frac{\check\Phi_0(E_\pi^{(l)})}
            {\varepsilon}\right)\right|}+
      t_{v,\pi}(E_\pi^{(l)})\right)\,(1+o(1)).
  \end{equation}
  The factor $\Lambda_n(V)$ is positive, and depends only on $V$
  and on $n$ (see section~\ref{sec:defin-analyt-prop}).\\
  In~\eqref{eq:19}, $o(1)$ tends to $0$ when $\varepsilon$ tends to
  $0$, uniformly in $E\in\check I_\pi^{(l)}$ and $l$ such that, for
  any $l'$, $I_\pi^{(l)}\cap I_0^{(l')}=\emptyset$.
\end{Th}
\noindent The fact that each of the intervals $\check I_\pi^{(l)}$
does contain some spectrum follows from
\begin{Th}
  \label{th:tib:sp:1a}   
  Let $dN_{\varepsilon}(E)$ denote the density of states measure of
  $H_{z,\varepsilon}$. In the case of Theorem~\ref{th:tib:sp:1}, for
  any $l$, one has
  \begin{equation*}
    \int_{\check I_\pi^{(l)}}dN_{\varepsilon}(E)=\frac\varepsilon{2\pi}.
  \end{equation*}
\end{Th}
\smallpagebreak{\it ``Level repulsion''.\/} \ Let $E_0$ be the point
in the sequence $(E_0^{(l')})_{l'}$ closest to $E_\pi:=E_\pi^{(l)}$.
Analyzing formulae~\eqref{eq:24} and~\eqref{eq:19}, one notices a
repulsion between the intervals $\check I_0$ and $\check I_\pi$.\\
Indeed, consider the second term in the right hand side
of~\eqref{eq:24}. As $\check\Phi_\pi'(E)>0$, this term has the same
sign as $\tan\left(\frac{\check\Phi_0(E_\pi)}{\varepsilon}\right)$.
Assume that $E_0$ and $E_\pi$ are sufficiently close to each other.
As, by definition, $\D\frac1\varepsilon\check\Phi_0(E_0)=\frac\pi2$
mod $\pi$ and as $\check \Phi_0'(E)<0$, the second term in the right
hand side of~\eqref{eq:24} is negative (resp. positive) if $E_\pi$ is
to the left (resp. right) of $E_0$. So, there is a repulsion between
$\check I_0$ and $\check I_\pi$. As the distance from $E_\pi$ to
$E_0$ controls the factor
\begin{equation*}
  \label{eq:25}
  \cos\left(\frac{\check\Phi_0(E_\pi)}{\varepsilon}\right),
\end{equation*}
the smaller this distance, the larger the repulsion.
\subsection{The Lyapunov exponent and the nature  of the spectrum in $J$}
\label{sec:nature-du-spectre}
Here, we discuss the nature of the spectrum in the interval $\check
I_\pi^{(l)}$. Therefore, we define
\begin{equation}
  \label{lambda_pi}
  \lambda_\pi(E)=
  \frac{t_{v,\pi}(E)}{t_h(E)}\,\dist\left(E,\bigcup_{l'}\{E_0^{(l')}\}\right),
\end{equation}
where, for a set $A$, $\dist(E,A)$ denotes the Euclidean distance from
$E$ to $A$.
\subsubsection{The Lyapunov exponent}
\label{sec:lyapunov-exponent}
We prove
\begin{Th}
  \label{th:gamma-pi}
  On the interval $\check I_\pi^{(l)}$, the Lyapunov exponent has the
  following asymptotic
  \begin{equation}
    \label{gamma-pi}
    \Theta(E,\varepsilon)=\frac\varepsilon{2\pi}
    \log^+\lambda_\pi(E_\pi^{(l)})+o(1),
  \end{equation}
  where $o(1)$ tends to $0$ when $\varepsilon$ tends to $0$, uniformly
  in $E\in\check I_\pi^{(l)}$ and $l$ such that, for any $l'$,
  $I_\pi^{(l)}\cap I_0^{(l')}=\emptyset$. Here, $\log^+=\max(0,\log)$.
\end{Th}
\subsubsection{Sharp drops of the Lyapunov exponent due to the resonance 
  interaction}
\label{sec:sharp-drops-lyapunov}
Formula~\eqref{gamma-pi} shows that the Lyapunov exponent becomes
``abnormally small'' on the interval $\check I_\pi^{(l)}$ when it
becomes close to one of the points $\{E_0^{(l')}\}$. Let us discuss this
in more details.\\
Assume that $(S_h-S_{v,\pi})(E_\pi^{(l)})>0$. If
$\dist\left(E_\pi^{(l)},\bigcup_{l}\{E_0^{(l')}\}\right)
\geq\varepsilon ^N$ (where $N$ is a fixed positive integer) then,
Theorem~\ref{th:gamma-pi} and formula~\eqref{lambda_pi} imply that
\begin{equation*}
  \label{eq:10}
  \Theta(E,\varepsilon)=\frac1{2\pi}(S_h-S_{v,\pi})
  (E_\pi^{(l)})+o(1)\text{ when }\varepsilon\to0.
\end{equation*}
On the other hand, when $E_\pi^{(l)}$ is only at a distance of size
$e^{-\delta/\varepsilon}$ (for $0<\delta<(S_h-S_{v,\pi})^+$) from the
set of energies $\{E_0^{(l')}\}$, on $\check I_\pi^{(l)}$, one has
\begin{equation*}
 \Theta(E,\varepsilon)=\frac1{2\pi}\left[(S_h-S_{v,\pi})
  (E_\pi^{(l)})-\delta\right]+o(1)\text{ when }\varepsilon\to0.
\end{equation*}
Hence, the value of $\Theta$ on $\check I_\pi^{(l)}$ drops sharply
when $E_\pi^{(l)}$ approaches the sequence $(E_0^{(l')})_{l'}$.
\subsubsection{ Singular spectrum} 
\label{sec:singular-spectrum}
As a natural consequence of Theorem~\ref{th:gamma-pi} and the
Ishii-Pastur-Kotani Theorem~\cite{MR94h:47068}, we obtain the
\begin{Cor}
  \label{cor:tib:sp:3}
  Fix $c>0$. For $\varepsilon$ sufficiently small, if $I_\pi^{(l)}$ is
  non-resonant and if $\varepsilon \log \lambda_\pi(E_\pi^{(l)})>c$,
  then, the interval $\check I_\pi^{(l)}$ defined in
  Theorem~\ref{th:tib:sp:1} only contains singular spectrum.
\end{Cor}
\subsubsection{ Absolutely continuous spectrum}
\label{sec:absol-cont-spectr-1}
If $\lambda_\pi$ is small on the interval $\check I_\pi^{(l)}$, most
of this interval is made of absolutely continuous spectrum; one shows
\begin{Th}
  \label{th:tib:sp:2}
  For $c>0$, there exists $\eta$, a positive constant, and a set of
  Diophantine numbers $D\subset (0,1)$ such that
  \begin{itemize}
  \item asymptotically, $D$ has total measure i.e.
    \begin{equation}
      \label{eq:9}
      \frac{\mes(D\cap(0,\varepsilon))}{\varepsilon}=
      1+e^{-\eta/\varepsilon}o\left(1\right).
    \end{equation}
  \item for $\varepsilon\in D$ sufficiently small, if $\check
    I_\pi^{(l)}$ is non-resonant and if $\varepsilon\log
    \lambda_\pi(E_\pi^{(l)})<-c$, then, one has
    \begin{equation}
      \label{eq:2}
      \frac{\mes(\check I_\pi^{(l)}\cap \Sigma_{\rm ac})}
        {\mes(\check I_\pi^{(l)})}=1+o(1),
    \end{equation}
    and $\Sigma_{ac}$ denotes the absolutely continuous spectrum of
    $H_{z,\varepsilon}$.\\
    In~\eqref{eq:9} and~\eqref{eq:2}, $o(1)$ tends to $0$ when
    $\varepsilon$ tends to $0$, uniformly in $E\in\check I_\pi^{(l)}$
    and $l$ such that, for any $l'$, $I_\pi^{(l)}\cap
    I_0^{(l')}=\emptyset$.
  \end{itemize}
\end{Th}
\subsubsection{A remark} 
\label{sec:remark}
The nature of the spectrum depends on the interplay between the values
of the actions $S_h$, $S_{v,0}$, $S_{v,\pi}$. So, when analyzing our
results, it is helpful to keep in mind the following observation. As
underlined at the end of section~\ref{sec:une-descr-gross}, choosing
$\varepsilon$ carefully, one can arrange that the distance between the
sequences of energies of type $0$ and $\pi$ be arbitrarily small;
moreover, this can be done in any compact subinterval of $J$ of length
at least $C\varepsilon$ (if $C$ is chosen sufficiently large).  On
such an interval, the actions $E\mapsto S_h(E)$, $E\mapsto S_{v,0}(E)$
and $E\mapsto S_{v,\pi}(E)$ vary at most of $C'\varepsilon$. Hence, at
the expense of choosing $\varepsilon$ sufficiently small in the right
way, we may essentially suppose that there exists an energy of type
$0$ and one of type $\pi$ at an arbitrarily small distance from each
other such that, on an $\varepsilon$-neighborhood of these points, the
triple $E\mapsto(S_h(E),S_{v,0}(E),S_{v,\pi}(E))$ takes any of its
possible values on $J$. This means that one can pick the values of
$E_\pi^{(l')}-E_0^{(l)}$ and $(S_h(E),S_{v,0}(E),S_{v,\pi}(E))$
essentially independently of each other.
\smallpagebreak Now, let us discuss two new spectral phenomena that
can occur under the hypothesis (TIBM).
\subsubsection{Transitions due to the proximity to a resonance}
\label{sec:trans-dues-lappr} 
The nature of the spectrum on the intervals defined in
Theorem~\ref{th:tib:sp:1} depends on their distance to the intervals
of the other family. The interaction can be strong enough to actually
change the nature of the spectrum. Let us consider a simple example.
Assume the interval $J$ satisfies:
\begin{gather}
 \label{eq:33}
  \min_{E\in J} S_h(E)>\max_{\nu\in\{0,\pi\}}\max_{E\in
  J}S_{v,\nu}(E),\\
  \intertext{and}
  \label{eq:33a}
 \frac32  \min_{\nu\in\{0,\pi\}}\min_{E\in J}S_{v,\nu}(E)>\max_{E\in J} S_h(E).
\end{gather}
Condition~\eqref{eq:33} guarantees that
$\D\delta_0=\frac12\,\min_{\nu\in\{0,\pi\}}\min_{E\in J}S_{v,\nu}(E)$.
Hence, there exists $c>0$ such that, for $E\in J$ and
$\nu\in\{0,\pi\}$,
\begin{equation}
  \label{eq:33b}
    S_h(E)-S_{v,\nu}(E)-\delta_0<-c<0.
\end{equation}
Consider now $I_0^{(l')}$ and $I_\pi^{(l)}$ both non resonant located
in $J\cap V_*$. Then,
\begin{itemize}
\item if the two intervals are distant of at least $\varepsilon^N$
  (where $N$ is a fixed integer) from each other,
  condition~\eqref{eq:33} guarantees that, on these intervals, the
  spectrum is controlled by Corollary~\ref{cor:tib:sp:3} and its
  analogue for the intervals of type $0$.
\item if the two intervals are distant of
  $C\exp\left(-\delta_0/\varepsilon\right)$ from each other, then
  condition~\eqref{eq:33b} guarantees that, on these intervals, the
  spectrum is controlled by Theorem~\ref{th:tib:sp:2} and its analogue
  for the intervals of type $0$.
\end{itemize}
\vskip.1cm That intervals $J$ where both~\eqref{eq:33}
and~\eqref{eq:33a} hold exist can be checked numerically, see
section~\ref{sec:numer-comp}. Thus, not only does the location of the
spectrum depend of the distance separating intervals of type $0$ for
neighboring intervals of type $\pi$, but so does also the nature of
the spectrum. Transition can occur due to this interaction phenomenon:
spectrum that would be singular were the intervals sufficiently
distant from each other can become absolutely continuous when they
are close to each other (see Fig.~\ref{fig:res_tun}).
\subsubsection{Alternating spectra}
\label{sec:alternance-des-types}
\begin{figure}[htbp]
  \centering \subfigure[Almost resonant transitions]{
    \includegraphics[bbllx=71,bblly=676,bburx=218,bbury=721,width=6cm]{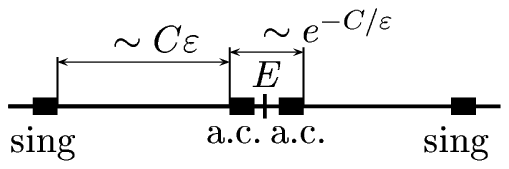}
  \label{fig:res_tun}}
\hskip3cm \subfigure[Alternating spectra]{
  \includegraphics[bbllx=71,bblly=676,bburx=218,bbury=721,width=6cm]{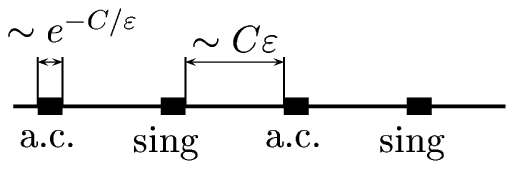}
  \label{fig:alt_spec}}
  \caption{Two new spectral phenomena}
  \label{fig:contours}
\end{figure}

To describe this phenomenon, to keep things simple, assume that, in
$V_*\cap J$, the distance between the points $\{E_0^{(l)}\}$ and the
points $\{E_\pi^{(l')}\}$ is larger than $\varepsilon^N$ (for some
fixed $N$); hence, all energies are non-resonant in $V_*\cap J$.
Taking Theorem~\ref{th:tib:sp:2} and Corollary~\ref{cor:tib:sp:3} into
account, we see that, on $\check I_0^{(l)}$ (resp. $\check
I_\pi^{(l')}$), the nature of the spectrum is determined by the size
of the ratio $t_{v,0} (E_0^{(l)})/t_{h}(E_0^{(l)})$ (resp.
$t_{v,\pi}(E_\pi^{(l')}) /t_{h}(E_\pi^{(l')})$). So, if for some
$\delta>0$, one has
\begin{equation}
  \label{eq:22}
  \forall E\in J\cap V_*,\quad S_{v,\pi}(E)-S_h(E)>\delta\quad\text{ et
  }\quad S_{v,0}(E)-S_h(E)<-\delta,
\end{equation}
then, in $V_*\cap J$, the sequences of type $0$ and $\pi$ contain
spectra of ``opposite'' nature: the spectrum in the intervals of type
$0$ is singular, and that in the intervals of type $\pi$ is, mostly,
absolutely continuous. This holds under the Diophantine condition on
$\varepsilon$ spelled out in Theorem~\ref{th:tib:sp:2}.  Hence, one
obtains an interlacing of intervals containing spectra of ``opposite''
types, see Fig.~\ref{fig:alt_spec}. In this case, the number of
Anderson transitions in $V_*\cap J$ is of order $1/\varepsilon$.\\
One can check numerically that the condition~\eqref{eq:22} is
fulfilled for some energy region $V_*$ and some values of $\alpha$
(see section~\ref{sec:numer-comp}).
%
%
\subsection{Numerical computations}
\label{sec:numer-comp}
We now turn to numerical results showing that the multiple phenomena
described in sections~\ref{sec:trans-dues-lappr}
and~\ref{sec:alternance-des-types} do occur.\\
%
\begin{figure}[h]
  \centering
  \includegraphics[width=14cm]{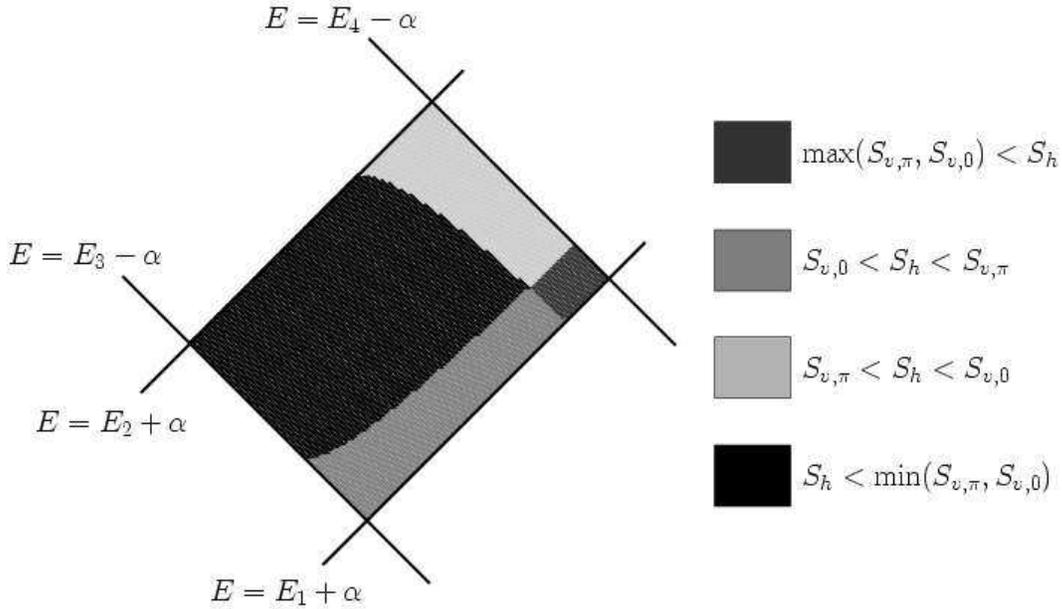}
  \caption{Comparing the actions}
  \label{fig:action}
\end{figure}
%
\noindent All these phenomena only depend on the values of the actions
$S_h$, $S_{v,0}$, $S_{v,\pi}$. For special potentials $V$, they are
quite easy to compute numerically.
\smallpagebreak We pick $V$ to be a two-gap potential; for these
potentials, the Bloch quasi-momentum $k$ (see
section~\ref{sec:le-quasi-moment}) is explicitly given by a
hyper-elliptic integral (\cite{MR52:11181,Ke-Mo:75}). The actions then
become easily computable. As the spectrum of $H_0=-\Delta+V$ only has
two gaps, we write $\sigma(H_0)=[E_1,E_2]\cup[E_3,E_4]\cup
[E_5,+\infty[$. In the computations, we take the values
\begin{equation*}
  E_1=0,\ E_2=3.8571429,\ E_3=6.8571429,\ E_4=12.100395,\text{
    and }E_5=100.70923.
\end{equation*}
On the figure~\ref{fig:action}, we represented the part of the
$(\alpha,E)$-plane where the condition (TIBM) is satisfied for $n=1$.
Its boundary consists of the straight lines $E=E_1+\alpha$,
$E=E_2+\alpha$, $E=E_3-\alpha$ and $E=E_4-\alpha$. Denote it by
$\Delta$.\\
The computations show that~(T) is satisfied in the whole of $\Delta$.
As $n=1$, one has $E_{2n-2}=-\infty$. So, it suffices to check~(T) for
$\zeta_{2n+3}=\zeta_5$. (T) can then be understood as a consequence of
the fact that $E_5-E_4$ is large.\\
On the figure~\ref{fig:action}, one sees that, for non-resonant
intervals,
\begin{itemize}
\item the zones where one has alternating spectral types (see
  section~\ref{sec:alternance-des-types}) are those where either
  $S_{v,0}<S_h<S_{v,\pi}$ or $S_{v,\pi}<S_h<S_{v,0}$
\item the transitions due to the proximity of the resonant situation
  (see section~\ref{sec:trans-dues-lappr}) take place in the part of
  region $\{S_h>\max(S_{v,\pi},S_{v,0})\}$ sufficiently close to
  $\{S_h<\min(S_{v,\pi},S_{v,0})\}$.
\end{itemize}
\subsection{Comments, generalizations and remarks }
\label{sec:comm-gener-remarks}
About assumption (T), its purpose is to select which tunneling
coefficients play the main role in the spectral behavior of
$H_{z,\varepsilon}$ in the interval $J$; this assumptions guarantees
that it is the tunneling coefficients associated to the loops defined
in section~\ref{sec:contours} that give rise to the principal terms in
the asymptotics of the monodromy matrix that we describe in
section~\ref{sec:MM}.
\par In the present paper, we restricted ourselves to perturbations of
$H_0$ of the form $\alpha\cos$. As will be seen from the proofs, this
is not necessary. The essential special features of the cosine that
were used are the simplicity of its reciprocal function (that is
multivalued on $\C$). More precisely, the assumption that is really
needed is that the geometry of the objects of the complex WKB method
that is used to compute the asymptotics of the monodromy matrix be as
simple as that for the cosine. This geometry does not only depend on
the perturbation; it also depends on the interval $J$ under
consideration and on the Bloch quasi-momentum of $H_0$. The precise
assumptions needed to have our analysis work are requirements on the
conformal properties of the complex momentum.
\par The methods developed
in~\cite{MR2002h:81069,MR2003f:82043,Fe-Kl:01b,Fe-Kl:03e,Fe-Kl:03f}
are quite general; using them, one can certainly analyze more
complicated situations i.e., more general adiabatic perturbations of
$H_0$.  Nevertheless, the computations may become much more
complicated than those found in the present paper.
\subsection{Asymptotic notations}
\label{COo}
We now define some notations that will be used throughout the paper.
Below $C$ denotes different positive constants independent of
$\varepsilon$, $E$ and $E_\pi$.\\
When writing $f=O(g)$, we mean that there exits $C>0$ such that
$|f|\le C\,|g|$ for all $\varepsilon$, $z$, $E$ in
consideration.\\
When writing $f=o(g)$, we mean that there exists $\varepsilon\mapsto
c(\varepsilon)$, a function such that
\begin{itemize}
\item $|f|\le c(\varepsilon) |g|$ for all $\varepsilon$, $z$, $E$ in
  consideration;
\item $c(\varepsilon)\to0$ when $\varepsilon\to 0$.
\end{itemize} 
When writing $f\asymp g$, we mean that there exists $C>1$ such that
$C^{-1} |g|\le |f|\le C|g|$ for all $\varepsilon$, $z$, $E$ in
consideration.\\
When writing error estimates, the symbol $O(f_1, f_2, \dots f_n)$
denotes functions satisfying the estimate
\begin{equation}
  \label{OOO}
  |O(f_1, f_2, \dots f_n)|\le C(|f_1|+|f_2|+\dots |f_n|),
\end{equation}
with a positive constant $C$ independent of $z$, $E$ and $\varepsilon$
under consideration.

%

%
\section{The monodromy matrix}
\label{sec:MM}
In this section, we consider the quasi-periodic differential equation
\begin{equation}
  \label{G.2z}
  -\frac{d^2}{dx^2}\psi(x)+(V(x-z)+\alpha\cos(\varepsilon x))\psi(x)=
   E\psi(x),\quad x\in\R,
\end{equation}
and recall the definition of the monodromy matrix and of the monodromy
equation for~\eqref{G.2z}. We also recall how these objects are
related to the spectral theory of the operator $H_{z,\varepsilon}$
defined in~\eqref{family}. Finally, we describe two monodromy matrices
for~\eqref{G.2z}.
\subsection{The monodromy matrices and the monodromy equation}
\label{sec:monodr-matr-monodr}
We now follow~\cite{MR2002h:81069,MR2003f:82043}, where the reader can
find more details, results and their proofs.
\subsubsection{The definition of the monodromy matrix}
\label{sec:Monodromy-matrix}
For any $z$ fixed, let $(\psi_j(x,\,z))_{j\in\{1,2\}}$ be two linearly
independent solutions of equation~\eqref{G.2z}.  We say that they form
a {\it consistent basis} if their Wronskian is independent of $z$,
and, if for $j\in\{1,2\}$ and all $x$ and $z$,
\begin{equation}
  \label{consistency}
  \psi_j(x,\,z+1)=\psi_j(x,\,z).
\end{equation}
As $(\psi_j(x,\,z))_{j\in\{1,2\}}$ are solutions to
equation~\eqref{G.2z}, so are the functions
$((x,z)\mapsto\psi_j(x+2\pi/\varepsilon,z+2\pi/\varepsilon))_{j\in\{1,2\}}$.
Therefore, one can write
\begin{equation}
  \label{monodromy}
  \Psi\,(x+2\pi/\varepsilon,z+2\pi/\varepsilon)= M\,(z,E)\,\Psi\,(x,z),\quad
  \Psi(x,z)=\begin{pmatrix}\psi_{1}(x,\,z)\\ \psi_{2}(x,z)\end{pmatrix},
\end{equation}
where $M\,(z,E)$ is a $2\times 2$ matrix with coefficients independent of
$x$. The matrix $M(z,E)$ is called {\it the monodromy matrix}
corresponding to the basis $(\psi_j)_{j\in\{1,2\}}$. To simplify the
notations, we often drop the $E$ dependence when not useful.
\smallpagebreak For any consistent basis, the monodromy matrix
satisfies
\begin{equation}
  \label{Mproperties}
  \det M\,(z)=1,\quad
  M\,(z+1)=M\,(z),\quad \forall z.
\end{equation}
\subsubsection{The monodromy equation and the link with the spectral
  theory of $H_{z,\varepsilon}$}
\label{sec::Mon-eq}
Set
\begin{equation}\label{h}
h=\frac{2\pi}{\varepsilon}\,{\rm mod}\, 1.
\end{equation}
Let $M$ be the monodromy matrix corresponding to the consistent basis
$(\psi_{j})_{j=1,2}$. Consider the {\it monodromy equation}
\begin{equation}
  \label{Mequation}
  F(n+1)=M\,(z+nh)F(n),\quad\text{where}\quad F(n)\in \C^2,\quad
  \forall n\in\Z.
\end{equation}
The spectral properties of $H_{z,\varepsilon}$ defined
in~\eqref{family} are tightly related to the behavior of solutions
of~\eqref{Mequation}. For now we will give a simple example of this
relation; more examples will be given in the course of the paper.
\smallpagebreak Recall the definition of the Lyapunov exponent for a
matrix cocycle.  Let $z\mapsto M(z)$ be an $SL(\C,\,2)$-valued
$1$-periodic function of the real variable $z$.  Let $h$ be a positive
irrational number. The Lyapunov exponent for the {\it matrix cocycle}
$(M,h)$ is the the limit (when it exists)
\begin{equation}
  \label{eq:44}
  \theta(M,h)=
  \lim_{L\to+\infty}\frac{1}{L}\log\Vert M(z+Lh)\cdot M(z+(L-1)h)\cdots
  M(z+h)\cdot M(z)\Vert.
\end{equation}
Actually, if $M$ is sufficiently regular in $z$ (say, if it belongs to
$L^\infty$), then $\theta(M,h)$ exists for almost every $z$ and does
not depend on $z$, see e.g.~\cite{MR94h:47068}.
\smallpagebreak One has
\begin{Th}[\cite{MR2002h:81069}]
  \label{Lyapunov}
  Let $h$ be defined by~\eqref{h}.  Let $z\mapsto M(z,E)$ be a
  monodromy matrix for equation~\eqref{G.2z} corresponding to basis
  solutions that are locally bounded in $(x,z)$ together with their
  derivatives in $x$.\\
  The Lyapunov exponents $\Theta(E,\varepsilon)$ and
  $\theta(M(\cdot,E),h)$ satisfy the relation
\begin{equation}
  \label{eq:Lyapunov}
  \Theta(E,h)=\frac\varepsilon{2\pi}\theta(M(\cdot,E),h).
\end{equation}
\end{Th}
\subsection{Asymptotics of two monodromy matrices}
\label{sec:mon-mat-as}
\smallpagebreak Recall that, in the interval $J$, the spectrum of
$H_{z,\varepsilon}$ is contained in two sequences of subintervals of
$J$, see Theorem~\ref{thr:2}. So, we consider two monodromy matrices,
one for each sequence; for $\nu\in\{0,\pi\}$, the monodromy matrix
$M_\nu$ is used to study the spectrum located near the points
$(E_{\nu}^{(l)})_l$.\\
In this section, we first describe the monodromy matrix $M_\pi$ in
detail. Then, we briefly discuss the monodromy matrix $M_0$.
\smallpagebreak Fix $\nu\in\{0,\pi\}$. The monodromy matrix $M_\nu$ is
analytic in $z$ and $E$ and has the following structure:
\begin{equation}
  \label{Tform}
  M_\nu=\begin{pmatrix} A_\nu   &  B_\nu  \\ B_\nu^* &
  A_\nu^*\end{pmatrix}.
\end{equation}
where, for $(z_1,\cdots,z_n)\mapsto g(z_1,\cdots,z_n)$, an analytic
function, we have defined
\begin{equation}
  \label{star}
  g^*(z_1,\cdots,z_n)=\overline{g(\overline{z_1},\cdots,\overline{z_n})}.
\end{equation}
When describing the asymptotics of the monodromy matrices,
we use the following notations:
\begin{itemize}
\item for $Y>0$, we let
\begin{equation}
  \label{TY}
  T_Y=e^{-2\pi Y/\varepsilon};
\end{equation}
\item we put
\begin{equation}
  \label{p}
  p(z)=e^{2\pi |\im z|}.
\end{equation}
\end{itemize}
One has
\begin{Th}
  \label{th:M-matrices}
  There exists $V_*$, a complex neighborhood of $E_*$, such that, for
  sufficiently small $\varepsilon$, the following holds.  Let
  \begin{equation}
    \label{Y's}
    Y_m=\frac1{2\pi}\inf_{E\in J\cap V_*}
    \max(S_{v,0}(E),S_{v,\pi}(E)),\quad \quad 
    Y_M=\frac1{2\pi}\sup_{E\in J\cap V_*}\max(S_{v,0}(E),S_{v,\pi}(E),S_h(E)).
  \end{equation}
  There exists $Y> Y_M$ and a consistent basis of solutions
  of~\eqref{G.2z} for which the monodromy matrix $(z,E)\mapsto
  M_\pi(z,E)$ is analytic in the domain $\left\{z\in \C:\,|\im
    z|<\frac{Y}\varepsilon\right\}\times V_*$
  and has the form~\eqref{Tform}.\\
  Fix $0<y<Y_m$. Let $V_*^\varepsilon= \left\{E\in V_*:\,\, |\im E|<
    \varepsilon\right\}$. In the domain
  \begin{equation}
    \label{analyticity:dom:b}
    \left\{z\in \C:\,|\im z|<\frac{y}\varepsilon\right\}\times V_*^\varepsilon,
  \end{equation}
  the coefficients of $M_\pi$ admit the asymptotic representations:
  \begin{equation}
    \label{A-pi:rough}
    A_\pi=2\,\frac{\alpha_\pi
      e^{\frac{i\check\Phi_\pi}\varepsilon}
      \,C_0}{T_h}+
    \frac12\,e^{\frac{i(\check\Phi_\pi-\check\Phi_0)}\varepsilon}\,
    \left(\frac1\theta+\theta\right) 
    +O\left(T_h,\,\frac{T_Yp(z)}{T_h},\,T_{v,0}p(z),\,T_{v,\pi}p(z)\right)
  \end{equation}
  and
  \begin{equation}
    \label{B-pi:rough}
    B_\pi=2\,\frac{\alpha_\pi e^{\frac{i\check\Phi_\pi}\varepsilon}\,C_0}{T_h}+
    \frac12\,e^{\frac{i\check\Phi_\pi}\varepsilon}\,
    \left(\frac1\theta\,e^{\frac{i\check\Phi_0}\varepsilon}+\theta\,
      e^{-\frac{i\check\Phi_0}\varepsilon}\right)+
    O\left(T_h,\,\,\frac{T_Yp(z)}{T_h},\,\,T_{v,0}p(z),\,\,T_{v,\pi}p(z)\right)
  \end{equation}
  with
  \begin{equation}\label{C-def}  
    C_0=\frac12\, \left(
  \alpha_0\,e^{\frac{i\check\Phi_0}\varepsilon}+
  \alpha_0^*\,e^{-\frac{i\check\Phi_0}\varepsilon}\right).  
  \end{equation}
  In these formulae, for $\nu\in\{0,\pi\}$,
  $(z,E)\mapsto\alpha_\nu(z,E)$ is an analytic function and is
  $1$-periodic in $z$; it admits the asymptotics
  \begin{equation}
    \label{alpha:as}
    \alpha_\nu =1+ T_{v,\nu} e^{2\pi
      i(z-z_\nu(E))}+O\left(T_Y\, p(z)\right).
  \end{equation}
  The quantities $E\mapsto\check\Phi_\nu(E)$, $E\mapsto T_{v,\nu}(E)$,
  $E\mapsto T_h(E)$, $E\mapsto\theta(E)$ and $E\mapsto z_\nu(E)$ are
  real analytic functions; they are independent of $z$; for $E\in
  V^\varepsilon_*$, they admit the asymptotics:
  \begin{itemize}
  \item
    \begin{gather}
      \label{check-Phi:as}
      \check\Phi_\nu(E)=\Phi_\nu(E)+o(\varepsilon), \\
      \label{T:as}
      T_h(E)=t_h(E)(1+o(1)),\quad T_{v,\nu}(E)=t_{v,\nu}(E)\,(1+o(1)),
    \end{gather}
    where $\Phi_\nu$ and $t_h$, $t_{v,\nu}$ are the phase integrals
    and the tunneling coefficients defined in
    section~\ref{sec:iso-energy-curve};
  \item
    \begin{equation}
      \label{theta:as}
      \theta(E)=\theta_n(V)\,(1+o(1)),
    \end{equation}
    where $\theta_n(V)$ is the constant defined in
    section~\ref{sec:Omega}; it is positive and depends only on $n$
    and $V$;
  \item
    \begin{gather}
      \label{z-nu:as}
      z_\pi(E)-z_0(E)=
      \frac{\check\Phi_\pi(E)-\check\Phi_0(E)}{2\pi\varepsilon}
      -\frac{\pi}{\varepsilon}+ o(1).\\
      \label{z-nu-prime}  
      z_\nu'(E)=O(1).
    \end{gather}
  \end{itemize}
\end{Th}
\noindent Note that the terms containing $\theta$ in the
asymptotics~\eqref{A-pi:rough} and~\eqref{B-pi:rough} are bounded
independently of $\varepsilon$.  So, with exponentially high accuracy,
the coefficients $A_\pi$ and $B_\pi$ are proportional.
\begin{Rem} The description of the monodromy matrix $M_0$ is similar
  to that of $M_\pi$: in Theorem~\ref{th:M-matrices}, one has to
  change
  \begin{enumerate}
  \item the indexes $0$ and $\pi$ by respectively $\pi$ and $0$;
  \item the quantity $\theta$ by $1/\theta$;
  \item $z_0(E)$ by $z_0(E)+h$ in formulae~\eqref{alpha:as}.
  \end{enumerate}
  Most of the analysis used to construct $M_0$ is the same as that for
  $M_\pi$. The differences are described in section~\ref{sec:MM:demo}.
\end{Rem}
\smallpagebreak Theorem~\ref{th:M-matrices} is the central technical
result of the paper. In the next two sections, we use
Theorem~\ref{th:M-matrices} to study the spectrum of
$H_{z,\varepsilon}$, and the remainder of the paper is devoted to its
proof.
\subsubsection{Useful observations}
\label{sec:useful-observations}
We now turn to a collection of estimates used when deriving the
results of sections~\ref{sec:une-descr-gross},~\ref{sec:la-descr-prec}
and~\ref{sec:nature-du-spectre} from Theorem~\ref{th:M-matrices}.  We
begin with
\begin{Le}
  \label{le:Phi'}
  Let $J_*\subset\R$ be a compact interval inside $V_*$.  There exists
  a neighborhood of $J_*$, say $\tilde V_*$, and $C>0$ such that, for
  sufficiently small $\varepsilon$, for $E\in\tilde V_*$ and
  $\nu\in\{0,\pi\}$, one has
  
  \begin{equation}
    \label{Phi':up}
    |\check\Phi_\nu'(E)|+|\check\Phi_\nu''(E)|\le C,
  \end{equation}
  and 
  \begin{equation}
    \label{Phi':down}
    \frac1C\le |\check\Phi_\nu'(E)|.
  \end{equation}
\end{Le}
\demo Recall that the phase integrals $\Phi_\nu$ are independent of
$\varepsilon$, analytic in a neighborhood of $J$, and, on $J$, the
derivatives $\Phi_\nu'(E)$ are bounded away from zero,
see~\eqref{eq:21}. Therefore, the statements of Lemma~\ref{le:Phi'}
follow from~\eqref{check-Phi:as} and the Cauchy estimates for the
derivatives of analytic functions ($o(\varepsilon)$
in~\eqref{check-Phi:as} is analytic in the domain $V_*$, and,
therefore, on any its fixed compact, one has the uniform estimates:
$\frac d{dE} o(\varepsilon)=o(\varepsilon)$ and $\frac{d^2}{dE^2}
o(\varepsilon)=o(\varepsilon)$). This completes the proof of
Lemma~\ref{le:Phi'}.
\qed\\
We also prove
\begin{Le}
  \label{obs:1} For sufficiently small $\varepsilon$, for
  $\nu\in\{0,\pi\}$, in the domain~\eqref{analyticity:dom:b}, one has
\begin{gather}
    \label{alpha:est}
    \alpha_\nu =1+ O(T_{v,\nu}  p(z))=1+o(1),\\
    \label{T:est:3}
    p(z)|T_{v,\nu}(E)|=o(1),\\
    \label{check-phi:est}
    \left|e^{i\check\Phi_\nu(E)/\varepsilon}\right|\asymp1, \\
    \label{T:est:2}
    |T_h(E)|+|T_{v,\nu}(E)|+T_Y\le Ce^{-2\delta_0/\varepsilon}, \\
   \label{T:est:1}
   T_Y=o( T_h(E))\quad\text{and}\quad T_Y=o(T_{v,\nu}(E)), \\
    \label{T:est}
    Ce^{-2\pi Y_M/\varepsilon}\leq |T_h(E)|\quad\text{and}\quad
    \frac1Ce^{-2\pi Y_M/\varepsilon}\leq |T_{v,\nu}(E)|\leq Ce^{-2\pi
    Y_m/\varepsilon}, \\
    \label{theta:est}
    |\theta(E)|\asymp1,\\
    \label{z-nu:est}
    |e^{2\pi i z_\nu(E)}|\asymp1.
  \end{gather}
  All the above estimates are uniform.
\end{Le}
\demo As $z_\nu$ is real analytic, estimate~\eqref{z-nu:est} follows
from~\eqref{z-nu-prime} and the definition of $V_*^\varepsilon$.
Estimate~\eqref{theta:est} follows from~\eqref{theta:as} as
$\theta_n(V)$ is a positive constant depending only on $n$ and $V$.
The estimates~\eqref{T:est} follow from~\eqref{T:as} and the
definitions of the tunneling coefficients, of the domain
$V_*^\varepsilon$ and numbers $Y_m$ and $Y_M$.
Estimates~\eqref{T:est:1} follow from~\eqref{T:est} as $Y>Y_M$.
Estimates~\eqref{T:est:2} follow from~\eqref{T:est:1},~\eqref{T:as}
and the definition of $\delta_0$. Estimate~\eqref{T:est:3} follows
from~\eqref{T:est} as in the domain~\eqref{analyticity:dom:b}, one has
$|\im z|\le y/\varepsilon$, and $y<Y_m$.
Estimate~\eqref{check-phi:est} follows from~\eqref{Phi':up}, the
definition of the domain $V_*^\varepsilon$ and from the real
analyticity of the phase integrals. The inequalities
in~\eqref{alpha:est} follow from~\eqref{T:est:1} and~\eqref{T:est:3}.
This completes the proof of Lemma~\ref{obs:1}. \qed
\section{Rough characterization of the spectrum and a new monodromy matrix}
\label{sec:rough-char-spectr}
In this section, we first obtain a rough description of the location
of the spectrum of $H_{z,\varepsilon}$ i.e., we prove
Theorem~\ref{thr:2}. Then, we change the consistent basis so that, in
a neighborhood of the spectrum, the new monodromy matrix have a form
more convenient for the spectral study.
\subsection{The scalar equation}
\label{sec:scalar-equation}
Our analysis of the spectrum is based on the analysis of solutions of
the monodromy equation with the monodromy matrices described in the
previous section. A monodromy equation is a first order finite
difference 2-dimensional system of equations, see~\eqref{monodromy}.
Instead, of working with this system, we study an equivalent scalar
second order finite difference equation. To derive this equation, we
use the following elementary observation
\begin{Le}
  \label{le:3}
  Let $M:\,z\mapsto M(z)$ be a $SL(2,\C)$-valued matrix function of
  the real variable $z$, and let $h$ be a real number. Assume that
  $M_{12}(z)\ne 0$ for all $z$. Define
  \begin{equation}
    \label{rho,v}
    \rho(z)=M_{12}(z)/M_{12}( z-h),\quad
    v(z)=M_{11}(z)+\rho( z)\,M_{22}( z).
  \end{equation}
  A fucntion $\Psi_1:\Z\to\C$ is the first component of a vector
  function $\Psi:\Z\to\C^2$ satisfying the equation
  \begin{equation*}
    \label{matrix-eq}
    \Psi(k+1)=M(hk+ z)\Psi(k),\quad \forall k\in\Z,
  \end{equation*}
  if and only if it satisfies the equation
  \begin{equation}
    \label{scalar-eq}
    \Psi_1(k+1)+\rho(hk+ z)\Psi_1(k-1)=v(hk+z)\Psi_1(k),\quad\forall
    k\in\Z.
  \end{equation}
\end{Le}
\noindent The reduction from the monodromy equation to the scalar
equations~\eqref{scalar-eq} has already been used in~\cite{MR96m:47060}
and~\cite{MR2003f:82043}. To characterize the location of the spectrum
of~\eqref{family}, we use
\begin{Pro}
  \label{resolvent-set} 
  Fix $E$ in equation~\eqref{G.2z}. Let $f$ and $g$ form a consistent
  basis in the space of the solutions of~\eqref{G.2z}, and let $M$ be
  the corresponding monodromy matrix.\\
  Assume that the functions $(x,z)\mapsto f(x,z)$, $(x,z)\mapsto
  g(x,z)$, $(x,z)\mapsto\partial_x f(x,z)$ and $(x,z)\mapsto\partial_x
  g(x,z)$ are continuous on $\R^2$.\\
  Suppose that $\D\min_{z\in\R}|M_{12}(z)|>0$.
  In terms of $M$, define the functions $\rho$ and $v$
  by~\eqref{rho,v} and define $h$ by~\eqref{h}.
  Let
  \begin{equation}
    \label{resolvent-set:cond}
    \max_{z\in\R} |\rho(z)|< 
    \left(\frac12\,\min_{z\in\R} |v(z)|\right)^2,\quad
    \ind\rho=\ind v=0,
  \end{equation}
  where $\ind g$ is the index of a continuous periodic function $g$.
  \smallpagebreak Then, $E$ is in the resolvent set of~\eqref{family}.
\end{Pro}
\noindent The proof of this proposition immediately follows from
Proposition~4.1 and Lemma~4.1 in~\cite{MR2003f:82043} based on the
analysis in~\cite{MR96m:47060}.
\begin{Rem}
  \label{rem:1}
  This proposition is very effective if the coefficient $M_{12}$ of
  the monodromy matrix is close to a constant. Then, it roughly says
  that the spectrum is located in the intervals where the absolute
  value of the trace of the monodromy matrix is larger than 2. This is
  the condition one meets in the classical theory of the periodic
  Schr{\"o}dinger operator (\cite{Eas:73}).
\end{Rem}
\subsection{Rough characterization of the location of the spectrum}
\label{sec:rough-char-place}
We now prove Theorem~\ref{thr:2}.\\
Pick $E_*\in J$. Let $V_*$ be as in Theorem~\ref{th:M-matrices}.
Consider the sequences $(E_\pi^{(l)})_l$ and $(E_0^{(l')})_{l'}$
defined by the quantization conditions~\eqref{eq:11}.
\smallpagebreak Introduce $\delta_0$ by~\eqref{eq:6}.  Let $J_*$ be a
compact subinterval of $J\cap V_*$. One has
\begin{Le}
  \label{sp-place-rough}
  Pick $0<\alpha<1$. For $\varepsilon$ sufficiently small, in $J_*$, the
  spectrum of $H_{z,\varepsilon}$ is contained in the
  $\varepsilon^\alpha e^{-\delta_0/\varepsilon}$-neighborhood of the
  points $(E_\pi^{(l)})_l$ and $(E_0^{(m)})_m$ defined by the
  quantization conditions~\eqref{eq:11}.
\end{Le}
\noindent Lemma~\ref{sp-place-rough} implies Theorem~\ref{thr:2} at
the possible expense of reducing $V_*$ somewhat.
\demo Define
\begin{equation}
  \label{V-rough}
  V_{\rm rough}=\{ E\in J_*: \,\,\,|E-E_{0}^{(m)}|\ge
  \varepsilon^\alpha e^{-\delta_0/\varepsilon},\ \forall m\}.
\end{equation}
We shall prove that, for $\varepsilon$ small enough, the spectrum of
$H_{z,\varepsilon}$ in $V_{\rm rough}$ is contained in the
$\varepsilon^\alpha e^{-\delta/\varepsilon}$-neighborhood of the
points $(E_\pi^{(l)})_l$.\\
In the remainder of this proof, we assume that $\varepsilon$ is
sufficiently small for the statements of Theorem~\ref{th:M-matrices}
and Lemma~\ref{le:Phi'} to hold.
\smallpagebreak The proof consists of the following steps.\\
{\bf 1.} \ We prove that, for $\varepsilon$ sufficiently small, 
\begin{equation*}
  \label{eq:1}
  \inf_{E\in V_{\rm rough}}\left|\cos\left(\frac{\check\Phi_0(E)}
      {\varepsilon}\right)\right|\ge e^{-\delta_0/\varepsilon}.
\end{equation*}
\smallpagebreak This follows from~\eqref{V-rough}, from the definition
of the set $\{E_0^{(l)}\}$, and from~\eqref{Phi':down}.
\smallpagebreak{\bf 2.} \ We check that, for $E\in J_*$, and for
$z\in \R$, each of the functions $A_\pi$ and
$B_\pi$ has the form
\begin{equation*}
  \frac{2}{T_h}\,\left[e^{i\check\Phi_\pi/\varepsilon}\,
  \cos(\check\Phi_0/\varepsilon)+O(e^{-2\delta_0/\varepsilon})\right].
\end{equation*}
Indeed, by the first inequality from~\eqref{alpha:est}, for
$\nu\in\{0,\pi\}$, $z\in\R$ and $E\in J_*$, one has
\begin{equation*}
  \alpha_\nu=1+O(T_{v,\nu}).
\end{equation*}
By means of this estimate and of~\eqref{check-phi:est}
and~\eqref{theta:est}, we transform the right hand sides both
in~\eqref{A-pi:rough} and~\eqref{B-pi:rough} to the form
\begin{equation*}
  \frac{2}{T_h}\,\,\left(e^{i\check\Phi_\pi/\varepsilon}
    \,\cos(\check\Phi_0/\varepsilon)+
    O(T_{v,0},\,T_{v,\pi},\,T_h,\,T_Y)\right).
\end{equation*}
This and~\eqref{T:est:2} imply that $A_\pi$ and $B_\pi$ have the
requested form.
\smallpagebreak {\bf 3.} \ Let $(z,E)\mapsto\rho(z,E)$ be the function
defined by~\eqref{rho,v} for $M=M_\pi(z,E)$. The previous two steps
imply that there exists $C>0$ such that, for $\varepsilon$
sufficiently small, one has
\begin{equation*}
  \label{rho:rough}
  \sup_{z\in\R}\sup_{E\in V_{\rm rough}}|\rho(z,E)-1|\leq 
  Ce^{-\delta_0/\varepsilon}.
\end{equation*}
{\bf 4.} \ Let $(z,E)\mapsto v(z,E)$ be the function defined
by~\eqref{rho,v} for $M=M_\pi(z,E)$. The previous three steps imply
that, for $\zeta\in\R$ and $E\in V_{\rm rough}$, one has
\begin{equation*}
\begin{split}
  v(z,E)&=A_\pi+A_\pi^*+(\rho(z,E)-1) A_\pi^*\\
  &=\frac{2}{T_h}\,\left(
  \left[2\cos(\check\Phi_\pi/\varepsilon)\,\cos(\check\Phi_0/\varepsilon)
    +O(e^{-2\delta_0/\varepsilon})\right]\right.\\
  &\hskip7cm \left.+\left[\left(e^{i\check\Phi_\pi/\varepsilon}\,
      \cos(\check\Phi_0/\varepsilon)+O(e^{-2\delta_0/\varepsilon})\right)\,
    O(e^{-\delta_0/\varepsilon})\right]\right)\\
  &=\frac{4}{T_h}\,\cos(\check\Phi_0/\varepsilon)\,
  \left(\cos(\check\Phi_\pi/\varepsilon)+
    O(e^{-\delta_0/\varepsilon})\right).
\end{split}
\end{equation*}
{\bf 5.} \ There exists $C>0$ such that, for $\varepsilon$
sufficiently small, if $E\in \sigma(H_{z,\varepsilon})\cap V_{\rm
  rough}$, then
\begin{equation}
  \label{rough:spectre}
  \left|\cos\frac{\check\Phi_\pi(E)}\varepsilon\right|\leq
    C\,\left(e^{-\delta_0/\varepsilon}+
    \frac{T_h}{\left|\cos\frac{\check\Phi_0(E)}\varepsilon\right|}\right).
\end{equation}
Indeed, by steps 1 and 2, for sufficiently small $\varepsilon$, for
$E\in V_{\rm rough}$, one has
\begin{equation*}
   \min_{z\in\R}|B_\pi(z,E)|>0,\quad \ind B_\pi(\cdot,E)=0.
\end{equation*}
Moreover, by steps 3 and 4, there exists $C>0$ such that, for
$\varepsilon$ sufficiently small, for $E\in V_{\rm rough}$, if
\begin{equation*}\dsize
  \left|\cos\frac{\check\Phi_\pi(E)}\varepsilon\right|\geq C\,\left(e^{-\delta_0/\varepsilon}+
   \frac{T_h}{\left|\cos\frac{\check\Phi_0(E)}\varepsilon\right|}\right),
\end{equation*}
then, one has
\begin{equation*}
  \min_{z\in\R} |v(z,E)|^2> 4\max_{z\in\R}|\rho(z,E)| ,\quad \ind
  v(\cdot,E)=0.  
\end{equation*}
These two observations and Proposition~\ref{resolvent-set} complete
the proof of~\eqref{rough:spectre}.\\
{\bf 6.} \ In view of~\eqref{T:est:2} and of the first step,
inequality~\eqref{rough:spectre} implies that
\begin{equation*}
  |\cos(\check\Phi_\pi(E)/\varepsilon)|\le Ce^{-\delta_0/\varepsilon}.
\end{equation*}
By the definition of $(E_\pi^{(l)})_l$ and Lemma~\ref{le:Phi'}, this
implies that there exists $l$ such that $|E-E_\pi^{(l)}|\le
C\varepsilon e^{-\delta_0/\varepsilon}$. This completes the proof of
Lemma~\ref{sp-place-rough}.\qed
\subsection{A new monodromy matrix}
\label{sec:new-monodromy-matrix}
As said, to study the spectrum, instead of working with the monodromy
equation itself, it is more convenient to work with the equivalent
scalar equation~\eqref{scalar-eq}. The use of this equation is very
effective when $M_{12}$, the element of the monodromy matrix, is close
to a constant, and $M_{11}$ (or/and its derivative in $E$) is much
larger than $M_{22}$. To satisfy these requirements for $E$ near the
points $(E_\pi^{(l)})_l$, we introduce a new monodromy matrix.
Therefore, we make the following simple observation:
\begin{Le}
  \label{gauge-le}
  Recall that $h$ is defined by~\eqref{h}. Let $M$ be a monodromy
  matrix for equation~\eqref{G.2z}, and let $U:\,z\mapsto U(z)\in
  SL(2,\C)$ be a $1$-periodic matrix function.  Then,
  \begin{equation}
    \label{M-gauge}
    M^U(z)=U(z+h)\,M(z)\,U(z)^{-1}
  \end{equation}
  is also a monodromy matrix for equation~\eqref{G.2z}.
\end{Le}
\demo Let $f_1$ and $f_2$ be the solutions of~\eqref{G.2z} that form a
consistent basis for which $M$ is the monodromy matrix. The components
of the vector
\begin{equation}
  \label{U-transformation}
  {\mathcal F}(x,z)=U(z) F(x,z),\quad
  F(x,z)=\begin{pmatrix}f_1(x,z)\\ f_2(x,z)\end{pmatrix},
\end{equation}
are also solutions of~\eqref{G.2z}; they form a consistent basis, and
$M^U$ is the corresponding monodromy matrix.\qed
\smallpagebreak For $(z,E)$ in the domain~\eqref{analyticity:dom:b},
we define the new monodromy matrix $M^U$ choosing $M=M_\pi(z,E)$, the
matrix described in Theorem~\ref{th:M-matrices}, and
\begin{equation}
  \label{U}
  U(z)=\frac12\begin{pmatrix}1&1\\ -i & i\end{pmatrix}\,
  \begin{pmatrix}\gamma(z) & 0\\ 0& \gamma^*(z)\end{pmatrix},\quad
  \text{where}\quad
  \gamma(z+h)=\sqrt{{\dsize\frac{\alpha_\pi^*(z)}{\alpha_\pi(z)}}}\,
  e^{-i\check\Phi_\pi/\varepsilon}.
\end{equation}
Recall that, for $(z,E)$ being in the
domain~\eqref{analyticity:dom:b}, by Lemma~\ref{obs:1}, one has
$\alpha=1+o(1)$ when $\varepsilon$ tends to $0$. So, we define a
branch of $\gamma$ analytic in this domain by the condition
$\sqrt{\frac{\alpha_\pi^*(z)}{\alpha_\pi(z)}}=1+o(1)$.\\
Then, one proves
\begin{Th} 
  \label{new-M}
  In the case of Theorem~\ref{th:M-matrices}, in the
  domain~\eqref{analyticity:dom:b}, the monodromy matrix $(z,E)\mapsto
  M^U(z,E)$ is real analytic and admits the representation:
  \begin{equation}
    \label{M-new:as}
    M^U(z,E)= P(z,E)+Q(z,E)+O\left(T_h,\,p(z)\frac{T_Y}{T_h},\,
     p(z)T_{v,0},\,p(z)T_{v,\pi}\right),
  \end{equation}
  where
  \begin{equation}
    \label{P}
    P(z,E)=\frac4{T_h}\,
    \begin{pmatrix} \tilde C_\pi(z,E)\,C_0(z,E) & -S_\pi(z,E) C_0(z,E)\\0 &
      0\end{pmatrix},
  \end{equation}
  \begin{equation}
    \label{Q}{\dsize
    Q(z,E)=\begin{pmatrix} \frac1\theta\,\cos\frac{\check\Phi_\pi-
        \check\Phi_0}\varepsilon\,+\theta\,\cos\frac{\check\Phi_\pi}
        \varepsilon\,\cos\frac{\check\Phi_0}\varepsilon& 
        -\frac1\theta\,\sin\frac{\check\Phi_\pi-\check\Phi_0}
        \varepsilon\,- \theta\,\sin\frac{\check\Phi_\pi}\varepsilon
        \,\cos\frac{\check\Phi_0}\varepsilon \\
      -\theta \sin\frac{\check\Phi_0}\varepsilon\, \tilde C_\pi(z,E) &
      \theta\,\sin\frac{\check\Phi_\pi}\varepsilon\,
      \sin\frac{\check\Phi_0}\varepsilon
    \end{pmatrix}.}
  \end{equation}
In these formulae
\begin{equation}               
    \label{CS-pi}
    \tilde C_\pi=\frac12\,\left[\tilde\alpha_\pi
      e^{i\check\Phi_\pi/\varepsilon}+ \tilde\alpha_\pi^*
      e^{-i\check\Phi_\pi/\varepsilon}\right],\quad
    S_\pi=\frac1{2i}\,\left[\tilde\alpha_\pi
      e^{i\check\Phi_\pi/\varepsilon}- \tilde\alpha_\pi^*
      e^{-i\check\Phi_\pi/\varepsilon}\right],
\end{equation}
and $(z,E)\mapsto \tilde\alpha(z,E)$ is an analytic function that
admits the asymptotics:
\begin{equation}\label{tilde-alpha:as}
 \tilde\alpha_\pi=1+T_{v,\pi}\,\left[\cos(2\pi(z-z_\pi))+i\sin(2\pi(z-h-z_\pi))\right]+O(p^2(z)T_{v,\pi}^2,\,p(z)T_Y).
\end{equation}
All the above estimates are uniform in the domain~\eqref{analyticity:dom:b}.
\end{Th}
\demo The monodromy matrix $M^U$ is analytic in the
domain~\eqref{analyticity:dom:b} as $M_\pi$ and $U$ are. As the
consistent basis in Theorem~\ref{th:M-matrices} consists of a pair of
solutions of the form $f_1=f$ and $f_2=f^*$, for $U$ given
by~\eqref{U}, formula~\eqref{U-transformation} defines two consistent
solutions of~\eqref{G.2z}, say $f^U_1$ and $f^U_2$, such that, for $x$
fixed, $(z,E)\mapsto f^U_1(x,z,E)$ and $(z,E)\mapsto f^U_2(x,z,E)$ are
real analytic.  So, the new monodromy matrix $(z,E)\mapsto M^U(z,E)$
is also real analytic.\\
Compute $M^U_{11}$. By~\eqref{U} and~\eqref{M-gauge},
\begin{equation}
  \label{new-M-11:1}
  M^U_{11}= \frac{S+S^*}{2}\quad\text{where}\quad
  S=\gamma^*(z)\left[\gamma(z+h) A_\pi(z)+
    \gamma^*(z+h) B_\pi^*(z)\right].
\end{equation}
The definition of $\gamma$ yields
\begin{equation*}
  S=\frac{\gamma(z+h)}{\gamma(z)}\,\left[A_\pi(z)+
    e^{2i\check\Phi_\pi}\frac{\alpha_\pi(z)}{\alpha_\pi^*(z)}
    B_\pi^*(z)\right].
\end{equation*}
Substituting the asymptotic representations~\eqref{A-pi:rough}
and~\eqref{B-pi:rough} into this expression, and using the real analyticity of 
$T_h$, $\check \Phi_0$, $\check \Phi_\pi$, $\theta$ and $C_0$, we get
\begin{equation}
  \label{eq:12}
  \begin{split}
    \D S&=\frac4{T_h}\,\tilde\alpha_\pi(z)\,
    e^{i\check\Phi_\pi/\varepsilon}C_0(z)
    +\frac{\gamma(z+h)}{\gamma(z)}\,
    \frac{e^{i(\check\Phi_\pi-\check\Phi_0)/\varepsilon}}2\,
    \left(\theta+\frac1\theta\right)\\ &\hskip5cm
    +\frac{\alpha_\pi(z)}{\alpha_\pi^*(z)}\,
    \frac{\gamma(z+h)}{\gamma(z)}\,
    \frac{e^{i\check\Phi_\pi/\varepsilon}}2\,
    \left(\frac1\theta\,e^{-i\check\Phi_0/\varepsilon}+\theta
      \,e^{i\check\Phi_0/\varepsilon}\right)\\
    &\hskip6cm+\frac{\gamma(z+h)}{\gamma(z)}\,{\mathcal O}+
   e^{2i\check\Phi_\pi/\varepsilon} 
    \frac{\gamma(z+h)}{\gamma(z)}\,
    \frac{\alpha_\pi(z)}{\alpha_\pi^*(z)}\,{\mathcal O},
  \end{split}
\end{equation}
where $\tilde
\alpha_\pi(z)={\dsize\frac{\gamma(z+h)}{\gamma(z)}}\,\alpha_\pi(z)$,
and $\mathcal O$ denotes $O(T_h,\,T_Y
p(z)/T_h,\,T_{v,0}p(z),\,T_{v,\pi }p(z))$.  By the estimates of
Lemma~\ref{obs:1}, from~\eqref{eq:12}, one obtains
\begin{equation*}
  S=\frac4{T_h}\,\tilde\alpha_\pi(z)\,e^{i\check\Phi_\pi/\varepsilon}C_0(z)
  +e^{i\check\Phi_\pi/\varepsilon}\left(\frac1\theta e^{-i\check\Phi_0/
      \varepsilon}+\theta\cos(\check\Phi_0/\varepsilon)\right)+\mathcal O.
\end{equation*}
Substituting this result into~\eqref{new-M-11:1}, we get the formula
announced for $M^U_{11}$ in Theorem~\ref{new-M}.\\
The other coefficients of the matrix $M^U$ are computed
analogously; so, we omit the details.\\
To complete the proof of Theorem~\ref{new-M}, it remains only to
check~\eqref{tilde-alpha:as}. Put $\alpha_{\pi,1}=\alpha_\pi-1$. By
Lemma~\ref{obs:1}, one has $\alpha_{\pi,1}=O(p T_{v,\pi})$.
Therefore,
\begin{equation}
  \label{eq:13}
  \begin{split}
  \tilde\alpha_\pi(z)&={\dsize\frac{\gamma(z+h)}{\gamma(z)}}\,\alpha_\pi(z)
  =\left(\frac{\alpha_\pi^*(z)\alpha_\pi(z)\alpha_\pi(z-h)}{\alpha_\pi^*(z-h)}\right)^{\frac12}\\
  &=1+\frac12\left(\alpha_{\pi,1}^*(z)+\alpha_{\pi,1}(z)+\alpha_{\pi,1}(z-h)-
    \alpha_{\pi,1}^*(z-h)\right)+O(p^2(z)T_{v,\pi}^2).
\end{split}                              
\end{equation}
In view of~\eqref{alpha:as}, one has $\alpha_{\pi,1}= T_{v,\pi}
e^{2\pi i(z-z_\pi(E))}+O\left(T_Y\, p(z)\right)$. Substituting this
in~\eqref{eq:13} yields~\eqref{tilde-alpha:as}. This completes the
proof of Theorem~\ref{new-M}. \qed
\smallpagebreak Finally, we note that, similarly to~\eqref{alpha:est},
one proves that
\begin{Le} 
  \label{le:4}
  Uniformly in $(z,E)$ in the domain~\eqref{analyticity:dom:b}, one
  has
  \begin{equation*}
    \label{tilde-alpha:est}
    \tilde\alpha_\pi =1+ O(T_{v,\pi}
    p(z))=1+O(e^{-2\pi(Y_m-y)/\varepsilon})
    =1+o(1).
  \end{equation*}
\end{Le}
\section{The spectrum in the  ``non-resonant'' case}
\label{sec:pl-sp}
\noindent We now prove the results on the spectrum of
$H_{z,\varepsilon}$ formulated in
Theorems~\ref{th:tib:sp:1},~\ref{th:tib:sp:1a},~\ref{th:gamma-pi},~\ref{th:tib:sp:2}
and Corollary~\ref{cor:tib:sp:3}.\\
Pick $E_*\in J$. Let $V_*$ be as in Theorem~\ref{th:M-matrices}. 
Let $J_*\subset V_*\cap\R$ be a compact interval centered at $E_*$.\\
We always assume that  $\varepsilon$ is so small that the statements of 
Theorem~\ref{new-M} and  Lemma~\ref{le:Phi'}  hold.  \\
Let $E_\pi$ be one of the points of $(E_\pi^{(l)})_l$ in $J_*$. We
assume that $E_\pi$ satisfy the non resonant condition
\begin{equation}
  \label{non-res}
  \inf_{m}|E^{(m)}_0-E_\pi|\ge 2e^{-\delta_0/\varepsilon}.
\end{equation}
In this section, we fix  $\alpha$ satisfying
\begin{equation*}
  0<\alpha<1.
\end{equation*}
and study the spectrum in the $\varepsilon^\alpha
e^{-\delta_0/\varepsilon}$-neighborhood of $E_\pi$.\\
Our main tool will be the scalar equation~\eqref{scalar-eq}; recall
that we consider the one associated to the monodromy matrix $M^U$
described in Theorem~\ref{new-M}.
\par In the sequel, we use the notations defined in
section~\ref{COo}. Now, all the symbols are uniform in $E_\pi$.
\subsection{Coefficients of the scalar equation}
\label{sec:coeff-scal-equat}
Here, we analyze the coefficients of the scalar equation for energies
$E$ satisfying
\begin{equation}
  \label{neighb}
  |E-E_\pi|<\varepsilon^\alpha e^{-\delta_0/\varepsilon}.
\end{equation}
\subsubsection{The results}
\label{sec:results}
We now define a few objects that we shall use throughout the analysis.
Let
\begin{equation}
  \label{sc-eq:1}
  \sigma_\pi=-\sin\left(\frac{\check\Phi_\pi(E_\pi) }\varepsilon\right).
\end{equation}
As $E_\pi\in\{E_\pi^{(l)}\}$, one has either $\sigma_\pi=+1$ or
$\sigma_\pi=-1$.\\
Let
\begin{equation}
  \label{sc-eq:3}
  F_\pi(E)=\sigma_\pi\,\left\{
    \frac{4}{T_h(E_\pi)}\,
    \cos\left(\frac{\check\Phi_0(E_\pi)}\varepsilon\right)\,
    \frac{\check\Phi_\pi'(E_\pi)}{\varepsilon}\,(E-E_\pi)-
    2\Lambda_n(V)\,\sin\left(\frac{\check\Phi_0(E_\pi)}
      \varepsilon\right)\right\}.
\end{equation}
The factor $\Lambda_n(V)$ is defined in~\eqref{Lambda-n-V}. The
coefficient $F_\pi(E)$ will play the role of an ``effective spectral parameter''.\\
Also, we define the factor
\begin{equation}
  \label{sc-eq:4}
  \lambda_\pi=4\sigma_\pi\, \frac{T_{v,\pi}(E_\pi)}{T_h(E_\pi)}\,
  \cos\left(\frac{\check\Phi_0(E_\pi)}\varepsilon\right).
\end{equation}
This factor will play the role of an ``effective coupling constant''.
Finally, we let
\begin{equation}
  \label{delta1}
  \delta_1=\min\left\{\delta_0,\,\left(2\pi Y-\max_{E\in J\cap V_*}
      S_h(E)\right)\right\}
\end{equation}
where $\delta_0$ is defined by~\eqref{eq:6} and $Y$ is the constant
from Theorem~\ref{th:M-matrices}.  We note that
\begin{equation*}
  \label{delta1:prop}
  0<\frac{\delta_1}{2\pi}\leq\frac{Y_m}2.
\end{equation*}
These inequalities follow from the inequalities $Y>Y_M$ and
$\delta_0\le \pi Y_m$ in which $Y_M$ and $Y_m$ are the numbers defined
by~\eqref{Y's}.\\
We prove
\begin{Pro}
  \label{new:rho-and-v}
  Let $\rho^U$ and $v^U$ be the the coefficients $\rho$ and $v$ of the
  scalar equation~\eqref{scalar-eq} corresponding to the monodromy
  matrix $M^U$.\\
  Assume that  the condition~\eqref{non-res} is satisfied.\\
  Fix $0<y<\delta_1/(2\pi)$. Then, the strip $\{|\im z|\le
  y/\varepsilon\}$, for $E$ satisfying~\eqref{neighb}, one has
  $M_{12}\ne 0$, and the coefficients $\rho^U$ and $v^U$ admit the
  following asymptotic representations

  \begin{gather}
    \label{new:rho} 
    \rho^U(z,E)=1+O\left(p(z)\varepsilon
      e^{-\delta_0/\varepsilon}\right),\\
    \label{new:v}
    \begin{split}
      v^U(z,E)&=\left\{F(E)+ \lambda_\pi\sin(2\pi(z-h-z_\pi(E_\pi)))
        +o(p(z)\lambda_\pi(E))+ O\left(p(z)
          e^{-\delta_1/\varepsilon}\right)\right\}\\
      &\hskip9.5cm\cdot\left(1+O\left(p(z)\varepsilon
          e^{-\delta_0/\varepsilon}\right)\right).
    \end{split}
  \end{gather}
  Here, the function $E\mapsto F(E)$ is independent of $z$; $F(E)$ and
  $F'(E)$ admit the asymptotic representations:
  \begin{equation}
    \label{F-F0}
    F(E)=F_\pi(E)(1+o(1))+o(1),\quad{\rm and}\quad
      F'(E)=F_\pi'(E)(1+o(1)).
  \end{equation}
\end{Pro}
\smallpagebreak We often shall use simplified versions
of~\eqref{new:rho} and~\eqref{new:v}, namely
\begin{Cor}
  \label{Cor:v:simple} 
  In the case of Proposition~\eqref{new:rho-and-v}, one has
  \begin{gather}
    \label{v:simple}
    \rho^U(z,E)=1+o(1),\\
    \label{eq:20} v^U(z,E)=
    \left\{F_\pi(E)+\lambda_\pi\cos(2\pi(z-h-z_\pi(E_\pi)))+
      o(\lambda_{\pi}p(z))+ o(1)\right\}\,(1+o(1)).
  \end{gather}
\end{Cor}
\demo For $0<y<\delta_1/(2\pi)$ and $|\im z|\le y/\varepsilon$, one
has
\begin{equation}
  \label{p-delta}
  p(z)e^{-\delta_0/\varepsilon}+p(z)e^{-\delta_1/\varepsilon}
  \leq e^{-(\delta_1-2\pi y)/\varepsilon}.
\end{equation}
Representation~\eqref{v:simple} is obtained from~\eqref{new:rho} by
means of~\eqref{p-delta}.  Representation~\eqref{eq:20} is obtained
from~\eqref{new:v} by means of~\eqref{p-delta} and~\eqref{F-F0}. \qed
\subsubsection{Proof of Proposition~\ref{new:rho-and-v}}
\label{sec:proof-prop-refn}
Let us begin with three lemmas. First, we collect simple observations
following from Taylor's formula. One has
\begin{Le}
  \label{new:v:factors}
  For $\varepsilon$ sufficiently small, for all $E$
  satisfying~\eqref{neighb}, for $\nu\in\{0,\pi\}$, one has
  \begin{gather}
 \label{cos-Phi-0:2}
    \left|\cos\left(\check\Phi_\pi(E)/\varepsilon\right)\right|
    \leq C\varepsilon^{\alpha-1}e^{-\delta_0/\varepsilon},\\
  \label{cos-Phi-pi}
   \cos\left(\check\Phi_\pi(E)/\varepsilon\right)
   =\sigma_\pi\,\varepsilon^{-1}\,\check\Phi_\pi'(E_\pi)\,
   (E-E_\pi)[1+O(\varepsilon^{\alpha-1}e^{-\delta_0/\varepsilon})],\\
 \label{sin-Phi-0}
    \sin\left(\check\Phi_\nu(E)/\varepsilon\right)=
    \sin\left(\check\Phi_\nu(E_\pi)/\varepsilon\right)+
    O(\varepsilon^{\alpha-1}e^{-\delta_0/\varepsilon}),\\
    \label{cos-Phi-0:3}
    \left|\cos\left(\check\Phi_0(E)/\varepsilon\right)\right|
    \geq  C\varepsilon^{-1}e^{-\delta_0/\varepsilon}\\
    \label{cos-Phi-0:1}
    \cos(\check\Phi_0(E)/\varepsilon)=
    \cos(\check\Phi_0(E_\pi)/\varepsilon)
    \,\left(1+O(\varepsilon^\alpha)\right),\\
  \label{pl-sp:5}
  T_h(E)=T_h(E_\pi)(1+O(\varepsilon^{\alpha-1}
  e^{-\delta_0/\varepsilon})),\quad T_{v,\nu}(E)=T_{v,\nu}(E_\pi)
  (1+O(\varepsilon^{\alpha-1}e^{-\delta_0/\varepsilon})).
  \end{gather}
\end{Le}
\demo These results follow from the Taylor formula. When proving the
first five results, one uses~\eqref{Phi':up} and~\eqref{Phi':down} and
has to keep in mind the definitions of $E_0$ and $E_\pi$. We omit the
elementary details.\\
The two estimates~\eqref{pl-sp:5} are proved in one and the same way.
We prove only the first one. Therefore, one uses the Taylor formula
for $\log T_h\,(E)$ in the neighborhood~\eqref{neighb} of $E_\pi$.
By~\eqref{T:as} and the definition of $t_h$, one has $\log
T_h(E)=-\frac1{2\varepsilon}S_h(E)+g(E)$, where $g(E)=o(1)$ uniformly
in $V_*$. The estimates $|S_h'(E)|\le C$ and $\frac{dg}{dE}=o(1)$ hold
uniformly on any fixed compact of $V_*$ (the last estimate follows
from the Cauchy estimates). This implies that, for $E$ in a fixed
compact of $V_*$,
\begin{equation}
  \label{T_h-prime}
  |T_h'(E)|\le C \varepsilon^{-1}|T_h(E)|,
\end{equation}
and this estimate implies the estimate for $T_h$ from~\eqref{pl-sp:5}.
This completes the proof of Lemma~\ref{new:v:factors}.\qed\\
We also prepare simplified representations for factors $C_0$, $S_\pi$
and $\tilde C_\pi$ defined in~\eqref{C-def} and~\eqref{CS-pi}. We
prove
\begin{Le}
  \label{CCS:simple} Fix $y$ as in Proposition~\ref{new:rho-and-v}. 
  Under condition~\eqref{non-res}, for $|\im z|\le y/\varepsilon$ and
  $E$ satisfying~\eqref{neighb}, one has
\begin{gather}
  \label{c0:1}
  C_0=\cos\left(\check\Phi_0(E)/\varepsilon\right)
  (1+O\left(p(z)\varepsilon e^{-\delta_0/\varepsilon}\right)),\\
  \label{cpi} \tilde C_\pi=\cos\left(\check\Phi_\pi(E)/\varepsilon\right)+\sigma_\pi
  T_{v,\pi}(E_\pi)\,
  \sin(2\pi(z-h-z_\pi(E_\pi))+ o(pT_{v,\pi}(E_\pi)),\\
  \label{spi}
  S_\pi=\sin\left(\check\Phi_\pi(E)/\varepsilon\right)
  (1+O(pe^{-2\delta_0/\varepsilon})).
\end{gather}
\end{Le}
\demo The definitions of $C_0$ and $S_\pi$,~\eqref{C-def}
and~\eqref{CS-pi}, and~\eqref{check-phi:est},~\eqref{alpha:est} imply
that
\begin{equation}
  \label{c0:1:1} 
  C_0=\cos\left(\check\Phi_0(E)/\varepsilon\right)+O(p T_{v,0})
  \quad{\rm and}\quad
  S_\pi=\sin\left(\check\Phi_\pi(E)/\varepsilon\right)+O(p T_{v,\pi}).
\end{equation}
Representation~\eqref{c0:1} follows from~\eqref{c0:1:1}, from
estimate~\eqref{cos-Phi-0:3} and from~\eqref{T:est:2}. Similarly,
\eqref{spi} follows from~\eqref{c0:1:1},~\eqref{cos-Phi-0:2}
and~\eqref{T:est:2}.\\
Prove~\eqref{cpi}. The definition of $\tilde C_\pi$,~\eqref{CS-pi},
and representation~\eqref{tilde-alpha:as} imply that
\begin{equation*}
   \dsize \tilde C_\pi=\cos\left(\frac{\check\Phi_\pi(E)}\varepsilon
   \right)+T_{v,\pi}(E)\,\left[ \cos\left(\frac{\check\Phi_\pi(E)}
       \varepsilon\right)\,c(z)-\sin\left(\frac{\check\Phi_\pi(E)}
       \varepsilon\right)\,s(z)\right]+O(p^2T_{v,\pi}^2,\,pT_Y). 
\end{equation*}
where $s(z)= \sin(2\pi(z-h-z_\pi(E))$ and
$c(z)=\cos(2\pi(z-z_\pi(E))$. Now, representation~\eqref{cpi} follows
from~\eqref{cos-Phi-0:2}, from~\eqref{pl-sp:5}
and from estimates~\eqref{T:est:1} and~\eqref{T:est:3}.\qed
\smallpagebreak Turn to the proof of Proposition~\ref{new:rho-and-v}.
Compute $\rho^U$. By~\eqref{M-new:as}, we have
\begin{equation}\label{rho-U:1}
M^U_{12}= P_{12}+Q_{12}+R_{12},\quad R_{12}=O\left(T_h,\,p(z)\frac{T_Y}{T_h},\,
     p(z)T_{v,0},\,p(z)T_{v,\pi}\right).
\end{equation}
Show that, for $E$ satisfying~\eqref{neighb} and $|\im z|\le y/\varepsilon$, one has  
\begin{equation}\label{rho-U:2}
P_{12}=\frac{4}{T_h}\,\sin(\check\Phi_\pi(E)/\varepsilon)
  \cos(\check\Phi_0(E)/\varepsilon)\,
  \left(1+O(p(z)\varepsilon e^{-\delta_0/\varepsilon})\right),\quad 
  |Q_{12}|+|R_{12}|\le  C.
\end{equation}
The estimate for $P_{12}$ follows from Lemma~\ref{CCS:simple}. The
estimate for $Q_{12}$ follows from~\eqref{theta:est}
and~\eqref{check-phi:est}. Check the estimate for $R_{12}$.
By~\eqref{T:est:2} and the definition of $\delta_1$, one has $|
R_{12}|\le C\,p(z) e^{-\delta_1/\varepsilon}$.  Recall that $p=e^{2\pi
  |\im z|}$. As $y<\delta_1/(2\pi)$, for $|\im z|\le y/\varepsilon$ we
get
\begin{equation}
  \label{rho-U:4}
  p(z) e^{-\delta_1/\varepsilon}\le  e^{-(\delta_1-2\pi
  y)/\varepsilon} \le C.
\end{equation}
This implies the announced estimate for $R_{12}$ and completes the
proof of~\eqref{rho-U:2}.\\
For $E$ satisfying~\eqref{neighb}, as $E_\pi$ satisfies~\eqref{eq:11},
for $\varepsilon$ sufficiently small, one has
$|\sin(\check\Phi_\pi(E)/\varepsilon)|\ge 1/2$ ;
taking~\eqref{T:est:2} and~\eqref{cos-Phi-0:3} into account, we get
\begin{equation*}
  \left|\frac{4}{T_h}\,\sin(\check\Phi_\pi(E)/\varepsilon)
    \cos(\check\Phi_0(E)/\varepsilon)\right|^{-1}\le C\varepsilon
    e^{-\delta_0/\varepsilon}.
\end{equation*}
From this,~\eqref{rho-U:1} and~\eqref{rho-U:2}, one deduces
\begin{equation}
  \label{rho-U:5}
  M^U_{12}=\frac{4}{T_h}\,\sin(\check\Phi_\pi(E)/\varepsilon)
  \cos(\check\Phi_0(E)/\varepsilon)\,
  \left(1+O(p(z)\varepsilon e^{-\delta_0/\varepsilon})\right).
\end{equation}
In view of~\eqref{rho-U:4}, there exists $\varepsilon_0>0$ such that,
for $0<\varepsilon<\varepsilon_0$, the error term in~\eqref{rho-U:5}
be smaller than $1/2$. From now on, we assume that
$0<\varepsilon<\varepsilon_0$.  Then, we get $M_{12}^U\ne 0$, and, as
$\rho^U(z)=M^U_{12}(z)/M^U_{12}(z-h)$, the
representation~\eqref{rho-U:5} implies~\eqref{new:rho}.
\smallpagebreak Now, let us compute $v^U$.  Note that $
v^U(z,E)=M^U_{11}(z,E)+M^U_{22}(z-h,E)+(\rho^U(z,E)-1)
M^U_{22}(z-h,E)$.  Using the representations~\eqref{M-new:as},
\eqref{P} and~\eqref{Q}, we transform this expression to
\begin{gather}
  \label{v:1}
  v^U(z,E)=P_{11}(z,E)+(Q_{11}(E)+Q_{22}(E))+R(z,E),\\
  R=(\rho^U(z,E)-1) (Q_{22}(E)+r_1(z,E))+ r_2(z,E),\\
  \dsize r_j(z,E)=O\left(T_h,\,p(z)\frac{T_Y}{T_h},\,p(z)T_{v,0},
    \,p(z)T_{v,\pi}\right)\quad \text{ for }j\in\{1,2\}.
\end{gather} 
We now show that
\begin{gather}
  \label{v:2}\dsize
  P_{11}(z,E)=\left(\tilde F(E)+\lambda_\pi
    s(z)+o(p\lambda_\pi)\right)\,\left(1+g(z,E)\right), \\
  \intertext{where}
  \label{v:3}
  \tilde F(E)=\frac{4\cos\frac{\check\Phi_0(E)}\varepsilon\,
    \cos\frac{\check\Phi_\pi(E)}\varepsilon}{T_h(E)},\quad
  s(z)=\sin(2\pi(z-h-z_\pi(E_\pi))),\quad |g(z,E)|\le C\,
  p(z)\varepsilon e^{-\delta_0/\varepsilon},\\
  \intertext{and that}
  \label{v:4}\dsize
  |Q_{11}(E)|+|Q_{22}(E)|\le C,\quad\quad |R(z,E)|\le C\,p(z)
  e^{-\delta_1/\varepsilon},\\
  \label{v:4a}
  |g(z,E)|\le C\varepsilon,\quad |R(z,E)|\le C.
\end{gather}
Lemma~\ref{CCS:simple} implies that
\begin{equation}
  \label{eq:14}
  \begin{split}
    \dsize
    P_{11}(z,E)=& \dsize \frac{4C_0(z,E)\tilde C_\pi(z,E)}{T_h(E)} \\
    =& \dsize \frac{4}{T_h(E)}
    \cdot\cos\frac{\check\Phi_0(E)}\varepsilon\,
    \left(1+O\left(p(z)\varepsilon
        e^{-\delta_0/\varepsilon}\right)\right)\,
    \left(\cos\frac{\check\Phi_\pi(E)}\varepsilon+\sigma_\pi
      T_{v,\pi}s(z)
      +o(p(z)T_{v,\pi})\right)\\
    =& \dsize\left( \tilde F(E)+\tilde\lambda(E)s(z)+o(p(z)\tilde
      \lambda(E))\right)\, \left(1+O\left(p(z)\varepsilon
        e^{-\delta_0/\varepsilon}\right)\right),
  \end{split}                              
\end{equation}
where
\begin{equation*}
  T_{v,\pi}=T_{v,\pi}(E_\pi), \quad 
  \tilde\lambda(E)=\frac{4\sigma_\pi T_{v,\pi}}{T_h(E)}
  \cdot\cos\frac{\check\Phi_0(E)}\varepsilon.
\end{equation*}
In view of~\eqref{cos-Phi-0:1} and~\eqref{pl-sp:5}, we have
$\tilde\lambda(E)=\lambda_\pi(1+o(1))$. This and~\eqref{eq:14}
imply~\eqref{v:2} and~\eqref{v:3}.\\
The first estimate in~\eqref{v:4} is proved in the same way as the
second estimate in~\eqref{rho-U:2}.\\
Prove the second estimate in~\eqref{v:4}. As when proving the third
estimate in~\eqref{rho-U:2}, one checks that, for $j\in\{1,2\}$,
$|r_j|\le C\,p e^{-\delta_1/\varepsilon}$ and $ | r_j|\le C$. Recall
that $|Q_{22}|\le C$. These observations and~\eqref{new:rho} imply
that $|R|\le C\,|\rho^U(z,E)-1|+ |r_2|\le C\,p
e^{-\delta_1/\varepsilon} $.\\
The ``rough'' estimates~\eqref{v:4a} follow from the already obtained
and~\eqref{rho-U:4}. This completes the proof of~\eqref{v:2}
--~\eqref{v:4a}.\\
Now, assume that $\varepsilon$ is so small that $|g(z,E)|<1/2$ for all
$z$ and $E$ in the case of Proposition~\ref{new:rho-and-v}.  This is
possible in view of~\eqref{v:4a}. Then, substituting
representation~\eqref{v:2} into~\eqref{v:1}, and taking into
account~\eqref{v:4}, we get
\begin{align*}
  v^U &=\left[\tilde F(E)+\lambda_\pi s(z)+o(p\lambda_\pi)+
    \frac{Q_{11}(E)+Q_{22}(E)+R(z,E)}{1+g(z,E)}
  \right]\,\left(1+g(z,E)\right)\\
  &=\left[F(E)+\lambda_\pi s(z)+o(p\lambda_\pi)+ O(R(z,E))+
    O(g(z,E))\right]\,\left(1+g(z,E)\right)
\end{align*}
with
\begin{equation*}
  F(E)=\tilde F(E)+(Q_{11}(E)+Q_{22}(E)).
\end{equation*}
In view of~\eqref{v:3} and~\eqref{v:4}, this implies~\eqref{new:v}.\\
Now, we only have to check~\eqref{F-F0} to complete the proof of
Proposition~\ref{new:rho-and-v}. For sufficiently small $\varepsilon$,
the representation for $F$ in~\eqref{F-F0} follows from
\begin{gather}
  \label{eq:18}
  \tilde F(E)=4\,\sigma_\pi\,(T_h(E_\pi))^{-1}\,
  \cos\left(\frac{\check\Phi_0(E_\pi)}\varepsilon\right)\,
  \frac{\check\Phi_\pi'(E_\pi)}{\varepsilon}\,(E-E_\pi)\,(1+o(1)),\\
    \label{eq:23}
    Q_{11}(E)+Q_{22}(E)=
    -2\sigma_\pi\,\Lambda_n(V)\,\sin\left(\frac{\check\Phi_0(E_\pi)}
      \varepsilon\right)+o(1).
\end{gather}
The formula~\eqref{eq:18}~follows
from~\eqref{cos-Phi-0:1},~\eqref{cos-Phi-pi} and~\eqref{pl-sp:5}.  To
prove~\eqref{eq:23}, we note that, by~\eqref{Q},
\begin{equation*}
  Q_{11}(E)+Q_{22}(E)=(\theta+1/\theta)\,\cos\left((\check\Phi_\pi-\check\Phi_0)/
  \varepsilon\right).
\end{equation*}
This in conjunction
with~\eqref{sin-Phi-0},~\eqref{cos-Phi-0:2},~\eqref{theta:as},~\eqref{Lambda-n-V}
and~\eqref{sc-eq:1}  yields~\eqref{eq:23}.\\
Finally, the asymptotics for $F'$ in~\eqref{F-F0} follows from
\begin{equation}
  \label{F':est}
  \tilde F'(E)=F'_\pi\,(1+o(1)),\quad |F_\pi'|\ge C\varepsilon^{-2}
  e^{\delta_0/\varepsilon},\quad |Q'_{11}(E)|+|Q'_{22}(E)|\le
  C\varepsilon^{-1}.
\end{equation} 
Prove the first of these estimates. It follows from
Lemma~\ref{le:Phi'} and
estimates~\eqref{T_h-prime},~\eqref{cos-Phi-0:2}
and~\eqref{cos-Phi-0:3} that
\begin{equation*}
  \tilde F'(E)=-\frac{4\cos\frac{\check\Phi_0(E)}\varepsilon\,
    \sin\frac{\check\Phi_\pi(E)}\varepsilon}{T_h(E)}\,
  \frac{\check\Phi_\pi'(E)}\varepsilon\,(1+o(\varepsilon^\alpha)).
\end{equation*}
Now, using~\eqref{cos-Phi-0:1},~\eqref{sin-Phi-0} for
$\nu=\pi$,~\eqref{pl-sp:5} and the estimate
$\check\Phi_\pi'(E)=\check\Phi_\pi'(E_\pi)(1+o(1))$ (following from
Lemma~\ref{le:Phi'}), we get
\begin{equation*}
  \tilde
  F'(E)=\frac{4\sigma_\pi\cos\frac{\check\Phi_0(E_\pi)}\varepsilon}
  {T_h(E_\pi)}\,\frac{\check\Phi_\pi'(E_\pi)}\varepsilon\,(1+o(1)).
\end{equation*}
This and the definition of $F_\pi$ imply the representation for $F'$
in~\eqref{F':est}. The estimate for $|F_\pi'|$ follows from the
definition of $F_\pi$ and the estimates~\eqref{cos-Phi-0:3},
\eqref{T:est:2} and~\eqref{Phi':down}. The last estimate
in~\eqref{F':est} follows from~\eqref{Phi':up},~\eqref{theta:est}
and the Cauchy estimates for $E\mapsto\theta(E)$.\\
This completes the proof of Proposition~\ref{new:rho-and-v}.\qed
\subsection{The location of the spectrum}
\label{sec:place-spectrum}
We now prove Theorem~\ref{th:tib:sp:1}. Therefore, we apply
Proposition~\ref{resolvent-set} to the scalar equation with the
coefficients $\rho^U$ and $v^U$ computed in
section~\ref{sec:coeff-scal-equat}.\\ 
Let $J^\varepsilon_*$ the subinterval of $J$ described
by~\eqref{neighb}. One has
\begin{Le}
  \label{le:place}
  The spectrum of $H_{z,\varepsilon}$ in $J^\varepsilon_*$ is
  contained in the interval described by
    \begin{equation}
      \label{place:0}
      |F_\pi(E)|\leq\left(2+
        \left|\lambda_{\pi}\right|\right)\,(1+o(1)),
    \end{equation}
    where $o(1)$ is independent of $E$ and $E_\pi$
    (satisfying~\eqref{non-res}).
\end{Le}
\demo First, we find $r$, a subset of $J^\varepsilon_*$, where $M^U$,
$v^U$ and $\rho^U$ satisfy the assumptions of
Proposition~\ref{resolvent-set}. Hence, $r$ is in the resolvent set
of~\eqref{family}.
\\
Recall that $(z,E)\mapsto\rho^U(z,E)$ and $(z,E)\mapsto v^U(z,E)$ are
real analytic as the matrix $(z,E)\mapsto M^U(z,E)$ is. Therefore, the
equalities $\ind\rho^U(\cdot,E)=0$ and $\ind v^U(\cdot,E)=0$
automatically follow from the inequalities $\D\min_{z\in\R}
|\rho^U(z,E)|>0$ and $\D\max_{z\in\R} |\rho^U(z,E)|<\frac14
\left(\min_{z\in\R} |v^U(z,E)|\right)^2$.\\
Furthermore, by~\eqref{v:simple}, the first of these inequalities is
satisfied for all $E\in J^\varepsilon_*$. So, in $J_*^\varepsilon$,
the assumptions of Proposition~\ref{resolvent-set} are satisfied if
and only if $\D\max_{z\in\R} |\rho^U(z,E)|^{1/2}<
\frac12\,\min_{z\in\R} |v^U(z,E)|$.\\
Corollary~\ref{Cor:v:simple} yields
\begin{gather}
\max_{z\in\R} |\rho^U(z)|^{1/2}\le 1+o(1),\\
\frac12\,\min_{z\in\R} |v^U(z)|\ge
\frac{1+o(1)}2\,\left(\min_{x\in\R}\left|F_\pi(E)+
    \lambda_\pi\sin(x)\right|+o(1)(1+|\lambda_\pi|)\right),
\end{gather}
where $o(1)$ is independent of $E$ and $E_\pi$. So, $v^U$ and $\rho^U$
satisfy the assumptions of Proposition~\ref{resolvent-set} if $E$
satisfies the inequality of the form $|F_\pi(E)|\ge
\left(2+|\lambda_{\pi}|\right)\,(1+o(1))$, where $o(1)$ is independent
of $E$ and $E_\pi$.  Now, Proposition~\ref{resolvent-set} implies the
statement of Lemma~\ref{le:place}.\qed
\smallpagebreak Lemma~\ref{le:place} and the definitions of
$\lambda_\pi$ and $F_\pi$, namely~\eqref{sc-eq:3} and~\eqref{sc-eq:4},
imply that, in $J^\varepsilon_*$, the spectrum of $H_{z,\varepsilon}$
is contained in $\check I_\pi$, the interval described by
\begin{equation*}
  \left|\frac2\varepsilon\,\check\Phi_\pi'(E_\pi)\,(E-E_\pi)-
    \Lambda_n(V)\,T_h(E_\pi)\,\tan\left(\frac{\check\Phi_0(E_\pi)}
      \varepsilon\right)\right|\le
  2\left(\frac{T_h(E_\pi)}{2\left|\cos\left(\frac{\check\Phi_0(E_\pi)}
          \varepsilon\right)\right|}+T_{v,\pi}(E_\pi)\right)\,(1+o(1)),
\end{equation*}
where $o(1)$ depends only on $\varepsilon$. The interval $\check
I_\pi$ is centered at the point
\begin{equation}\label{centre}
  \check E_\pi=E_\pi+\frac{\varepsilon\,\Lambda_n(V)\,T_h(E_\pi)}
  {2\check\Phi_\pi'(E_\pi)}
  \,\tan\left(\frac{\check\Phi_0(E_\pi)}\varepsilon\right),
\end{equation}
and it has the length
\begin{equation}
  \label{length}
  |\check I_\pi|= \frac{2\varepsilon}{\check\Phi_\pi'(E_\pi)}\,
  \left(\frac{T_h(E_\pi)}{2\left|\cos
        \left(\frac{\check\Phi_0(E_\pi)}\varepsilon\right)\right|}+ 
    T_{v,\pi}(E_\pi)\right)\,(1+o(1)).
\end{equation}
This completes the proof of Theorem~\ref{th:tib:sp:1}.\qed\\
Note that
\begin{equation}\label{check}
  |E_\pi-\check E_\pi|+|\check I_\pi|\le C\varepsilon e^{-\delta_0/\varepsilon}.
\end{equation}
These estimates follow from~\eqref{centre},~\eqref{length} and
estimates~\eqref{Phi':down},~\eqref{T:est:2}
and~\eqref{cos-Phi-0:3}.\\
Finally, we note that, using~\eqref{centre}, one can
rewrite~\eqref{sc-eq:3} as
\begin{equation}
  \label{F0:simple}
  F_\pi(E)=
  \frac{4\sigma_\pi}{T_h(E_\pi)}\,\cos\left(\frac{\check\Phi_0(E_\pi)}
    \varepsilon\right)\,                           
  \frac{\check\Phi_\pi'(E_\pi)}{\varepsilon}\,(E-\check E_\pi).
\end{equation}
\subsection{Computation of the integrated density of states}
\label{sec:comp-integr-dens}
We now compute the increment of the integrated density of states on
the intervals described in Theorem~\ref{th:tib:sp:1} and, thus, prove
Theorem~\ref{th:tib:sp:1a}. We use the approach developed
in~\cite{MR2003f:82043}. One has
\begin{Pro}
  \label{pro:ids:1} 
  Pick two points $a<b$ of the real axis. Let $\gamma$ be a
  continuous curve in $\C_+$ connecting $a$ and $b$.\\
  Assume that, for all $E\in\gamma$, one can construct a consistent basis
  such that the corresponding monodromy matrix is continuous in
  $(z,E)\in \R\times \gamma$ and satisfies the
  conditions
  \begin{equation}
    \label{resolvent-set:cond:1}
    \min_{z\in\R} |M_{12}|>0,\quad \max_{z\in\R} |\rho(z)|< 
    \left(\frac12\,\min_{z\in\R} |v(z)|\right)^2,
  \end{equation}
  where $\rho$ and $v$ are defined by~\eqref{rho,v} with  $h$ from~\eqref{h}.
  Assume in addition that 
  the coefficients of $M$ are real for real $E$ and $z$.
  Then, one has
  \begin{equation}
    \label{DeltaN}
    N(b)-N(a)=-\left.\frac\varepsilon{2\pi^{2}} \int_0^1\arg
      v(z,E)\,dz\right|_{\gamma},
  \end{equation}
  where $f|_{\gamma}$ denotes the increment of $f$ when going from $a$
  to $b$ along $\gamma$.
\end{Pro}
\demo In~\cite{MR2003f:82043}, we proved a more general result; we
assumed that, for all $(z,E)\in \R\times\gamma$, the monodromy matrix
satisfies the conditions of Lemma~\ref{resolvent-set} and got the
formula
\begin{equation}
  \label{DeltaN0}
  N(b)-N(a)=-\left.\frac\varepsilon{2\pi^{2}} \int_0^1\arg
    G(z,E)\,dz\right|_{\gamma},
\end{equation}
where $G$ is the continued fraction
\begin{equation}\label{cf:G}
  G(z)=   v\,(z) -\frac{\dsize \rho(z)}   {\dsize
    v\,(z-h) -\frac{\dsize  \rho(z-h)} {\dsize
      v\,(z-2h) -\frac{\dsize \rho(z-2h)}{\dsize
        \quad\cdots}}}.
\end{equation}
Such continued fractions were studied in~\cite{MR96m:47060}. It was
proved that, if the functions $z\mapsto\rho(z)$ and $z\mapsto v(z)$
are continuous and $1$-periodic and if they satisfy the
conditions~\eqref{resolvent-set:cond}, then
\begin{itemize}
\item the continued fraction $z\mapsto G(z)$ converges to a continuous
  $1$-periodic function uniformly in $\R$;
\item if $\rho$ and $v$ depend on a parameter $E$, if they are
  continuous in $(z,E)$ in some domain $D$, and if, for all $(z,E)\in
  D$, they satisfy conditions~\eqref{resolvent-set:cond}, then
  $(z,E)\mapsto G(z,E)$ is also continuous in $D$.
\item for $ z\in\R$, one has
  \begin{equation}
    \label{Gestimate:0}
    |G(z)-v(z)|<\frac12\, \min_{z\in\R}|v(z)|-
    \sqrt{\left(\frac12\,\min_{z\in \R}|v(z)|\right)^2
      -\max_{z_\in\R}|\rho(z)|};
  \end{equation}
\end{itemize}
Now, turn to the proof of~\eqref{DeltaN}. As, in our case, $v(z,E)$
and $\rho(z,E)$ are real for real $z$ and $E$, we conclude that (1)
$\ind v=\ind \rho=0$ (which follows
from~\eqref{resolvent-set:cond:1}); (2) the right hand sides in
both~\eqref{DeltaN} and~\eqref{DeltaN0} belong to
$\varepsilon/2\pi\Z$. The first observation
and~\eqref{resolvent-set:cond:1} imply that, for all $(z,E)\in
\R\times\gamma$, the monodromy matrix satisfies the conditions of
Lemma~\ref{resolvent-set}.  In view of the second observation,
formula~\eqref{DeltaN} follows from~\eqref{DeltaN0}, the continuity of
$(z,E)\mapsto G(z,E)/v^U(z,E)$ and the inequality $\sup_{z\in
  \R}\frac{|G(z,E)-v^U(z,E)|}{|v^U(z,E)|}<1$ valid for all
$E\in\gamma$. And, the last one follows from~\eqref{Gestimate:0} and
the second condition from~\eqref{resolvent-set:cond:1}:
\begin{equation*}                                
\sup_{z\in \R, \ E\in \gamma}\frac{|G(z,E)-v^U(z,E)|}{|v^U(z,E)|}<  
2\,\frac{\max_{z\in\R}|\rho^U(z,E)|}{\min_{z\in\R}|v^U(z,E)|^2} <1.
\end{equation*}
This  completes the  proof of Proposition~\ref{pro:ids:1}.\qed  
\subsubsection{The computation}
\label{sec:computation}
Let $E_\pi$ be as in the beginning of section~\ref{sec:pl-sp} and, in
particular, be such that~\eqref{non-res} is satisfied. As above,
let $J^\varepsilon_*$, be the subinterval of $J$ described by~\eqref{neighb}.\\
As seen in the previous section, in $J^\varepsilon_*$, the spectrum of
$H_{z,\varepsilon}$ is contained in $\check I_\pi$, the interval
centered at $\check E_\pi$ (see~\eqref{centre}) of length
$|\check I_\pi|$ (see~\eqref{length}).\\
To compute the increment of the integrated density of states on
$\check I_\pi$, we use Proposition~\ref{pro:ids:1} and choose:
\begin{equation*}
  \gamma=\left\{E\in\C_+: \ |E-\check E_\pi|=\frac12
  \varepsilon^{\alpha}e^{-\delta_0/\varepsilon}\right\}.
\end{equation*}
Let $a<b$ be the ends of $\gamma$. Then, by~\eqref{check}, one has
$\check I_\pi\subset(a,b)$. We prove
\begin{Le}
  \label{le:ids:1}
  On $\gamma$,  the monodromy matrix $M^U$ and the functions
  $\rho^U$ and $v^U$ satisfy the conditions~\eqref{resolvent-set:cond:1}.
\end{Le}
\smallpagebreak Recall that the integrated density of states of
$H_{z,\varepsilon}$ is constant outside the spectrum of
$H_{z,\varepsilon}$. So, its increment on $\check I_\pi$ is equal to
its increment between the ends of the semi-circle $\gamma$.  And, in
view of Lemma~\ref{le:ids:1}, the latter is given by the
formula~\eqref{DeltaN}. In view of this formula, to prove
Theorem~\ref{th:tib:sp:1a}, it suffices to check that
$\left.\int_0^1\arg v^U(z,E)\,dz \right|_{\gamma}=-\pi$. This follows
from
\begin{Le}
  \label{le:ids:3} For $(z,E)\in\R\times \gamma$, one has
  \begin{equation}
    \label{eq:34}
    v^U(z,E)=F_\pi(E)(1+o(1)).
  \end{equation}
\end{Le}
\noindent Indeed, note that for $z\in\R$ and $E\in\R$, the
functions $F_\pi$ and $v^U$ take real values. Therefore, the estimate
of Lemma~\ref{le:ids:3} implies that $\left.\int_0^1\arg
  v^U(z,E)\,dz\right|_{\gamma}=\left.\arg F_\pi(E) \right|_{\gamma}$.
In view of~\eqref{F0:simple}, the last quantity is equal to
$\left.(E-\check E_\pi) \right|_{\gamma}=-\pi$.  So, to complete the
proof of Theorem~\ref{th:tib:sp:1a}, we have only to check
Lemmas~\ref{le:ids:1} and~\ref{le:ids:3}. They will follow from
\begin{Le}
  \label{le:ids:2} For $(z,E)\in\R\times \gamma$, one has
\begin{equation}
  \label{ids:3}
  |F_\pi(E)|\ge C\varepsilon^{\alpha-2}.
\end{equation}
\end{Le}
\demo The lower bound for $|F_\pi(E)|$ follows from~\eqref{F0:simple},
the definition of $\gamma$ and the estimates~\eqref{T:est:2},
\eqref{cos-Phi-0:3} and~\eqref{Phi':down}.\qed\\
\smallpagebreak {\it Proof of Lemmas~\ref{le:ids:3}.\/} Prove the
asymptotic representation for $v^U$.  Therefore, we first derive an
upper bound for the ratio $\lambda_\pi/F_\pi(E)$.  By~\eqref{sc-eq:4}
and~\eqref{F0:simple}, we get $|\lambda_\pi/F_\pi(E)|=
\frac{\varepsilon T_{v,\pi}(E_\pi)}{\check\Phi_\pi'(E_\pi)|E-\check
  E_\pi|}$.  Now, the definition of $\gamma$ and the
estimates~\eqref{T:est:2} and~\eqref{Phi':down} imply that
\begin{equation}\label{lambda-F}
  |\lambda_\pi/F_\pi(E)|\le
  C\,\varepsilon^{1-\alpha}\,e^{-\delta_0/\varepsilon}. 
\end{equation}
So, the ratio is small when $\varepsilon$ tends to $0$. The
representation~\eqref{eq:34} follows
from~\eqref{eq:20},~\eqref{lambda-F} and~\eqref{ids:3}. This completes
the proof of Lemma~\ref{le:ids:2}.\qed
\smallpagebreak {\it Proof of Lemmas~\ref{le:ids:1}.\/} 
By Proposition~\ref{new:rho-and-v}, for sufficiently small
$\varepsilon$, for $E\in\gamma$ and
$z\in\R$, one has $M_{12}^U(z,E)\ne 0$. 
Finally, for sufficiently small $\varepsilon$, for all $E\in\gamma$,
from~\eqref{ids:3} and~\eqref{new:rho}, it follows that
$\D\max_{z\in\R}|\rho(z,E)|<\frac14\min_{z\in\R}|v(z,E)|^2$. This
completes the proof of Lemma~\ref{le:ids:1}.\qed
\subsection{Computation of the Lyapunov exponent}
\label{sec:comp-lyap-expon}
We now derive the asymptotics of the Lyapunov on the interval $\check
I_\pi$, i.e., prove formula~\eqref{gamma-pi}, and, thus, prove
Theorem~\ref{th:gamma-pi}.
\subsubsection{Preliminaries}
\label{sec:preliminaries-1}
To compute $\Theta(E,\varepsilon)$, we use Theorem~\ref{Lyapunov} and
compute the Lyapunov exponent of the matrix cocycle defined by the
monodromy matrix $M^U(\cdot,E)$. It appears to be difficult to compute
directly $\theta(M^U(\cdot,E),h)$: one can obtain only rough results.
However, using the scalar equation with the coefficients $v^U$ and
$\rho^U$, one can construct another matrix cocycle that has the same
Lyapunov exponent as $(M^U(\cdot,E),h)$ and for which the computations
become much simpler.
\subsubsection{The Lyapunov exponent and the scalar equation}
\label{sec:lyap-expon-scal}
In this section, we assume $z\mapsto M(z)$ to be a $1$-periodic,
$SL(2,\R)$-valued, bounded measurable function of the real variable
$z$. Let $h$ is a positive irrational number. We check the following
simple
\begin{Le}
  \label{le:Le:1} 
 Assume that there exists $A>1$ such that
  \begin{equation}
    \label{Le:1}
    \forall z\in\R,\quad A^{-1} \le M_{12}(z)\le A.
  \end{equation}
  In terms of $M$ and $h$, construct $v$ and $\rho$ by formulae~\eqref{rho,v}. Set
  \begin{equation}
    \label{Le:2}
    N(z)=
    \begin{pmatrix} v(z)/\sqrt{\rho(z)} & -\sqrt{\rho(z)} \\
      1/\sqrt{\rho(z)} & 0
    \end{pmatrix}.
  \end{equation}
  Then, the Lyapunov exponents for the matrix cocycles $(M,h)$ and
  $(N,h)$ are related by the formula
  \begin{equation}\label{Le:3}
    \theta(M,h)=\theta(N,h).
  \end{equation}
\end{Le}
\demo Let
\begin{equation*}
  H(z)=\frac1{M_{12}(z)}\,\begin{pmatrix} M_{12}(z)& 0 \\
    M_{22}(z) &-1\end{pmatrix}
\end{equation*}
One has
\begin{equation}
  \label{Le:4}
  M(z)=e^{l(z)}H(z) N(z) H^{-1}(z-h),\quad
  l(z)=\frac12\,\log\rho(z).
\end{equation}
Note that, under the condition~\eqref{Le:1},
\begin{equation*}
  |l(z)|\le \log A<\infty,\quad \forall z\in\R,
\end{equation*}
and that $l(z)$ is $1$-periodic. As $h$ is irrational, by Birkhoff's
Ergodic Theorem (\cite{MR94h:47068}), one has
\begin{equation}
  \label{Le:5}
  \lim_{L\to\infty}\frac1L\sum_{j=1}^Ll(z+jh)=\int_0^1l(z)dz
\end{equation}
for almost all $z\in\R$. As $2l(z)=\log\rho(z)=\log M_{12}(z)-\log
M_{12}(z-h)$, the integral in~\eqref{Le:5} vanishes. This, the
definition of the Lyapunov exponent~\eqref{eq:44},
relation~\eqref{Le:4} imply relation~\eqref{Le:3}. This completes the
proof of Lemma~\ref{le:Le:1}. \qed
\smallpagebreak Now, for $\rho=\rho^U$ and $v=v^U$, we construct $N^U$
by formula~\eqref{Le:2}.  Relations~\eqref{eq:Lyapunov}
and~\eqref{Le:3} imply that the Lyapunov exponent for the operator
$H_{z,\varepsilon}$ is given by the formula
\begin{equation}
  \label{Le:6}
  \Theta(E,\varepsilon)=\frac{\varepsilon}{2\pi}\theta(N^U(\cdot,E),h).
\end{equation}
In the next two subsections, we prove a lower and an upper bound for
$\theta(N^U(\cdot,E),h)$. They will coincide up to error terms, and,
thus, yield the asymptotic formula for $\Theta(E,\varepsilon)$.
\subsubsection{The lower bound for the Lyapunov exponent}
\label{sec:lower-bound-lyapunov}
Here, we prove that, in the case of Theorem~\ref{th:tib:sp:1}, for
$E\in\check I_\pi$, the Lyapunov exponent admits the lower bound:
\begin{equation}
  \label{Le:7}
  \theta(N^U(\cdot,E),h)\ge \log^+|\lambda_\pi|+o(1).
\end{equation}
Therefore, we use the following construction.\\
Assume that a matrix function $M:\C\to SL(2,\C)$ is $1$-periodic in
$z$ and depends on a parameter $\varepsilon>0$. One has
\begin{Pro}
  \label{le:Le:2}
  Let $\varepsilon_0>0$. Assume that there exist $y_{0}$ and $y_{1}$
  satisfying the inequalities $0<y_0<y_1<\infty$ and such that, for
  any $\varepsilon\in(0,\varepsilon_0)$ one has
  \begin{itemize}
  \item the function $z\to M(z,\varepsilon)$ is analytic in the strip
    $\{z\in\C;\ 0\le\im z\le y_1/\varepsilon\}$;
  \item in the strip $\{z\in\C;\ y_0/\varepsilon\le\im z\le y_1/\varepsilon \}$,
    $M(z,\varepsilon)$ admits the following uniform in $z$
    representation
    \begin{equation}\label{Le:form}
      M(z,\varepsilon)=\lambda(\varepsilon)e^{2\pi i m z}\cdot
      \left(\begin{pmatrix}1& 0\\0 & 0 \end{pmatrix}+o(1)\right),\quad
      \varepsilon\to 0,
    \end{equation}
    where $\lambda(\varepsilon)$ and $m$ are constant; $m$ is integer
    (independent of $\varepsilon$).
  \end{itemize}
  Then, there exit a $\varepsilon_1>0$ such that, if
  $0<\varepsilon<\varepsilon_1$, one has
  \begin{equation}
    \label{Le:8}
    \theta(M,h)>\log|\lambda(\varepsilon)|+o(1);
  \end{equation}
  the number $\varepsilon_1$ and the error estimate in~\eqref{Le:8}
  depend only on $\varepsilon_0$, $y_0$, $y_1$ and the norm of the
  term $o(1)$ in~\eqref{Le:form}.
\end{Pro}
\smallpagebreak This proposition immediately follows from
Proposition~10.1 from~\cite{MR2003f:82043}. Note that the proof of the
latter is based on the ideas of~\cite{MR93b:81058} generalizing
Herman's argument~\cite{MR85g:58057}.
\smallpagebreak We apply Proposition~\ref{le:Le:2} to the matrix
$N^U(z,E)$.  Therefore, we fix $y_2$ and $y_1$ so that
$0<y_2<y_1<y<\delta_1/(2\pi)$, where $\delta_1$ is the constant from
the Proposition~\ref{new:rho-and-v}. Then, the estimate~\eqref{Le:7}
follows from Proposition~\ref{le:Le:2} and
\begin{Le}
  \label{le:Le:3} 
  Assume that $\lambda_\pi\ge 1$.  In the strip $y_2\le \im z\le y_1$,
  for $E\in\check I_\pi$, the functions $\rho^U$
  satisfies~\eqref{v:simple} and $v^U$ admit the asymptotics:
  \begin{gather}
    \label{Le:v} 
    v^U(z,E)=\lambda_\pi\,e^{-2\pi i(z-z_\pi(E_\pi))}\,(1+o(1)).
  \end{gather}
  These asymptotics are uniform in $E_\pi$
  (satisfying~\eqref{non-res}), $E$ and $z$.
\end{Le}
\noindent We postpone the proof of this lemma and complete the
proof of the estimate~\eqref{Le:7}.  If $|\lambda_\pi|<1$, the
estimate~\eqref{Le:7} gives a trivial lower bound as the Lyapunov
exponent is always non-negative. So, it suffices to prove~\eqref{Le:7}
in the case $|\lambda_\pi|>1$.  Substituting~\eqref{v:simple}
and~\eqref{Le:v} into~\eqref{Le:2}, for $E\in \check I_\pi$ and
$y_2/\varepsilon\le \im z\le y_1/\varepsilon$, one obtains
\begin{equation*}
  N^U(z)=\lambda_\pi\, e^{-2\pi i(z-z_\pi(E_\pi))}\,
  \left[\begin{pmatrix} 1 & 0 \\
      0 & 0\end{pmatrix}+o(1)\right]
\end{equation*}
as $z_\pi$ is real and $|e^{2\pi i z}|\ge e^{2\pi y_1/\varepsilon}>1$.
Proposition~\ref{le:Le:2} then implies~\eqref{Le:7}.\qed
\smallpagebreak {\it Proof of Lemma~\ref{le:Le:3}.\/}
The first statement is taken from Corollary~\ref{Cor:v:simple}.
Let us prove~\eqref{Le:v}.  First, we recall that, as $E\in\check
I_\pi$, one has~\eqref{place:0}.  On the other hand, for $\im
z>y_2/\varepsilon>0$, one has
\begin{equation}
  \label{Le:11}
  \lambda_\pi\sin(2\pi(z-h-z_\pi(E_\pi)))=\frac{\lambda_\pi}{2i}\, e^{-2\pi i
  (z-h-z_\pi)} (1+o(1)). 
\end{equation}
Note that, as $|\lambda_\pi|\ge 1$ and $z_\pi\in\R$, the right hand
side is exponentially large as $\varepsilon\to 0$.  Then, in the strip
$y_2/\varepsilon\le \im z\le y_1/\varepsilon$, for $E\in\check
I_\pi$,~\eqref{eq:20},~\eqref{place:0} and~\eqref{Le:11}
imply~\eqref{Le:v}. This completes the proof of
Lemma~\ref{le:Le:3}.\qed
\subsubsection{The upper bound for the Lyapunov exponent}
\label{sec:upper-bound-lyapunov}
We now prove that, in the case of Theorem~\ref{th:tib:sp:1}, the
Lyapunov exponent admits the upper bound
\begin{equation}
  \label{Le:14}
  \theta(N^U(\cdot,E),h)\le \log^+|\lambda_\pi|+C.
\end{equation}
This upper bound follows from the definition of Lyapunov exponent for
matrix cocycles~\eqref{eq:44} and the estimate
\begin{equation*}
  \sup_{z\in\R}\sup_{E\in \check I_\pi}\|N^U(z,E)\|\le C (|\lambda_\pi|+ 1),
\end{equation*}
which follows from~\eqref{Le:2},~\eqref{v:simple} and the
estimate
\begin{equation*}
  \sup_{z\in\R}\sup_{E\in \check I_\pi}|v^U(z,E)|\le C(|\lambda_\pi|+ 1),
\end{equation*}
which follows from~\eqref{eq:20} and~\eqref{place:0}. This completes
the proof of~\eqref{Le:14}.\qed
\subsubsection{Completing the proof of Theorem~\ref{th:gamma-pi}}
\label{sec:compl-proof-theor}
Estimates~\eqref{Le:7} and~\eqref{Le:14} together with the
representation~\eqref{Le:6} imply the uniform representation
\begin{equation*}
  \forall E\in\check I_\pi,\quad\quad\Theta(E,\varepsilon)=
  \frac{\varepsilon}{2\pi}\log^+|\lambda_\pi(E_\pi)|+O(\varepsilon).
\end{equation*}
In view of~\eqref{sc-eq:4}, to complete the proof of Theorem~\ref{th:gamma-pi},
it suffices to check that
\begin{equation*}
  \left|\cos\left(\frac{\check\Phi_0(E_\pi)}\varepsilon\right)
  \right|\asymp\frac{C}\varepsilon\,\inf_{l}|E_\pi-E_0^{(l)}|,
\end{equation*}
which follows from the definition of the points $(E_0^{(l)})_l$ and
from~\eqref{Phi':up} and~\eqref{Phi':down}.\qed
\subsection{Absolutely continuous spectrum}
\label{sec:absol-cont-spectr}
We now turn to the proof of Theorem~\ref{th:tib:sp:2}. The idea is to
find a subset of $\check I_\pi$ where $E\mapsto\Theta(E,\varepsilon)$
vanishes.  Then, by the Ishii-Pastur-Kotani Theorem
(\cite{MR94h:47068}), this subset is contained in the absolutely
continuous spectrum of the ergodic family~\eqref{family}.\\
As before, we assume that $h$ is defined by~\eqref{h}, and that the
functions $\rho^U$ and $v^U$ are the coefficients of the scalar
equation equivalent to the monodromy equation with the matrix $M^U$.\\
As in the previous subsection, to analyze $\Theta(E,\varepsilon)$, we
use the matrix cocycle $(N^U(\cdot,E),h)$, the matrix $N^U$ being
defined by~\eqref{Le:2} for $M=M^U$. Recall that
$\Theta(E,\varepsilon)$ is related to $\theta(N^U(\cdot,E),h)$, the
Lyapunov exponent of this cocycle, by the formula~\eqref{Le:6}.\\
First, under the conditions of Theorem~\ref{th:tib:sp:2}, we check
that, up to error terms, $N^U$ is independent of $z$. This allows then
to characterize the subset of $\check I_\pi$ where $\theta(N^U,h)=0$
by means of a standard KAM construction found in~\cite{MR2003f:82043}.
\subsubsection{The asymptotic behavior of the matrix $N^U$}
\label{sec:asympt-behav-matr}
We need to control the behavior of the matrix $N^U$ for bounded $|\im
z|$ and $E$ near the interval $\check I_\pi$. One has
\begin{Le}
  \label{le:ac:3} 
  Fix $c>0$, $\varkappa>0$ and $r>0$. For $\varepsilon$ sufficiently small, 
  if $E_\pi$ satisfies~\eqref{non-res}, and if
  $\varepsilon\log\lambda_\pi(E_\pi)\le -c$, then
  \begin{equation}
    \label{ac:3}
    \begin{split}
      N^U(z,E)&=N_0(E)+N_1(z,E),\\
      \quad N_0(E)=\begin{pmatrix} F(E) & -1
        \\ 1 & 0\end{pmatrix},&\quad\quad
      \sup_{\substack{|E-\check E_\pi|\le \varkappa |\check I_\pi|\\
          |\im z|\le r}} \|N_1(z,E)\|\le C\,e^{-\eta/\varepsilon},
    \end{split}
  \end{equation}
  where the constant $\eta$ is defined by $\eta=\min\{c,\delta_1\}$,
  and $F$ is the function from~\eqref{new:v}.
\end{Le}
\demo It suffices to prove, that under the conditions of the lemma,
there exists $C>0$ such that, for $\varepsilon$ sufficiently small,
one has
\begin{gather}
  \label{ac:7}
  \sup_{\substack{|E-\check E_\pi|\le \varkappa |\check I_\pi|\\
      |\im z|\le r}}|\rho^U(z,E)-1|\le C\,e^{-\eta/\varepsilon},\quad
  \sup_{\substack{|E-\check E_\pi|\le \varkappa |\check I_\pi|\\
      |\im z|\le r}}|v^U(z,E)-F(E)|\leq
  Ce^{-\eta/\varepsilon},\\\label{eq:26} \sup_{|E-\check E_\pi|\le
    \varkappa |\check I_\pi|}|F(E)|\le C.
\end{gather}
Begin with the proof of~\eqref{eq:26}. Recall that, for $E$ in the
$\varepsilon^\alpha e^{-\delta_0/\varepsilon}$-neighborhood of
$E_\pi$, one has~\eqref{F-F0}. On the other hand, the interval $\check
I_\pi$ is located in the $(C\varepsilon
e^{-\delta_0/\varepsilon})$-neighborhood of $E_\pi$,
see~\eqref{check}. So, it suffices to prove~\eqref{eq:26} with $F$
replaced by $F_\pi$.\\
Recall that $\check I_\pi$ is centered at $\check E_\pi$,
see~\eqref{centre}, and that, by~\eqref{F0:simple}, one has
$F_\pi(\check E_\pi)=0$. The estimate~\eqref{place:0} is an estimate
for $F_\pi(E)$ on the interval $\check I_\pi$. As $E\mapsto F_\pi(E)$
is affine, it implies that
\begin{equation*}
  \sup_{|E-\check E_\pi|\leq\varkappa|\check I_\pi|}
  |F_\pi(E)|\le \varkappa (1+|\lambda_\pi|)(1+o(1)).
\end{equation*}
As $|\lambda_\pi|<1$, this implies~\eqref{eq:26}.\\
Let us prove~\eqref{ac:7}. The representation~\eqref{new:v} and
estimate~\eqref{eq:26} imply that, for some $C>0$,
\begin{align*}
\sup_{\substack{|E-\check E_\pi|\le \varkappa |\check I_\pi|\\ |\im z|\le r}}|v^U(z,E)-F(E)| &\leq
C\,\varepsilon e^{-\delta_0/\varepsilon} \,\sup_{|E-\check E_\pi|\leq\varkappa|\check I_\pi|} |F(E)|+
C\,\lambda_\pi+
C\,e^{-\delta_1/\varepsilon}\le \\
&\le C(\varepsilon  e^{-\delta_0/\varepsilon}+e^{-c/\varepsilon}+e^{-\delta_1/\varepsilon}).
\end{align*}
In view of~\eqref{delta1} and the definition of $\eta$, this
expression is bounded by $C\,e^{-\eta/\varepsilon}$.  This proves the
second estimate from~\eqref{ac:7}. The first one follows
from~\eqref{new:rho}, \eqref{delta1} and the definition of $\eta$.
Lemma~\ref{le:ac:3} is proved.\qed
\subsubsection{The KAM theory construction}
\label{sec:KAM}
Here, we formulate a corollary from the construction developed in
section~11 of~\cite{MR2003f:82043} that is based on standard ideas of
KAM theory (see~\cite{MR57:10076,Be-Li-Te:83}).
\smallpagebreak Let $I\subset \R$ be a bounded interval. Fix $r>0$.
Let $S_{r}$ be the strip $\{z\in\C;\ |\im z|\le r\}$. We consider
$\mathcal A$, the set of $2\times 2$-matrix valued functions $(z,\varphi)\in
S_r\times I\mapsto M(z,\varphi)$ that are
\begin{enumerate}
\item analytic and $1$-periodic in $z\in S_r$;
\item analytic in $\varphi$ in $V(I)$, a complex neighborhood of $I$;
\item of the form $\begin{pmatrix} a & b \\ b^* & a^*\end{pmatrix}$.
\end{enumerate}
Let $D= \begin{pmatrix} e^{i\varphi} & 0 \\ 0 & e^{-i\varphi}
\end{pmatrix}$, and let $A\in{\mathcal A}$ satisfy $\D
\lambda(A)=\sup_{\varphi\in V(I),\,z\in S_r}\|A(z,\varphi)\|
<\infty$.
\smallpagebreak Fix $0<h<1$. For $z\in\R\mapsto\psi(z)\in\C^2$, a
vector function, consider the equation
\begin{equation}
  \label{ac:1}
  \psi(z+h)=(D+ A)(z)\psi(z).
\end{equation}
Define
\begin{equation*}
  \label{diophantine}
  H(\mu):=\{h\in(0,1);\ \min_{l\in \N}|h-l/k|\geq\mu/k^3\text{ for }
  k=1,2,3\dots\}.
\end{equation*}
One has
\begin{Pro}
  \label{le:ac:1}
  Fix $\sigma\in(0,1)$. There exists $\lambda_0(r,\sigma,I)>0$ such
  that, for any $A$, $D$ and $h$ chosen as above and satisfying the
  conditions
  \begin{enumerate}
  \item $\det(D+A)=1$,
  \item $\lambda=\lambda(A)<\lambda_0(r,\sigma,I)$,
  \item $h\in H(\lambda^\sigma)$
  \end{enumerate}
  there exists $\Phi_\infty\subset I$, a Borel set of Lebesgue measure
  smaller than $\lambda^{\sigma/2}$ and such that, for all $\varphi\in
  I\setminus \Phi_\infty$, equation~\eqref{ac:1} has two linearly
  independent bounded solutions.
\end{Pro}
\noindent This proposition immediately follows from Proposition~11.1
of~\cite{MR2003f:82043}. The constant $\lambda_0(r,\sigma,I)$ depends
only on the length of $I$, but not of its position.
\smallpagebreak Proposition~\ref{le:ac:1} implies
\begin{Cor}
  \label{le:ac:2}
  In the case of Proposition~\ref{le:ac:1}, for all $\varphi\in
  I\setminus \Phi_\infty$, the Lyapunov exponent of the cocycle
  $(D+A,h)$ is zero.
\end{Cor}
\demo Let $\Psi(z)$ be the matrix the columns of which are the vector
solutions defined in Proposition~\ref{le:ac:1}. Then, $\Psi(z)$ is a
matrix solution of~\eqref{ac:1}. As the vector solutions are linearly
independent, $\det\Psi(z)\ne0$ for all $z\in\R$. For $l\in\Z$, put
$\chi(l)=\Psi(z+lh)$. Then, $\chi(l+1)=(D+A)(hl+z)\chi(l)$, and, as
$\Psi(z)$ is bounded, for $L\geq1$, we have
\begin{equation*}
  \|(D+A)(Lh+z)\cdots(D+A)(h+z)(D+A)(z)\|=\|\chi(L+1)\chi^{-1}(0)\|\leq
  C.
\end{equation*}
Now, the statement of the corollary follows from~\eqref{eq:44}, the
definition of the Lyapunov exponent.\qed
\subsubsection{The proof of Theorem~\ref{th:tib:sp:2}}
\label{sec:proof-theor-refth:t}
The idea is the following. Let $S$ be a constant matrix such that
$\det S \ne 0$.  Clearly,
\begin{equation}\label{ac:9}
\theta(N^U,h)=\theta(S^{-1}N^US,h).
\end{equation}
Recall that $N^U$ admits the representation~\eqref{ac:3}. We shall
choose $S$ so that the matrices
\begin{equation}
  \label{ac:10}
  D=S^{-1} N_0 S\quad {\rm and}\quad A=S^{-1}N_1 S
\end{equation}
satisfy the assumptions Proposition~\ref{le:ac:1}. Then, we apply
Corollary~\ref{le:ac:2} to the so constructed matrix $D+A$. We divide
the analysis into ``elementary'' steps.
\smallpagebreak{\it Diagonalization.\/} Let $E^0$ be a point of $\check
I_\pi$ such that
\begin{equation*}
  -1<F(E)<1.
\end{equation*}
Then, in $V^0$, a neighborhood of $E^0$, one can define an analytic branch of
the function $E\mapsto\varphi(E)$ solution to
\begin{equation}
  \label{ac:12}
  \cos\varphi(E)=F(E).
\end{equation}
In $V^0$, the exponentials $e^{\pm i\varphi(E)}$ are the eigenvalues of
the matrix $N_0(E)$ (see~\eqref{ac:3}); the columns of the matrix
\begin{equation*}
  S(E)=\begin{pmatrix} e^{i\varphi(E)} & e^{-i\varphi(E)}\\ 1&
  1\end{pmatrix}
\end{equation*}
are its eigenvectors. Define $D$ and $A$ by~\eqref{ac:10}. Clearly,
\begin{equation}
  \label{ac:14}
  D(E)=\begin{pmatrix} e^{i\varphi} & 0 \\ 0 &
    e^{-i\varphi}\end{pmatrix}.
\end{equation}
As $E\mapsto N_1(E)$ is real analytic, $A(z,E)$ has the form
\begin{equation*}
  A=\begin{pmatrix}a&b\\b^*&a^*\end{pmatrix}.
\end{equation*}
For some $C>0$, one has
\begin{equation}
  \label{ac:3a}
  \forall E\in V^0,\quad
  \sup_{z\in\R}\|A(z,E)\|\le C\,\frac{e^{2|\im\varphi(E)|}}
  {|\sin\varphi(E)|}\,\sup_{z\in\R}\|N_1(z,E)\|.
\end{equation}
{\it A change of variables: $E\to\varphi$.\/} Now, we change the
variable $E$ to $\varphi$, and check that, as a function of $\varphi$,
$A$ satisfies the conditions of Proposition~\ref{le:ac:1} and
Corollary~\ref{le:ac:2}. We use
\begin{Le}
  \label{le:ac:4}
  Fix $\varkappa<1$. There exists $\varepsilon_0>0$ such that, for
  $0<\varepsilon<\varepsilon_0$ the following holds. Let $E_\pi$
  satisfy~\eqref{non-res}. Let $I\subset \R$ be the interval centered
  at $\check E_\pi$ and of length $\varkappa |\check I_\pi|$. Then,
  \begin{itemize}
  \item in a neighborhood of $I$, there exists a real analytic branch
    of $\varphi(E)$; it is monotonous on $I$;
  \item there exists a positive $\Delta=\Delta(\varkappa)$ such that
    $\varphi(I)\subset (\Delta,\pi-\Delta)$;
  \item $\varphi\mapsto E(\varphi)$, the function inverse to
    $E\mapsto\varphi(E)$ is analytic in $V(I)$, the
    $\Delta/2$-neighborhood of the interval $\varphi(I)$, and maps
    $V(I)$ into the $(C|\check I_\pi|)$-neighborhood of $\check
    I_\pi$.
  \end{itemize}
\end{Le}
\noindent As $F(E)$ is real analytic, Lemma~\ref{le:ac:4} immediately
follows from~\eqref{ac:12} and
\begin{Le}
  \label{le:ac:5} 
  Fix $\varkappa_1\in(0,1)$. For $\varepsilon$ sufficiently small, the
  following holds. Let $E_\pi$ satisfy~\eqref{non-res} and define
  $B=\{E\in\C;\ |E-\check E_\pi|\le \frac12\varkappa_1|\check I_\pi|\}$.\\
  Then, $F$ bijectively maps $B$ onto $F(B)$, and one has
  \begin{equation}
    \label{ac:15}
    \sup_{E\in B}|F(E)|\leq\varkappa_1+o(1),\quad
    \text{ \ and, \ for \ } |E-\check E_\pi|=\frac{\varkappa_1}2\,|\check I_\pi|,\quad
    |F(E)|=\varkappa_1+o(1).
  \end{equation}
\end{Le}
\demo Fix $0<\alpha<1$. By~\eqref{check}, $B$ is contained in the
$\varepsilon^\alpha e^{-\delta_0/\varepsilon}$-neighborhood of
$E_\pi$.  Therefore, $F'(E)$ admits the representation~\eqref{F-F0}.
This implies that $F$ is a bijection of $B$ onto $F(B)$. Indeed,
assume that, in $B$ there exist $E_1$ and $E_2$ such that $E_1\ne E_2$
and $F(E_1)=F(E_2)$.  Then, one has
\begin{equation*}
  0= F(E_2)-F(E_1)=\int_{E_1}^{E_2} F'(E) dE=F_\pi'\,\int_{E_1}^{E_2}
  (1+o(1))dE=F_\pi'(E_2-E_1) (1+o(1))\ne 0.
\end{equation*}
So, we get a contradiction, and $F$ is a bijection.\\
Estimates~\eqref{ac:15} follow from the following facts:
\begin{enumerate}
\item the representation for $F$ from~\eqref{F-F0} holds on $B$ (as
  $B$ is contained in the $\varepsilon^\alpha
  e^{-\delta_0/\varepsilon}$-neighborhood of $E_\pi$);
\item $E\mapsto F_\pi(E)$ is affine, and vanishes at $\check E_\pi$,
  the center of $\check I_\pi$;
\item at the ends of $\check I_\pi$, one has $|F_\pi(E)|=1+o(1)$
  (by~\eqref{place:0}, which is the definition of $\check I_\pi$, and
  as $\lambda_\pi=O(e^{-\eta/\varepsilon})$).
\end{enumerate}
This completes the proof of Lemma~\ref{le:ac:5}.\qed
\smallpagebreak Now, turn to the matrices $D$ and $A$ defined
by~\eqref{ac:10}.  Make the change of variables $E\to\varphi$ so that
$E=E(\varphi)$. Consider these matrices as functions of $\varphi$ in
$V(I)$. Then, for $\varepsilon$ sufficiently small, they satisfy the
conditions of section~\ref{sec:KAM}:
\begin{itemize}
\item $z\mapsto A(z,\varphi)$ is analytic and $1$-periodic in $S_{r}$
  \ (as $z\mapsto N^U(z,E)$ is analytic in the strip $\{|\im z|\le
  y\}$);
\item $\varphi\mapsto A(z,\varphi)$ is analytic in $V(I)$ (as
  $\varphi\mapsto E(\varphi)$ is analytic in $V(I)$, \ $\varphi(V(I))$
  is in the $(C|\check I_\pi|)$-neighborhood of $\check E_\pi$, and as
  $E\mapsto N^U(z,E)$ is analytic in this neighborhood);
\item $A$ has the form $\begin{pmatrix} a & b \\ b^* & a^*
  \end{pmatrix}$ (as $\varphi\mapsto E(\varphi)$ is real analytic, and
  as $E\mapsto A(z,E)$ already had this form);
\item $D$ is given by~\eqref{ac:14};
\item $\lambda(A)\leq \frac{C}{\Delta}e^{-\eta/\varepsilon}$ \ 
  (by~\eqref{ac:3},~\eqref{ac:3a} and Lemma~\ref{le:ac:4});
\item $\det(D+A)=1$ as $D+A=S^{-1}N^US$ and $\det N^U=1$
  by~\eqref{Le:2}.
\end{itemize}
{\it The Diophantine condition on $\varepsilon$.\/} \ 
To apply Corollary~\ref{le:ac:2}, we have to impose a {\it
  Diophantine} condition on the number $2\pi/\varepsilon$. Fix two
positive numbers $a$ and $b$. Consider the set
\begin{equation*}
  D(a,b):=\left\{\varepsilon\in(0,1):\,\,\min_{l\in \N}
   \left|\frac{2\pi}\varepsilon-l/k\right|\geq
   \frac{a}{k^3}\,e^{-b/\varepsilon},\ k=1,2,3\dots\right\}.
\end{equation*}
It can be easily checked
\begin{equation}
  \label{D:mes}
  \frac{\mes(D(a,b)\cap(0,\varepsilon))}{\varepsilon}=
  1+o\left(e^{-b/\varepsilon}\right)\text{
    when }\varepsilon\to0.
\end{equation}
The derivation of~\eqref{D:mes} is similar to the estimates in
section~4.4.6 of~\cite{MR2003f:82043}.
\smallpagebreak Fix $0<\sigma<1$.  For $\varepsilon\in
D((C/\Delta)^\sigma,\,\sigma\eta)$, the number $h$ defined
by~\eqref{h} belongs to the class $H(\mu)$ with $\mu=(C/\Delta
e^{-\eta/\varepsilon})^\sigma$.
\smallpagebreak{\it Completing the proof of
  Theorem~\ref{th:tib:sp:2}.\/} \ Let $A$ and $D$ be as constructed
above and $\varepsilon\in D$.  Then, for the matrix cocycle $(D+
A,\,h)$, the conditions of Corollary~\ref{le:ac:3} are satisfied
provided $\varepsilon$ is sufficiently small. So, for $\varepsilon$
sufficiently small, there exists $\Phi_\infty$, a subset of $I$ of
measure uniformly small with $\lambda(A)\le \frac C\Delta\,
e^{-\eta/\varepsilon}$, such that, for all $\varphi\in
I\setminus\Phi_\infty$, the Lyapunov exponent
$\theta(D+ A,\,h)$ is zero.\\
By~\eqref{ac:9} and~\eqref{Le:6}, this implies that $\Theta(E)$, the
Lyapunov exponent for the family of equations~\eqref{G.2z}, is zero on
$\varphi(I)\subset\check I_\pi$ outside a set of Lebesgue measure
$\D m:=\int_{\Phi_\infty}\frac{d E}{d\varphi}d\varphi$.\\
The Cauchy estimates and Lemma~\ref{le:ac:4} imply that
$\left|\frac{dE}{d\varphi}(\varphi)\right|\le C|\check I_\pi|$ for
$\varphi\in \varphi(I)$. So, $m=o(|I|)$ where $|I|$ denotes the
length of $I$.\\
As $m$ is small with respect to $|I|$ and as $\varkappa$ in the
definition of $I$ can be chosen arbitrarily close to $1$, we conclude
that $\Theta(E,\varepsilon)$ is zero on $\check I_\pi$ outside a set
the measure of which becomes small with respect to $|\check I_\pi|$ as
$\varepsilon$ tends to zero in $D(\eta)$.  This completes the proof of
Theorem~\ref{th:tib:sp:2}. \qed
\section{Computing the monodromy matrices}
\label{sec:MM:demo}
In this section, we prove Theorem~\ref{th:M-matrices}. As we have
seen, to study the spectrum of~\eqref{family}, one has to compute the
coefficients of the monodromy matrix up to terms that are exponentially
small (in $\varepsilon$) whereas these coefficients are exponentially
large outside small ``resonant'' neighborhoods (where the points
$\{E_\pi{(l)}\}_l$ are exponentially close to $\{E_0{(l')}\}_{l'}$).
To achieve such an accuracy, we use a natural factorization of the
monodromy matrix into the product of two simple ``transition''
matrices and carry out a rather delicate analysis of the properties of
their coefficients.
\smallpagebreak Below, we always work in terms of the variables
\begin{equation}
  \label{new-variables}
  x:=x-z,\quad\text{and}\quad  \zeta=\varepsilon z.
\end{equation}
In these variables, equation~\eqref{G.2z} takes the form
\begin{equation}
  \label{G.2}
  -\frac{d^2}{dx^2}\psi(x)+(V(x)+\alpha\cos(\varepsilon
  x+\zeta))\psi(x)= E\psi(x), \quad x\in\R,
\end{equation}
The advantage of the new variables is that now we can study solutions
of~\eqref{G.2} analytic in $\zeta$.\\
In terms of variables~\eqref{new-variables}, the consistency
condition~\eqref{consistency} takes the form
\begin{equation}
  \label{consistency:1}
  \psi_j(x+1,\,\zeta)=\psi_j(x,\,\zeta+\varepsilon).
\end{equation}
The definition of the monodromy matrix,~\eqref{monodromy}, turns into
\begin{equation}
  \label{monodromy:1}
  \Psi\,(x,\zeta+2\pi)= M\,(\zeta,E)\,\Psi\,(x,\zeta),\quad
  \Psi(x,\zeta)=\begin{pmatrix}\psi_{1}(x,\,\zeta)\\ \psi_{2}(x,\zeta)\end{pmatrix},
\end{equation}
and, now, the monodromy matrix is $\varepsilon$-periodic:
\begin{equation*}
  M\,(\zeta+\varepsilon,E)=M\,(\zeta,E),\quad \forall \zeta.
\end{equation*}
\subsection{Transition matrices}
\label{sec:transition-matrices}
Here, we describe the factorization and the asymptotics of the
transition matrices.
\subsubsection{Factorization}
\label{sec:general-construction}
Here, we describe a natural factorization of the monodromy matrix
under the assumption (TIBM).
\smallpagebreak {\it Two consistent bases.} \ In section~\ref{S4}, we
pick a point $E_*$ in $J$ and show the existence of $V_*$, a
neighborhood of $E_*$, such that, for $E\in V_*$, there exists two
consistent bases which will be indexed by $\nu$ in $\{0,\pi\}$.  Let
us describe some properties of these bases; they will be central
objects in this section.\\
Fix $\nu\in\{0,\pi\}$. The corresponding basis consists of two
consistent  solutions to~\eqref{G.2}, say $(x,\zeta,E)\mapsto
f_\nu(x,\zeta,E)$ and $(x,\zeta,E)\mapsto f_\nu^*(x,\zeta,E)$; the
second solution is related to the first one by the
transformation~\eqref{star}. 
For any $x\in\R$, the function $(\zeta,E)\mapsto f_\nu(x,\zeta,E)$ is
analytic in the domain
\begin{equation}
  \label{analyticity:dom:1}
  \left\{\zeta\in \C:\,\,|\im \zeta|<Y\right\}\times V_*,
\end{equation}
where $Y$ satisfies the inequality $Y>Y_M$ (recall that $Y_M$ is
defined in~\eqref{Y's}).
\smallpagebreak {\it Definitions of the transition matrices.} \ 
As both pairs $(\{f_\nu,f_\nu^*\})_{\nu\in\{0,\pi\}}$ are bases of the
space of solutions of~\eqref{G.2}, one can write
\begin{equation}
  \label{Tmatrices}
  F_\pi(x,\zeta+2\pi,E)=T_\pi(\zeta,E)\,F_0(x,\zeta,E),\quad
  F_0(x,\zeta,E)=T_0(\zeta,E)\,F_\pi(x,\zeta,E),\quad F_\nu=\begin{pmatrix}
  f_\nu \\ f_\nu^*\end{pmatrix}.
\end{equation}
For $\nu\in\{0,\pi\}$, the $2\times2$-matrix valued function
$(\zeta,E)\mapsto T_\nu(\zeta,E)$ is independent of $x$.
We call it {\it a transition matrix}. \\
Discuss the basic properties of a transition matrix.  As the basis
$\{f_\nu,\,f_\nu^*\}$ is consistent, for all $E$, $\zeta\mapsto
T_\nu(\zeta,E)$ is $\varepsilon$-periodic. It is analytic in the
domain~\eqref{analyticity:dom:1}. Finally, as the consistent solutions
$f_\nu$ and $f^*_\nu$ are related by the transformation~\eqref{star},
$T_\nu$ enjoys the same symmetry property as the monodromy matrix
(see~\eqref{Tform}); we write
\begin{equation*}
 T_\nu=\begin{pmatrix} a_\nu & b_\nu \\ b_\nu^* &
   a_\nu^*\end{pmatrix}. 
\end{equation*}
{\it Factorization of the monodromy matrices.}  For $\nu\in\{0,\pi\}$,
we denote by $M_\nu$ the monodromy matrix corresponding to the base
$\{f_\nu,\,f_\nu^*\}$.  The definitions~\eqref{monodromy:1}
and~\eqref{Tmatrices} imply that
\begin{equation}\label{factorization}
  M_\pi(\zeta)=T_\pi(\zeta)\,T_0(\zeta),\quad
  M_0(\zeta)=T_0(\zeta+2\pi)\,T_\pi(\zeta).
\end{equation}
Clearly, the monodromy matrices share the basic properties of the
transition matrices: they are $\varepsilon$-periodic in $\zeta$,
analytic in the domain~\eqref{analyticity:dom:1}
and have the form~\eqref{Tform}. \\
Note that, once transformed back to the $z$-variable, the
monodromy matrices are  analytic in the domain $\left\{\zeta\in \C;\
  |\im \zeta|<Y/\varepsilon\right\}\times V_*$.\\
Finally, by~\eqref{Mproperties} and~\eqref{factorization}, one has
\begin{equation}\label{det}
\det T_0\,\det T_\nu=1.
\end{equation}
The motivation for considering the factorizations is the following.
The solutions $f_0$ and $f_\pi$ are constructed so that $f_0$ has a
simple asymptotic behavior in the strip $\{-\pi<\re\zeta<\pi\}$, and
$f_\pi$ has a simple asymptotic behavior in the strip
$\{0<\re\zeta<2\pi\}$. In result, formulae~\eqref{factorization} give
factorizations of the monodromy matrices in terms of factors with
simple asymptotic behavior.
\subsubsection{Asymptotics of the transition matrices}
\label{Tass}
We now describe the asymptotics of the transition matrices
$(T_\nu)_{\nu\in\{0,\pi\}}$. Therefore, we shall use the conventions
introduced in~\eqref{TY},~\eqref{p} and~\eqref{Y's} in
section~\ref{sec:mon-mat-as}. We need a few more notations.
\smallpagebreak {\it 1. Asymptotic notations.\/} \ We shall use all
the notations introduced in section~\ref{COo}.\\
{\it 2. ``Analytic'' notations.} Pick $z_0\in\R$ and let $V_0$ be a
complex neighborhood of $z_0$. Let $z\mapsto a(z)$ be an analytic
function defined and non vanishing in $V_0$. In $V_0$, we define two
real analytic functions $z\mapsto\babs{a}(z)$ and
$z\mapsto\varphi(a)(z)$ by
\begin{equation*}
  \label{polar}
  \begin{split}
    a(z)&=\babs{a}(z)\,\exp(i\varphi(a)(z))\\\text{ such that }
    \babs{a}(z)&=|a(z)|,\text{ and }\varphi(a)(z)=\arg
    a(z)\text{\ when \ } z\in V_0\cap \R.
  \end{split}
\end{equation*}
{\it 3. ``Fourier expansion'' notations.} \ The transition matrices
being $\varepsilon$-periodic, we represent their Fourier expansion in
the form
\begin{equation}\label{a,b:F}
  a_\nu(\zeta)= a_{\nu,-1}(\zeta)+a_{\nu,0}+a_{\nu,1}\,
  e^{2\pi\zeta/\varepsilon}+a_{\nu,2}(\zeta),\quad
  b_\nu(\zeta)= b_{\nu,-1}(\zeta)+b_{\nu,0}+b_{\nu,1}\,
  e^{2\pi\zeta/\varepsilon}+b_{\nu,2}(\zeta),
\end{equation}
where we single out the sum of Fourier terms with negative index, the
zeroth and the first Fourier terms and the sums of Fourier series
terms with index greater than $1$.
\smallpagebreak One has
\begin{Th}
  \label{th:T-matrices} Pick $E_*\in J$. There exists $V_*$, a complex
  neighborhood of $E_*$, and $Y>Y_M$ such that, for sufficiently small
  $\varepsilon$ and $\nu\in\{0,\pi\}$, there exists
  $\{f_\nu,f_\nu^*\}$, a consistent basis of solutions to~\eqref{G.2},
  having the following properties:
  \begin{itemize}
  \item the basis $\{f_\nu,f_\nu^*\}$ and the transition matrices
    $T_\nu$ are defined and analytic in the
    domain~\eqref{analyticity:dom:1};
  \item the determinant of $T_\nu$ is independent of $\zeta$ and
    $\varepsilon$; it is a non-vanishing analytic function of $E\in
    V_*$;
  \item one has
    \begin{equation}
      \label{abs-Fourier:1}
      \begin{split}
        \babs{a_{\nu,0}}&=\exp\left(\frac1\varepsilon
        S_{h,\nu}+O(1)\right),\quad\babs{b_{\nu,0}}=
      \exp\left(\frac1\varepsilon S_{h,\nu}+O(1)\right),\\
      \babs{a_{\nu,1}}&=\exp\left(\frac1\varepsilon
        (S_{h,\nu}-S_{v,\nu})+O(1)\right),\quad\babs{b_{\nu,1}}=
      \exp\left(\frac1\varepsilon(S_{h,\nu}-S_{v,\nu})+O(1)\right),
      \end{split}
    \end{equation}
    and
    \begin{gather}
      \label{phases-Fourier:1}
      \begin{split}
        \varphi(a_{0,0})=\frac1{2\varepsilon}\,(\Phi_\pi+\Phi_0)+
        O(1),\quad
        \varphi(b_{0,0})=\frac1{2\varepsilon}\,(-\Phi_\pi+\Phi_0)+ O(1),\\
        \varphi(a_{\pi,0})=\frac1{2\varepsilon}\,(\Phi_\pi+\Phi_0)+
        O(1),\quad
        \varphi(b_{\pi,0})=\frac1{2\varepsilon}\,(\Phi_\pi-\Phi_0)+
        O(1),
      \end{split}\\
      \label{phases-Fourier:2}
      \begin{split}
        \varphi(a_{0,1})=-\frac1{2\varepsilon}\,(\Phi_0-\Phi_\pi)+
        O(1),\quad\varphi(b_{0,1})=-\frac1{2\varepsilon}\,(\Phi_0+\Phi_\pi)+
        O(1),\\
        \varphi(a_{\pi,1})=-\frac1{2\varepsilon}\,(\Phi_\pi-\Phi_0-4\pi^2)+
        O(1),\quad
        \varphi(b_{\pi,1})=-\frac1{2\varepsilon}\,(\Phi_\pi+\Phi_0-4\pi^2)+
        O(1),
      \end{split}
    \end{gather}
    where $O(1)$ denotes functions real on $V_*\cap\R$ and analytic in
    $E\in V_*$;
  \item moreover,
    \begin{equation}
      \label{abs-Fourier}
      a_{\nu,-1}(\zeta)=o(a_{\nu,0}),\quad
      b_{\nu,-1}(\zeta)=o(b_{\nu,0}),\quad
      a_{\nu,2}(\zeta)=o(p(\zeta/\varepsilon)\,a_{\nu,1}),\quad
      b_{\nu,2}(\zeta)=o(p(\zeta/\varepsilon)\,b_{\nu,1}).
    \end{equation}
  \end{itemize}
  All the above estimates are uniform in $E$ and $\zeta$ in the
  domain~\eqref{analyticity:dom:1}.
\end{Th}
\noindent Theorem~\ref{th:T-matrices} is proved in
  sections~\ref{S4}~--~\ref{sec:fur}.
\smallpagebreak When studying the spectral properties
of~\eqref{family}, we always assume that $E$ satisfies
\begin{equation}
  \label{E}
  E\in V_*^\varepsilon:=V_*\cap\{ |\im E|\le  \varepsilon\}.
\end{equation}
One proves
\begin{Cor}
  \label{rough-est}
  Pick $\nu\in\{0,\pi\}$. For sufficiently small $\varepsilon$, in the
  case of Theorem~\ref{th:T-matrices}, for $E\in V_*^\varepsilon$, one
  has
  \begin{equation}
    \label{abs-Fourier:E}
      |a_{\nu,0}|\asymp \frac1{|t_{h,\nu}|},\quad
      |b_{\nu,0}|\asymp \frac1{|t_{h,\nu}|},\quad\quad
      |a_{\nu,1}|\asymp \frac{|t_{v,\nu}|}{|t_{h,\nu}|}, \quad
      |b_{\nu,1}|\asymp \frac{|t_{v,\nu}|}{|t_{h,\nu}|}.
  \end{equation}
  where all the tunneling coefficients are computed at the point $\re
  E$ instead of $E$.
\end{Cor}
\demo The functions $E\mapsto S_{h,0}(E)$, $E\mapsto S_{h,\pi}(E)$,
$E\mapsto \Phi_0(E)$ and $E\mapsto \Phi_\pi(E)$ are independent of
$\varepsilon$ and analytic in a neighborhood of $J$. So, for
sufficiently small $\varepsilon$, for $E\in V_*^\varepsilon$, one has
\begin{equation*} 
|t_{d,\nu}(E)|\asymp |t_{d,\nu}(\re E)|,\quad 
|e^{i\Phi_\nu(E)/\varepsilon}|\asymp |e^{i\Phi_\nu(\re E)/\varepsilon}|,
\end{equation*}
for $\nu\in\{0,\pi\}$ and for $d\in \{h,v\}$. As the phase integrals
are real analytic, one has $|e^{i\Phi_\nu(E)/\varepsilon}|\asymp 1$.
Estimates~\eqref{abs-Fourier:E} follow from these observations and
representations~\eqref{abs-Fourier:1}~---~\eqref{phases-Fourier:2}.
This completes the proof of Corollary~\ref{rough-est}.\qed
\subsection{Relations between the coefficients $a_\nu$ and $b_\nu$ of
  the matrix $T_\nu$}
\label{sec:relat-betw-coeff}
It appears that, with a great accuracy, the coefficients $a_\nu$ and
$b_\nu$ are proportional. This makes the
factorizations~\eqref{factorization} extremely effective. Recall that
$Y_m$ is defined in~\eqref{Y's}. Define
\begin{equation}
  \label{eq:27}
      R_\nu(\zeta,E)= \frac{b_{\nu}(\zeta,E)}{a_\nu(\zeta,E)}
\end{equation}
One has
\begin{Pro}
  \label{pro:r-nu}
  Pick $\nu\in\{0,\pi\}$.  Fix $0<y<Y_m$.  For $\varepsilon$
  sufficiently small, in the case of Theorem~\ref{th:T-matrices}, for
  $|\im\zeta|<y$ and $E\in V_*^\varepsilon$ one has
  \begin{gather}
    \label{R-nu}
    R_\nu(\zeta,E)
    =e^{i(\varphi(b_{\nu,0})-\varphi(a_{\nu,0}))}\,\left(1-\frac{\det
        T_\nu}{2a_{\nu,0}a_{\nu,0}^*}+O_\nu\right),\\
    \intertext{where}
     \label{O-nu}
     O_\nu=O(t_{h,\nu}^4,\, T_Y\,p(\zeta/\varepsilon),\, t_{h,\nu}^2
     t_{v,\nu} p(\zeta/\varepsilon)).
  \end{gather}
\end{Pro}
\demo In this proof, we assume that $E\in V_*^\varepsilon$. We set
\begin{equation*}
  Y_{v,\nu}(E)=\frac1{2\pi} S_{v,\nu}(\re E),
\end{equation*}
and note that 
\begin{equation}
  \label{R-nu:6}
  0<y<Y_m\le Y_{v,\nu}(E)\le Y_M<Y,
\end{equation}
and
\begin{equation}\label{R-nu:7}
|t_{v,\nu}(E)|\asymp  e^{- 2\pi Y_{v,\nu}(E)/\varepsilon}.
\end{equation}
The plan of the proof is the following. We first prove that, for $|\im
\zeta|\le y$,
\begin{equation}
  \label{R-nu:1}
  R_\nu=r_{\nu}\,\left[1+O\left(e^{-2\pi(Y-|\im\zeta|)/\varepsilon},
      e^{2\pi|\im\zeta|/\varepsilon}\,t_{v,\nu}\,t_{h,\nu}^2\right)\right],
\end{equation}
where $r_\nu$ is independent of $\zeta$ and $r_\nu\asymp 1$. Then, we
compute $r_\nu$ with high enough accuracy: we prove that
\begin{equation}\label{r-nu:1}
  r_\nu=e^{i(\varphi(b_{\nu,0})-\varphi(a_{\nu,0}))}\, 
        \left[1-\frac{\det T_\nu}{2a_{\nu,0}a_{\nu,0}^*}+
            O(t_{h,\nu}^4,\,e^{-2\pi Y/\varepsilon},\,
        t_{v,\nu}\,t_{h,\nu}^2) \right].
\end{equation}
Representations~\eqref{R-nu:1} and~\eqref{r-nu:1} imply
Proposition~\ref{pro:r-nu}.  Indeed, to get~\eqref{R-nu}, one has to
substitute~\eqref{r-nu:1} into~\eqref{R-nu:1} and to take into account
that, in~\eqref{r-nu:1}, the second and the third terms in the square
brackets are bounded by a constant independent of $E$, $\zeta$ and
$\varepsilon$. Note that, from the second point of
Theorem~\ref{th:T-matrices} and estimates from
Corollary~\ref{rough-est}, it follows that
\begin{equation}\label{r:0}
  \left|\frac{\det T_\nu}{a_{\nu,0}a_{\nu,0}^*}\right|\le C\,t_{h,\nu}^2(Re E).
\end{equation}
To prove~\eqref{R-nu:1}, we use the following observation.
\begin{Le}
  \label{zeros} Pick  $\nu\in\{0,\pi\}$. For sufficiently small
  $\varepsilon$, in the case of Theorem~\ref{th:T-matrices}, one has
  \begin{itemize}
  \item in the strip $|\im\zeta|<Y$, each of the functions
    $\zeta\mapsto a_\nu(\zeta,E)$ and $\zeta\mapsto b_\nu(\zeta,E)$
    has one zero per period;
  \item the imaginary part of the zeros have the asymptotics
    $-Y_{v,\nu}(E)+O(\varepsilon)$;
  \item for any zero of $a_\nu$, there exists a unique zero of $b_\nu$
    such that the distance between them is bounded by
    $C\,\varepsilon\,t_{h,\nu}^2(\re E)$.
  \end{itemize}
\end{Le}
\noindent We prove this lemma later. In view of the
first point of Lemma~\ref{zeros}, we can represent $R_\nu$ in the form
\begin{equation}
  \label{R-nu:3}
  R_\nu(\zeta)=\Pi_\nu(\zeta)\,\rho_\nu(\zeta)\quad\text{where}\quad
  \Pi_\nu(\zeta)=\frac{e^{2\pi i(\zeta-\zeta_b)/\varepsilon}-1}
  {e^{2\pi i(\zeta-\zeta_a)/\varepsilon}-1},
\end{equation}
where $\zeta_a$ (resp. $\zeta_b$) is one of the zeros of $a$ (resp.
$b$) in the strip $\{|\im\zeta|\le Y\}$, and $\rho_\nu$ is a
$\varepsilon$-periodic function analytic in this strip. The
representation~\eqref{R-nu:1} then follows from the representations:
\begin{gather}\label{R-nu:4}
  \Pi_\nu(\zeta)=1+O\left(e^{-2\pi\im\zeta/\varepsilon}\,t_{v,\nu}
    \,t_{h,\nu}^2\right)\quad\text{for}\quad|\im\zeta|\le y,\\
  \label{R-nu:5}
  \rho_\nu(\zeta)=\rho_{\nu,0}+O(e^{-2\pi(Y-|\im\zeta|)/\varepsilon})
  \quad\text{for}\quad|\im\zeta|\le
  Y\quad\text{and}\quad\rho_{\nu,0}\asymp 1,
\end{gather}
where $\rho_{\nu,0}$ is the $0$-th Fourier coefficient of $\rho$.
Indeed, to get~\eqref{R-nu:1}, one has just to
substitute~\eqref{R-nu:4} and~\eqref{R-nu:5} into~\eqref{R-nu:3} and
to take into account the fact that the error term in~\eqref{R-nu:5} is
uniformly small when
$|\im\zeta|\le y$. And the latter follows from~\eqref{R-nu:6}.\\
Check~\eqref{R-nu:4}. In view of the second and the third points of
Lemma~\ref{zeros}, and~\eqref{R-nu:6}, for sufficiently small
$\varepsilon$ and $|\im\zeta|\le y$, we get
\begin{equation*}
    \Pi_\nu(\zeta)-1= e^{2\pi
      i(\zeta-\zeta_b)/\varepsilon}\, \frac{1-e^{2\pi
        i(\zeta_b-\zeta_a)/\varepsilon}}{e^{2\pi i(\zeta-\zeta_a)/\varepsilon}-1}=
    O\left(e^{-2\pi(\im\zeta+Y_{v,\nu})/\varepsilon}\,t_{h,\nu}^2\right)
    =O\left(e^{-2\pi\im\zeta/\varepsilon}\,t_{v,\nu}\,t_{h,\nu}^2\right),
\end{equation*}
where, at the last step, we have used~\eqref{R-nu:7}.
This proves~\eqref{R-nu:4}. \\
Recall that $\rho_{\nu,0}$ be the zeroth Fourier coefficient of
$\rho$. To prove~\eqref{R-nu:5}, it suffices to check that,
\begin{equation}
  \label{rho:2} 
|\rho(\zeta)-\rho_{\nu,0}|\le C e^{-2\pi Y/\varepsilon}\,
e^{2\pi|\im\zeta|/\varepsilon}\quad\text{for}\quad|\im\zeta|\le
  Y\quad\text{and}\quad\rho_{\nu,0}\asymp 1.
\end{equation}
Both these estimates follow from the representations
\begin{equation}
  \label{rho:3}
   \rho(\zeta)=\frac{b_{\nu,1}}{a_{\nu,1}}\,(1+o(1))\text{ for }
   \im\zeta=-Y,\quad\quad
   \rho(\zeta)=\frac{b_{\nu,0}}{a_{\nu,0}}\,(1+o(1))\text{ for }  \im\zeta=Y. 
\end{equation}
Indeed, in view of Corollary~\ref{rough-est}, one has
$\left|\frac{b_{\nu,0}}{a_{\nu,0}}\right|,
\,\left|\frac{b_{\nu,1}}{a_{\nu,1}}\right|\ \asymp 1$. Therefore, any
of the representations~\eqref{rho:3} implies that $\rho_{\nu,0}\asymp
1$;~\eqref{rho:3} also implies that, for $|\im\zeta|= Y$, we have
$|\rho(\zeta)|\le C$. This bound and general properties of periodic
analytic functions imply~\eqref{rho:2}.  So, to complete the proof
of~\eqref{R-nu:5}, we need only to
check~\eqref{rho:3}.\\
We check only the first of the representations~\eqref{rho:3}; the
other one is proved similarly.  First, we note that, for sufficiently
small $\varepsilon$ and $\im\zeta=-Y$,
\begin{equation*}
  \Pi_\nu(\zeta)=\frac{e^{2\pi i(\zeta-\zeta_b)/\varepsilon}-1}
  {e^{2\pi i(\zeta-\zeta_a)/\varepsilon}-1}=1+o(1).
\end{equation*}
Indeed, this follows from the last two points of Lemma~\ref{zeros}
and~\eqref{R-nu:6}.  Now, in view of~\eqref{R-nu:3}, it suffices to
check that, for $\im\zeta=-Y$,
\begin{equation*}
  R_\nu(\zeta)=\frac{b_{\nu,1}}{a_{\nu,1}}\,(1+o(1)),
\end{equation*}
which follows from
\begin{equation*}
  a_\nu(\zeta)=a_{\nu,1}e^{2\pi i\zeta}(1+o(1))\quad
  \text{and}\quad b_\nu(\zeta)=b_{\nu,1}e^{2\pi i\zeta}(1+o(1)).
\end{equation*}
We prove only the first one; the second is proved similarly. By
Theorem~\ref{th:T-matrices}, Corollary~\ref{rough-est}
and~\eqref{R-nu:7}, for $\im\zeta=-Y$ and $E\in V_*^\varepsilon$, we
have
\begin{align*}
  a_\nu(\zeta)&=a_{\nu,1}e^{2\pi
    i\zeta}\left(1+o(1)+O\left(\frac{a_{\nu,0}}{a_{\nu,1}} e^{-2\pi
        Y/\varepsilon}\right)\right)=
  a_{\nu,1}e^{2\pi i\zeta}\left(1+o(1)+O\left((t_{v,\nu})^{-1} e^{-2\pi Y/\varepsilon}\right)\right)\\
  &= a_{\nu,1}e^{2\pi i\zeta}\left(1+o(1)+O\left(e^{-2\pi
        (Y-Y_{v,\nu})/\varepsilon}\right)\right)= a_{\nu,1}e^{2\pi
    i\zeta}\left(1+o(1)\right),
\end{align*}
where we have used~\eqref{R-nu:6}. This completes the proof
of~\eqref{R-nu:5} and, thus the proof of~\eqref{R-nu:1}.\\
Now, we compute the constant $r_\nu$ from~\eqref{R-nu:1}. First, we prove that  
\begin{equation}
  \label{r:1}
  r_\nu\,r_\nu^*=1-\frac{\det
  T_\nu}{a_{\nu,0}\,a_{\nu,0}^*}+O(t_{h,\nu}^2\,t_{v,\nu},\,e^{-2\pi
  Y/\varepsilon}).
\end{equation}
This relation follows from the relations
\begin{equation}
  \label{r:2}
  R_\nu\,R_\nu^*=1-\frac{\det T_\nu}{a_{\nu}\,a_{\nu}^*}.
\end{equation}
and from the fact that, for $\zeta\in\R$,
\begin{equation}\label{r:3}
  a_\nu(\zeta)=a_{\nu,0}(1+ O(t_{v,\nu})).
\end{equation}
Indeed, recall that all the functions we work with are
$\varepsilon$-periodic; substituting~\eqref{R-nu:1} and~\eqref{r:3}
into~\eqref{r:2} and integrating along $\R$ over a period, we get
\begin{equation*}
  r_\nu r_\nu^*(1+O(e^{-2\pi
  Y/\varepsilon},\,t_{v,\nu}\,t_{h,\nu}^2))=1-\frac{\det
  T_\nu}{a_{\nu,0}\,a_{\nu,0}^*}(1+ O(t_{v,\nu})).
\end{equation*}
In view of~\eqref{r:0}, this immediately implies~\eqref{r:1}. So, to
complete the proof of~\eqref{r:1}, we have only to prove the
relations~\eqref{r:2} and~\eqref{r:3}. The relation~\eqref{r:2}
follows from the equalities $\det T_\nu=a_\nu a_\nu^*-b_\nu b_\nu^*$
and~\eqref{eq:27}. To prove the relation~\eqref{r:3}, we
rewrite~\eqref{a,b:F} in the form
\begin{equation}
  \label{r:4}
  a_\nu=a_{\nu,0}\left[1+\frac{a_{\nu,-1}(\zeta)}{a_{\nu,0}}+
    \frac{a_{\nu,1}}{a_{\nu,0}} e^{2\pi
      i\zeta/\varepsilon}\left(1+ 
      \frac{a_{\nu,2}(\zeta)}{a_{\nu,1} e^{2\pi
          i\zeta/\varepsilon}}\right)\right]. 
\end{equation}
By~\eqref{abs-Fourier}, $\D\sup_{\zeta\in\R}\left|\frac{a_{\nu,2}
    (\zeta)} {a_{\nu,1}e^{2\pi i\zeta/\varepsilon}}\right|=o(1)$, and,
by Corollary~\ref{rough-est}, one has $\frac{a_{\nu,1}}{a_{\nu,0}}=
O(t_{v,\nu})$. Therefore, to prove~\eqref{r:3}, it suffices to check
that, for $\zeta\in\R$
\begin {equation}
  \label{r:5}
  g(\zeta):=\frac{a_{\nu,-1}(\zeta)}{a_{\nu,0}}=o(t_{v,\nu}).
\end{equation}
Let us check this. We know that
\begin{enumerate}
\item $g$ is analytic in the half plane $\{\im\zeta\le Y\}$ and tends
  to zero as $\im\zeta\to -\infty$ (as it is the sum of the Fourier
  series terms with the negative indexes of a function analytic in the
  strip $\{|\im\zeta|\le Y\}$ );
\item for $\im\zeta=Y$, one has $|g|\le C$ (by~\eqref{abs-Fourier}). 
\end{enumerate}
This implies that $|g|\le Ce^{-2\pi (Y-\im\zeta)/\varepsilon}$ in the
half plane $\{\im\zeta\le Y\}$. In view of~\eqref{R-nu:7}
and~\eqref{R-nu:6}, this implies~\eqref{r:5}, hence,~\eqref{r:1}.\\
Finally, we check that
\begin{equation}
  \label{r:6}
  \varphi(r_\nu)=\varphi(b_{\nu,0})-\varphi(a_{\nu,0})+
  O(t_{h,\nu}^2\,t_{v,\nu},\,e^{-2\pi Y/\varepsilon}).
\end{equation}
Therefore, for $\zeta\in\R$, we substitute the
representations~\eqref{R-nu:1} and~\eqref{r:3} in the relation
$b_\nu=R_\nu a_\nu$, and integrate the result over the period. As
$a_{\nu,0}$ is the zeroth Fourier coefficient of $a_\nu$, the mean
value of the error term in~\eqref{r:3} is zero. Hence, $b_{\nu,0}=
r_\nu a_{\nu,0} (1+O(t_{h,\nu}^2\,t_{v,\nu},\,e^{-2\pi
  Y/\varepsilon}))$ which implies~\eqref{r:6}.\\
Representations~\eqref{r:1},~\eqref{r:6} and estimate~\eqref{r:0}
imply~\eqref{r-nu:1}.  The proof of Proposition~\ref{pro:r-nu} is
complete.\qed
\smallpagebreak {\it Proof of Lemma~\ref{zeros}.\/} We check the first
and the second point for $a_\nu$; for $b_\nu$, the proof is similar.
Theorem~\ref{th:T-matrices} implies that, for $|\im\zeta|\le Y$,
$a_\nu$ admits the representation
\begin{equation}
  \label{zeros:1}
  a_\nu(\zeta)=a_{\nu,0}(1+g_0)+a_{\nu,1} e^{2\pi i\zeta/\varepsilon}(1+g_1)
  \quad\text{where}\quad|g_0|+|g_1|=o(1).
\end{equation}
Therefore, the possible zeros of $a_\nu$ in the strip $\{|\im\zeta|\le
Y\}$ are located in $o(\varepsilon)$-neighborhoods of the points
\begin{equation}
  \label{zeros:2}
  \frac\varepsilon{2\pi i}\ln \left(-a_{\nu,0}/a_{\nu,1}\right)+
  l\varepsilon,\quad l\in\Z.
\end{equation}
This, Corollary~\ref{rough-est} and the first point in
Lemma~\ref{zeros} imply the second point of Lemma~\ref{zeros}.\\
To prove the first point of Lemma~\ref{zeros}, we apply Rouch{\'e}'s
Theorem to the functions $f=a_{\nu,0}+a_{\nu,1} e^{2\pi i
  \zeta/\varepsilon}$ and $\delta f=a_{\nu,0}g_0+a_{\nu,1} e^{2\pi i
  \zeta/\varepsilon}g_1$.  Clearly, all the zeros of $f$ are simple
and they are all listed in~\eqref{zeros:2}. Let $\zeta_a$ be one of
them. One compares $f$ and $\delta f$ on the circle centered at
$\zeta_a$ of radius $c\varepsilon$ (where $c$ is a fixed positive
sufficiently small constant independent of $\varepsilon$). As
\begin{equation*}
  \frac{\delta  f(\zeta)}{f(\zeta)}=
  \frac{g_0}{1-u}+\frac{g_1}{1-1/u},\quad 
  u=e^{2\pi i (\zeta-\zeta_a)/\varepsilon},
\end{equation*}
then, on such a circle, one has $\delta f/f=o(1)$. This and Rouch{\'e}'s
Theorem imply that $a_\nu$ has a unique simple zero in
$c\varepsilon$-neighborhood of $\zeta_a$. This implies the first two
points of Lemma~\ref{zeros} for $a_\nu$.
\smallpagebreak To prove the last point of Lemma~\ref{zeros}, we
compare the zeros of the functions $b_\nu b_\nu^*$ and
$a_\nu\,a_\nu^*$ inside the strip $\{-Y\le \im\zeta\le 0\}$.  We use the
following observations:
\begin{itemize}
\item by the first two points of Lemma~\ref{zeros}, in the strip
  $\{-Y\le \im\zeta\le 0\}$, all the zeros $b_\nu\,b_\nu^*$ are zeros
  of $b_\nu$, and all the zeros $a_\nu\,a_\nu^*$ are zeros of $a_\nu$;
\item we know $a_\nu a_\nu^*-b_\nu b_\nu^*=\det T_\nu$ and that
  $T_\nu=O(1)$ (see the second point of Theorem~\ref{th:T-matrices}).
\end{itemize}
So, the zeros of $a_\nu a_\nu^*$ have to be exponentially close to
those of $a_\nu a_\nu^*-\det T_\nu$, i.e. to the zeros of $b_\nu
b_\nu^*$. To study the distance between the zeros of $a_\nu a_\nu^*$
and those of $a_\nu a_\nu^*-\det T_\nu$, we again use Rouch{\'e}'s
Theorem. Therefore, we pick $\zeta_a$, a zero of $a_\nu$ and compare
the functions $f=a_\nu a_\nu^*$ and $\delta f=\det T_\nu$ on $C_r$,
the circle centered at $\zeta_a$ of radius
\begin{equation*}
  r=\frac{r_0\varepsilon}{a_{\nu,0}a_{\nu,0}^*}.
\end{equation*}
where $r_0$ is a fixed positive constant, sufficiently large but
independent of $\varepsilon$. Note that, by Corollary~\ref{rough-est},
one has
\begin{equation}
  \label{r:est}
  |r|\asymp r_0\varepsilon\,t_{h,\nu}^2(\re E).
\end{equation}
When applying Rouch{\'e}'s theorem, we have to control $f$ on $C_r$.
Therefore, we use the relation
\begin{equation}
  \label{zeros:f-prime}
  f'(\zeta)=-\frac{2\pi
      i}\varepsilon\,a_{\nu,0}^* a_{\nu,0}(1+o(1))
    \quad\text{for}\quad |\zeta-\zeta_a|\le \varepsilon^2.
\end{equation}
We prove~\eqref{zeros:f-prime} later, and, now, we use it to complete
the proof of Lemma~\ref{zeros}. By means of~\eqref{zeros:f-prime}
and~\eqref{r:est}, for $|\zeta-\zeta_a|=r$, we get
\begin{equation*}
  |f(\zeta)|=\frac{2\pi}\varepsilon\,a_{\nu,0}^* a_{\nu,0}r(1+o(1))=
   2\pi r_0(1+o(1)).
\end{equation*}
As $\delta f=\det T_\nu=O(1)$, this implies that 
\begin{equation*}
  \max_{|\zeta-\zeta_a|=r}\left|\frac{\delta
  f(\zeta)}{f(\zeta)}\right|\le C/r_0.
\end{equation*}
So, if $r_0$ is fixed sufficiently large, then, for sufficiently small
$\varepsilon$, \ $f-\delta f$ has one simple zero inside the circle
$|\zeta-\zeta_a|=r$. As this is a zero of $b_\nu$, and as $r$ admits
the estimate~\eqref{r:est}, this implies the third point of
Lemma~\ref{zeros}.\\
To complete the proof of this lemma, we only have to
check~\eqref{zeros:f-prime}. Therefore, first, we note that,
by~\eqref{zeros:1}, for $-Y\le \im \zeta\le 0$, one has
\begin{equation*}
  a_\nu^* \dsize =a_{\nu,0}^*(1+o(1))+a_{\nu,1}^* e^{-2\pi
    i\zeta/\varepsilon}(1+o(1))=
  a_{\nu,0}^*\left(1+o(1)+o\left(\frac{a_{\nu,1}^*}{a_{\nu,0}^*}\right)\right)
=a_{\nu,0}^*(1+o(1)),
\end{equation*}
where we have used Corollary~\ref{rough-est} to estimate
$\frac{a_{\nu,1}^*}{a_{\nu,0}^*}$.  The result of this computation
and~\eqref{zeros:1} imply that, for $ -Y\le \im \zeta\le 0$,
\begin{equation}\label{zeros:3}
  f(\zeta)=a_{\nu,0}^* a_{\nu,0}(1+o(1))+
  a_{\nu,0}^* a_{\nu,1} e^{2\pi i \zeta/\varepsilon}(1+o(1)).
\end{equation}
The Cauchy estimates applied to $o(1)$, the functions
from~\eqref{zeros:3} give $\frac{\partial}{\partial \zeta}o(1)=o(1)$
in any fixed compact of the strip $\{-Y<\im\zeta<0\}$.  Therefore, for
$|\zeta-\zeta_a|=\varepsilon^2$, we get
\begin{equation*}
  f'(\zeta)= \frac{2\pi i}\varepsilon\,a_{\nu,0}^* a_{\nu,1} e^{2\pi i
    \zeta_a/\varepsilon}(1+o(1))+ 
  o(a_{\nu,0}^* a_{\nu,0})+o( a_{\nu,0}^* a_{\nu,1} e^{2\pi i
    \zeta_a/\varepsilon}). 
\end{equation*}
As $a_{\nu,1} e^{2\pi i \zeta_a/\varepsilon}=-a_{\nu,0}$, this
implies~\eqref{zeros:f-prime}. This completes the proof of
Lemma~\ref{zeros}.\qed
\subsection{Asymptotics of the coefficients of the monodromy matrix}
\label{sec:asympt-monodr-matr}
Using Theorem~\ref{th:T-matrices} and Proposition~\ref{pro:r-nu}, we
prove Theorem~\ref{th:M-matrices}. Actually, we compute only the
matrix $M_\pi$ corresponding to the consistent basis
$\{f_\pi,f_\pi^*\}$.  The asymptotic representations for the
coefficients of the matrix
$M_0$ are obtained similarly. The proof consists of two steps.\\
\subsubsection{Combinations of Fourier coefficients}
\label{sec:four-coeff-comb-1}
First, we define the functions $\alpha_\nu$ and the quantities
$\check\Phi_\nu$, $T_{v,\nu}$, $T_h$, $\theta$ and $z_\nu$ introduced
in~\eqref{A-pi:rough}~--~\eqref{alpha:as} in terms of the Fourier
coefficients of the transition matrices. The asymptotics of the
Fourier coefficient combinations met here are computed in terms of the
iso-energy curve $\Gamma$ in section~\ref{sec:four-coeff-comb}.\\
{\it 1. The phases.} \ The phases $\check \Phi_\nu$ are defined by the
formulae
\begin{equation}
  \label{vrai-phases}
  \begin{split}
    \check\Phi_0=\frac\varepsilon2\,
    \left(\varphi(a_{\pi,0})+\varphi(a_{0,0})
      -\varphi(b_{\pi,0})+\varphi(b_{0,0})\right),\\
    \check\Phi_\pi=
    \frac\varepsilon2\,\left(\varphi(a_{\pi,0})+\varphi(a_{0,0})
      +\varphi(b_{\pi,0})-\varphi(b_{0,0})\right).
  \end{split}
\end{equation}
In section~\ref{sec:vrai-phases:vrai-as}, we check that these phases
admit the representations~\eqref{check-Phi:as}. They imply in
particular~\eqref{check-phi:est}.\\
{\it 2. The constant $\theta$.} \ Let
\begin{equation}
  \label{theta}
  \theta=-\babs{\frac{a_{0,0}}{a_{\pi,0}}}\,\det T_{\pi}.
\end{equation}
Note that, in view of~\eqref{det}, one has 
\begin{equation*}
  \babs{\frac{a_{\pi,0}}{a_{0,0}}}\,\det T_{0}=-\frac1\theta.
\end{equation*}
In section~\ref{sec:constant-theta}, we prove that $\theta$ admits the
representations~\eqref{theta:as} which, in particular
imply~\eqref{theta:est}.\\
{\it 3. The coefficients $T_h$ and $T_{v,\nu}$.} \ Let
\begin{equation}
  \label{T:definitions}
  T_h=\babs{ a_{\pi,0}\,a_{0,0}}^{-1}\quad\text{and}\quad
  T_{v,\nu}=\babs{\frac{a_{\nu,1}}{a_{\nu,0}}}.
\end{equation}
Using computations analogous to those done in
section~\ref{sec:constant-theta}, one proves
representations~\eqref{T:as}. These show that, for $E\in
V^\varepsilon_*$,
\begin{equation}
  \label{eq:28} 
  |T_h|\asymp |t_h|=|t_{h,0} t_{h,\pi}|\quad\text{and}\quad
  ||T_{v,\nu}|\asymp|t_{v,\nu}|.
\end{equation} 
{\it 4.The constant $z_\nu$.} \ Let
\begin{equation}
  \label{z-nu:definition}
  z_\nu=-\frac1{2\pi}\,\varphi\left(\frac{a_{\nu,1}}{a_{\nu,0}}\right).
\end{equation}
Using computations analogous to those performed in
section~\ref{sec:vrai-phases:vrai-as}, one proves~\eqref{z-nu:as}.
Estimate~\eqref{z-nu-prime} is proved in the section~\ref{sec:z-nu}.
It implies~\eqref{z-nu:est}.\\
{\it 5. The functions $\alpha_\nu$.} \ Define $\alpha_\nu=
a_\nu/a_{\nu,0}$. One has
\begin{Le}
  \label{alpha-nu}
  Fix $0<y<Y_m$. For sufficiently small $\varepsilon$, in the case of
  Theorem~\ref{th:T-matrices}, for $|\im \zeta|<y$ and $E\in V_*$, one
  has~\eqref{alpha:as}.
\end{Le}
\demo Start with~\eqref{r:4} or, equivalently, with
\begin{equation}
  \label{r:4a}
  \alpha_\nu=1+g(\zeta)+T_{v,\nu} e^{2\pi
  i(\frac\zeta\varepsilon-z_\nu)}\left(1+\tilde
  g(\zeta)\right),\quad\quad 
  g(\zeta)=\frac{a_{\nu,-1}(\zeta)}{a_{\nu,0}},\quad \tilde
  g(\zeta)=\frac{a_{\nu,2}(\zeta)}{a_{\nu,1} e^{2\pi
  i\zeta/\varepsilon}}.
\end{equation}
When proving~\eqref{r:5}, we have seen that $|g|\le Ce^{-2\pi
  (Y-\im\zeta)/\varepsilon}$ in the half plane $\{\im\zeta\le Y\}$.
Similarly, one proves that $|\tilde g|\le Ce^{-2\pi
  (Y+\im\zeta)/\varepsilon}$ in the half plane $\{\im\zeta\ge -Y\}$.
In view of~\eqref{R-nu:7} and~\eqref{R-nu:6}, in the strip $|\im
\zeta|\le y$, one has $|T_{v,\nu} e^{2\pi i\zeta\varepsilon}|\le C$.
These three estimates imply~\eqref{alpha:as}. \qed\\
Note that~\eqref{alpha:as} can be simplified into~\eqref{alpha:est}.\\
{\it 6. Real analyticity.} \ Note that $\check\Phi_\nu$, $T_{v,\nu}$,
$T_h$, $\theta$ and $z_\nu$, regarded as functions of $E$, are real
analytic in $V_*^\varepsilon$ (this follows from the definitions of
$\babs{\cdot}$ and $\varphi(\cdot)$). Therefore, each of them is
invariant with respect to the operation $*$ (see~\eqref{star}).
\subsubsection{Computing the matrix $M_\pi$}
\label{sec:proof-theorem-ref}
The representation~\eqref{factorization} and the relation $b_\nu=R_\nu
a_\nu$ imply that
\begin{equation}
  \label{M:11}
  A_\pi=a_\pi a_{0}+R_\pi R_0^*a_\pi a_{0}^*\quad\text{and}\quad
  B_\pi=R_0\,a_\pi a_{0}+R_\pi\,a_\pi a_{0}^*.
\end{equation}
Now, for $\nu\in\{0,\pi\}$,
\begin{itemize}
\item in~\eqref{M:11}, we substitute the representation
  $a_\nu=\babs{a_{\nu,0}}\,e^{i\varphi(a_{\nu,0})}\,\alpha_\nu$;
\item in~\eqref{M:11}, we replace the functions $R_\nu$ by their
  representations~\eqref{R-nu};
\item we express the Fourier coefficient combinations we meet in terms of
  $\check\Phi_\nu$, $T_{v,\nu}$, $T_h$, $\theta$ and $z_\nu$;
\item we use det$T_0T_\pi=1$.
\item we use the invariance of $\check\Phi_\nu$, $T_{v,\nu}$, $T_h$,
  $\theta$ and $z_\nu$ with respect to the transformation $*$.
\end{itemize}
This leads to the formulae
\begin{equation}
  \label{A-pi}
  A_\pi=2\,\frac{\alpha_\pi e^{i\frac{\check\Phi_\pi}\varepsilon}\,C_0}{T_h}+
  \alpha_\pi\,\alpha_0^*\,e^{i\frac{\check\Phi_\pi-\check\Phi_0}\varepsilon}\,\left\{
    \frac{\theta+1/\theta}2+\frac{T_h}4+\frac{O_\pi+O_0^*}{T_h}+
    \frac{O_\pi/\theta+O_0^*\theta}2+\frac{O_\pi O_0^*}{T_h}\right\},
\end{equation}
and
\begin{equation}
  \label{B-pi}
  B_\pi\,e^{-i\Delta}=2\,\frac{\alpha_\pi
  e^{i\frac{\check\Phi_\pi}\varepsilon}\,C_0}{T_h}+
  \alpha_\pi\,e^{i\frac{\check\Phi_\pi}\varepsilon}\,\left\{\frac{\alpha_0
      e^{i\frac{\check\Phi_0}\varepsilon}/\theta+\alpha_0^* e^{-i\frac{\check\Phi_0}\varepsilon}\theta}2+
    \frac{\alpha_0e^{i\frac{\check\Phi_0}\varepsilon}O_0+\alpha_0^*e^{-i\frac{\check\Phi_0}\varepsilon}
      O_\pi}{T_h}\right\}.
\end{equation}
where $\Delta=\varphi(b_{0,0})-\varphi(a_{0,0})$. Furthermore, one
checks that, for $|\im\zeta|\le y$ and $E\in V_*^\varepsilon$, one has
\begin{equation}
  \label{A-pi:1}
    \alpha_\pi\,\alpha_0^*\,e^{i\frac{\check\Phi_\pi-\check\Phi_0}
      \varepsilon}\,\Big\{\cdots\Big\}= 
    \frac12 e^{i\frac{\check\Phi_\pi-\check\Phi_0}\varepsilon}\,
    \left(\theta+\frac1\theta\right)+
    O\left(pT_{v,0},\,pT_{v,\pi},\,T_h,\,pT_Y/T_h\right)
\end{equation}
and
\begin{equation}
  \label{B-pi:1}
  \alpha_\pi\,e^{i\frac{\check\Phi_\pi}\varepsilon}\,\Big\{\cdots\Big\}=
  \frac12\,e^{\frac{i\check\Phi_\pi}\varepsilon}\,
  \left(\frac1\theta\,e^{\frac{i\check\Phi_0}\varepsilon}+\theta\,
    e^{-\frac{i\check\Phi_0}\varepsilon}\right)+
  O\left(pT_{v,0},\,pT_{v,\pi},\,T_h,\,pT_Y/T_h\right).
\end{equation}
In~\eqref{A-pi:1} and~\eqref{B-pi:1}, the terms with the curly
brackets are the ones from~\eqref{A-pi} and~\eqref{B-pi} respectively,
and $p=p(\zeta/\varepsilon)$.  These two representations follow from
estimates~\eqref{alpha:est},~\eqref{T:est:3},~\eqref{check-phi:est},~\eqref{theta:est},~\eqref{O-nu}
and~\eqref{eq:28}. We omit the elementary details.\\
Finally, we ``kill'' the constant factor $e^{-i\Delta}$
in~\eqref{B-pi} by replacing the consistent basis $\{f_\pi,f_\pi^*\}$
with the consistent base $\{g,g^*\}$ where $g=e^{-i\Delta/2}f_\pi$:
for the monodromy matrix corresponding to $\{g,g^*\}$, the coefficient
with index $11$ is equal to $A_\pi$, and the coefficient with index
$12$ is equal to $B_\pi\,e^{-i\Delta}$. For the coefficients $M_{11}$
and $M_{12}$ of this new monodromy matrix, we keep the old notations
$A_\pi$ and $B_\pi$.  With this ``correction'', the asymptotic
representation~\eqref{A-pi:rough} follows from the
representations~\eqref{A-pi} and~\eqref{A-pi:1}, and the asymptotic
representation~\eqref{B-pi:rough} follows from the
representations~\eqref{B-pi} and~\eqref{B-pi:1} This completes the
proof of Theorem~\ref{th:M-matrices}.\qed
%


%
\section{Periodic Schr{\"o}dinger operators}
\label{S3}
\noindent In this section, we discuss the periodic Schr{\"o}dinger
operator~\eqref{Ho} where $V$ is a $1$-periodic, real valued,
$L^2_{loc}$-function. First, we collect well known results needed in
the present paper
(see~\cite{MR2002f:81151,Eas:73,Ma-Os:75,MR80b:30039,Ti:58}). In the
second part of the section, we introduce a meromorphic differential
defined on the Riemann surface associated to the periodic operator.
This object plays an important role for the adiabatic constructions
(see~\cite{Fe-Kl:03e}).
\subsection{Analytic theory of Bloch solutions}
\label{sec:analyt-theory-bloch}
\subsubsection{Bloch solutions}
\label{sec:bloch-solutions}
Let $\psi$ be a nontrivial  solution of the equation
\begin{equation}\label{PSE}
  -\frac{d^2}{dx^2}\psi\,(x)+ V\,(x)\psi\,(x)=\mathcal{E}\psi\,(x),
   \quad x\in\R,
\end{equation}
satisfying the relation $\psi\,(x+1)=\lambda\,\psi\,(x)$ for all
$x\in\R$ with $\lambda\in\C$ independent of $x$. Such a solution is
called a {\it Bloch solution}, and the number $\lambda$ is called the
{\it Floquet multiplier}. Let us discuss properties of
Bloch solutions (see~\cite{MR2002f:81151}).
\smallpagebreak As in section~\ref{sec:periodic-operator}, we denote
the spectral bands of the periodic Schr{\"o}dinger equation by
$[E_1,\,E_2]$, $[E_3,\,E_4]$, $\dots$, $[E_{2n+1},\,E_{2n+2}]$,
$\dots$. Consider $\mathcal{S}_\pm $, two copies of the complex plane
$\mathcal{E}\in\C$ cut along the spectral bands. Paste them together
to get a Riemann surface with square root branch points.  We denote
this Riemann surface by $\mathcal{S}$. In the sequel, $\pi_c:\ 
{\mathcal S}\mapsto\C$ is the canonical projection.
\smallpagebreak One can construct a Bloch solution
$\psi(x,\mathcal{E})$ of equation~\eqref{PSE} meromorphic on $\mathcal
S$.  For any $\mathcal{E}$, we normalize it by the condition
$\psi(1,\mathcal{E})= 1$. Then, the poles of
$\mathcal{E}\mapsto\psi(x,\mathcal{E})$ are projected by $\pi_c$
either in the open spectral gaps or at their ends. More precisely,
there is exactly one simple pole per open gap.  The position of the
pole is independent of $x$ (see~\cite{MR2002f:81151}).
\smallpagebreak Let $\hat{\cdot}:\ {\mathcal S}\mapsto {\mathcal S}$
be the canonical transposition mapping: for any point
$\mathcal{E}\in\mathcal{S}$, the point $\hat{\mathcal{E}}$ is the
unique solution to the equation $\pi_c(\mathcal{E})=E$ different
from $\mathcal E$ outside the branch points.\\
The function $x\mapsto \psi(x,{\hat{\mathcal E}})$ is one more Bloch
solution of~\eqref{PSE}. Except at the edges of the spectrum (i.e. the
branch points of $\mathcal{S}$), the functions $\psi(\cdot,{\mathcal
  E})$ and $\psi(\cdot,\hat {\mathcal E})$ are linearly independent
solutions of~\eqref{PSE}. In the spectral gaps, they are real valued
functions of $x$, and, on the spectral bands, they differ only by the
complex conjugation (see~\cite{MR2002f:81151}).
\subsubsection{The Bloch quasi-momentum}
\label{SS3.2}
Consider the Bloch solution $\psi(x,\mathcal{E})$. The corresponding
Floquet multiplier $\lambda\,(\mathcal{E})$ is analytic on
$\mathcal{S}$. Represent it in the form
$\lambda(\mathcal{E})=\exp(ik(\mathcal{E}))$. The function
$k(\mathcal{E})$ is the {\it Bloch quasi-momentum}.
\\ The Bloch quasi-momentum is an analytic multi-valued function of
$\mathcal{E}$. It has the same branch points as $\psi(x,\mathcal{E})$
(see~\cite{MR2002f:81151}).
\\ Let $D\in \C$ be a simply connected domain containing no branch
point of the Bloch quasi-momentum $k$. On $D$, fix $k_0$, a continuous
(hence, analytic) branch of $k$. All other branches of $k$ that are
continuous on $D$ are then given by the formula
\begin{equation*}
  \label{eq:55}
   k_{\pm ,l}({\mathcal E})=\pm k_0({\mathcal E})+2\pi l,\quad l\in\Z.
\end{equation*}
All the branch points of the Bloch quasi-momentum are of square root
type: let $E_l$ be a branch point, then, in a sufficiently small
neighborhood of $E_l$, the quasi-momentum is analytic as a function of
the variable $\sqrt{\mathcal{E}-E_l}$; for any analytic branch of $k$,
one has
\begin{equation*}
  \label{sqrt}
  k(\mathcal{E})=k_l+c_l\sqrt{\mathcal{E}-E_l}+O(\mathcal{E}-E_l),\quad c_l\not=0,
\end{equation*}
with constants $k_l$ and $c_l$ depending on the branch.\\
Let $\C_+$ be the upper complex half-plane. There exists $k_p$, an
analytic branch of $k$ that conformally maps $\C_+$ onto the quadrant
$\{k\in\C;\ \im k> 0,\,\,\re k> 0\}$ cut along compact vertical
intervals, say $\pi l+i I_l$ where $l\in\N^*$ and $I_l\subset\R$,
(see~\cite{MR2002f:81151}). The branch $k_p$ is continuous up to the
real line.  It is real and increasing along the spectrum of $H_0$; it
maps the spectral band $[E_{2n-1}, E_{2n}]$ on the interval
$[\pi(n-1),\pi n]$. On the open gaps, $\re k_p$ is constant, and $\im
k_p$ is positive and has exactly one maximum; this maximum is non
degenerate.\\
We call $k_p$ the {\it main} branch of the Bloch
quasi-momentum.\\
Finally, we note that the main branch can be analytically continued on
the complex plane cut only along the spectral gaps of the periodic
operator.
\subsection{Meromorphic differential $\Omega$}
\label{sec:Omega}
\subsubsection{The definition and analytic properties}
\label{sec:defin-analyt-prop}
On the Riemann surface $\mathcal S$, 
consider the function
\begin{equation}\label{omega}
\omega({\mathcal E})=
-\frac{\int_0^1 \psi(x,\hat {\mathcal E})\,\left(\dot\psi(x,{\mathcal
      E})-i\dot k({\mathcal E}) x\,\psi(x,{\mathcal E})\right)\,dx} 
{\int_0^1\psi(x,{\mathcal E})\psi(x,\hat {\mathcal E}) dx}.
\end{equation}
where $k$ is the Bloch quasi-momentum of $\psi$,
and the dot denotes the partial derivative with respect to ${\mathcal E}$. This
function was introduced in~\cite{Fe-Kl:03f} (the definition given in
that paper is equivalent to~\eqref{omega}). In~\cite{Fe-Kl:03f}, we
have proved that $\omega$ has the following properties:
\begin{enumerate}
\item the differential $\Omega=\omega\,d{\mathcal E}$ is meromorphic
  on $\mathcal S$; its poles are the points of $P\cup Q$, where $P$ is
  the set of poles of $\psi(x,{\mathcal E})$, and $Q$ is the set of
  points where $k'({\mathcal E})=0$;
\item all the poles of $\Omega$ are simple;
\item $\forall p\in P\setminus Q$, $\res_p \Omega=1$; $\forall q\in
  Q\setminus P$, $\res_q\Omega=-1/2$; $\forall r\in P\cap Q$,
  $\res_r\Omega=1/2$.
\item if $\pi_c({\mathcal E})$ belongs to a gap, then
  $\omega({\mathcal E})\in \R$;
\item if $\pi_c({\mathcal E})$ belongs to a band, then
  $\overline{\omega({\mathcal E})}=\omega(\hat {\mathcal E})$.
\end{enumerate}
The following quantities appeared in the description of the spectrum
of $H_{z,\varepsilon}$ (see sections~\ref{sec:la-descr-prec}
and~\ref{sec:nature-du-spectre})
\begin{gather}
  \label{Lambda-n-V}
  \Lambda_n(V)=\frac12\left(\theta_n(V)+\frac1{\theta_n(V)}\right),\\
  \intertext{where}
  \label{Lambda-omega} \theta_n(V)=\exp(l_n(V)),\quad\quad
  l_n(V)=\int_{g_n} \Omega({\mathcal E}),
\end{gather}
and $g_n$ is a simple closed curve on $\mathcal{S}$ such that
\begin{itemize}
\item $g_n$ is located on $\C\setminus \sigma(H_0)$, the sheet of the
  Riemann surface $\mathcal{S}$ where the Bloch quasi-momentum of
  $\psi(x,{\mathcal E})$ is equal to $k_p(\pi_c({\mathcal E}))$ for
  $\im\pi_c({\mathcal E})>0$;
\item $\pi_c(g_n)$ is a positively oriented loop going once around the
  $n$-th spectral gap of the periodic operator $H_0$.
\end{itemize}
We prove
\begin{Le}
  \label{Lambda:prop}  
  The integral  $l_n$ is  real valued.
\end{Le}
\demo Let ${\mathcal E}_0$ be a point that projects onto an internal
point of a spectral band. Let $U$ be a neighborhood of $\mathcal{E}_0$
where $\pi_c^{-1}$ is analytic. Let here ${\mathcal E}^*=
\pi_c^{-1}(\overline{\pi_c({\mathcal E})})$. By the fifth property of
$\omega$, for ${\mathcal E}\in U$, one has $\omega(\hat {\mathcal
  E})=\overline{\omega({\mathcal E}^*)}$. Consider $g_n$, the
integration contour for $l_n$. We can and do assume that $\pi_c(g_n)$
(as a set, but not as an oriented curve) is symmetric with respect to
the real line. As $\pi_c(g_n)$ intersects the real line at internal
points of spectral bands, starting from one of these intersections,
one can continue $\mathcal{E}\mapsto\mathcal{E}^*$ to $g_n$ as a
continuous function; then, one has $\omega(\hat{\mathcal
  E})=\overline{\omega({\mathcal E}^*)}$. Note that $g_n^*=-g_n$. One
has
\begin{equation}\label{eq:theta:1}
  \begin{split}
    \overline{\int_{g_n}\Omega({\mathcal E})}&= \overline{\int_{g_n}
      \omega({\mathcal E})\,d{\mathcal E}}= \int_{g_n^*}
    \overline{\omega({\mathcal E}^*)}\,d{\mathcal E}=
    \int_{g_n^*} \omega(\hat{\mathcal E})\,d{\mathcal E}\\
    &=-\int_{g_n} \omega(\hat {\mathcal E})\,d{\mathcal E}=
    -\int_{\hat g_n} \omega({\mathcal E})\,d{\mathcal E}= -\int_{\hat
      g_n}\Omega({\mathcal E}).
  \end{split}
\end{equation}
On $\mathcal S$, there are exactly two points, say $q$ and $\hat q$,
in $Q$ that project inside the $n$-th spectral gap of $H_0$.
Furthermore, on $\mathcal S$, there is exactly one point, say $p$, in
$P$ that projects inside the $n$-th spectral gap or at one of its
edges. On ${\mathcal S}\setminus\{q,\hat q, p\}$, up to homotopy, one
has
\begin{equation*}
  \hat g_n=-g_n+\sum_{{\mathcal E}\in \{q,\hat q, p\}} C({\mathcal
  E}),
\end{equation*}
where $C({\mathcal E})$ is a infinitesimally small, positively
oriented circle centered at $\mathcal E$.  This and the description of
the poles of $\Omega$ imply that
\begin{equation}
  \label{eq:theta:2}
  \int_{\hat g_n}\Omega({\mathcal E})=-\int_{g_n} \Omega({\mathcal
    E})+2\pi i \sum_{{\mathcal E}\in \{q,\hat q, p\}}\res_{\mathcal
    E}\Omega({\mathcal E})=-\int_{g_n}\Omega({\mathcal E}).
\end{equation}
Relations~\eqref{eq:theta:1} and~\eqref{eq:theta:2} imply that
$\overline{l_n}=l_n$.  This completes the proof of
Lemma~\ref{Lambda:prop}.  \qed
\\ Lemma~\ref{Lambda:prop} imply 
\begin{Cor}  One has  $\theta_n(V)>0$ and  $\Lambda_n(V)\geq1$. \end{Cor}
%


%
\section{The consistent solutions}
\label{S4}
\noindent In this section, we describe the consistent solutions
$(f_\nu)_{\nu\in\{0,\pi\}}$ used in
section~\ref{sec:transition-matrices}. Many of the results presented
here are taken from~\cite{Fe-Kl:03e}.
\smallpagebreak In~\cite{MR2002h:81069} and~\cite{Fe-Kl:03e}, we have
developed a new asymptotic method to study solutions to an
adiabatically perturbed periodic Schr{\"o}dinger equation i.e., to study
solutions of the equation
\begin{equation}
  \label{G.2a}
 -\frac{d^2}{dx^2}\psi(x,\zeta)+(V(x)+W(\varepsilon
  x+\zeta))\psi(x,\zeta)= E\psi(x,\zeta)
\end{equation}
in the limit $\varepsilon\to 0$. The function $\zeta\mapsto W(\zeta)$
is an analytic function that is not necessarily periodic. The main
idea of the method is to get the information on the behavior of the
solutions in $x$ from the study of their behavior on the complex plane
of $\zeta$. The natural condition allowing to relate the behavior in
$\zeta$ to the behavior in $x$ is the consistency
condition~\eqref{consistency:1}: one can construct solutions
to~\eqref{G.2a} satisfying this condition and having simple standard
asymptotic behavior on certain domains of the complex plane of $\zeta$.\\
We first describe the standard asymptotic behavior of the solutions
studied in the framework of the complex WKB method. The domains where
these solutions have the standard behavior are described in terms of
Stokes lines. So, next, we describe the Stokes lines for $V$, $W$ and
$E$ considered in this paper. Finally, we describe $f_0$ and $f_\pi$,
the solutions used to construct the consistent bases and transitions
matrices of Theorem~\ref{th:T-matrices}.
\subsection{Standard behavior of consistent solutions}
\label{sec:stand-behav-solut}
We now discuss two analytic objects central to the complex WKB method,
the complex momentum defined in~\eqref{complex-mom} and the canonical
Bloch solutions defined below. For $\zeta\in\mathcal{D}(W)$, the
domain of analyticity of the function $W$, we define
\begin{equation}
  \label{eq:29}
  {\mathcal E}(\zeta)=E-W(\zeta)
\end{equation}
The complex momentum and the canonical Bloch solutions are the Bloch
quasi-momentum and particular Bloch solutions of the equation
\begin{equation}
  \label{eq:52}
  -\frac{d^2}{dx^2}\psi+V\psi={\mathcal E}(\zeta)\psi.
\end{equation}
considered as functions of $\zeta$.
\subsubsection{The complex momentum}
\label{sec:kappa}
For $\zeta\in\mathcal{D}(W)$, the domain of analyticity of the
function $W$, the complex momentum is given by the formula
$\kappa(\zeta)=k(\mathcal{E}(\zeta))$ where $k$ is the Bloch
quasi-momentum of~\eqref{Ho}. Clearly, $\kappa$ is a multi-valued
analytic function; a point $\zeta$ such that $W'(\zeta)\ne0$ is a
branch point of $\kappa$ if and only if
\begin{equation}
  \label{bp}
  E_j+W(\zeta)=E\quad\text{ for some }\quad j\in\N^*.
\end{equation}
All the branch points of the complex momentum are of square root type.
\smallpagebreak A simply connected set $D\subset\mathcal{D}(W)$
containing no branch points of $\kappa$ is called {\it regular}. Let
$\kappa_p$ be a branch of the complex momentum analytic in a regular
domain $D$. All the other branches analytic in $D$ are described by
\begin{equation}
  \label{allbr}
  \kappa_m^\pm  =\pm  \kappa_p+2\pi m\quad\text{where}\quad m\in\Z.
\end{equation}
\subsubsection{Canonical Bloch solutions}
\label{sec:canon-bloch-solut}
To describe the asymptotic formulae of the complex WKB method, one
needs Bloch solutions of equation~\eqref{eq:52} analytic in $\zeta$ on
a given regular domain. We build them using the 1-form
$\Omega=\omega\,d\mathcal{E}$ introduced in section~\ref{sec:Omega}.
\smallpagebreak Pick $\zeta_0$, a regular point. Let
$\mathcal{E}_0=\mathcal{E}(\zeta_0)$. Assume that
$\mathcal{E}_0\not\in P\cup Q$ (the sets $P$ and $Q$ are defined in
section~\ref{sec:Omega}). In $U_0$, a sufficiently small neighborhood
of $\mathcal{E}_0$, we fix $k$, a branch of the Bloch quasi-momentum,
and $\psi_\pm(x,\mathcal{E})$, two branches of the Bloch solution
$\psi(x,\mathcal{E})$ such that $k$ is the Bloch quasi-momentum of
$\psi_+$.  Also, in $U_0$, consider $\Omega_\pm$, the two
corresponding branches of $\Omega$, and fix a branch of the function
$\mathcal{E}\mapsto q(\mathcal{E})=\sqrt{k'(\mathcal{E})}$.  Assume
that $V_0$ is a neighborhood of $\zeta_0$ such that ${\mathcal
  E}(V_0)\subset U_0$.  For $\zeta\in V_0$, we let
\begin{equation}
  \label{canonicalBS}
  \Psi_\pm (x,\zeta)=
  q(\mathcal{E})\,e^{\int_{\mathcal{E}_0}^\mathcal{E} \Omega_\pm}
  \psi_\pm (x,\mathcal{E}),\quad\text{where}\quad
  \mathcal{E}=\mathcal{E}(\zeta).
\end{equation}
The functions $\Psi_\pm$ are the {\it canonical Bloch solutions
  normalized at the point $\zeta_0$}. Its quasi-momentum is
$\kappa(\zeta)=k(E-W(\zeta))$.
\smallpagebreak The properties of the differential $\Omega$ imply that
the solutions $\Psi_\pm$ can be analytically continued from
$V_0$ to any regular domain $D$ containing $V_0$. \\
One has (see~\cite{MR2002h:81069})
\begin{equation}
  \label{Wcanonical}
  w(\Psi_+(\cdot ,\zeta),\Psi_-(\cdot ,\zeta))=w(\Psi_+(\cdot ,\zeta_0),\Psi_-(\cdot ,\zeta_0))=
  k'(\mathcal{E}_0)w(\psi_+(\cdot,\mathcal{E}_0),\psi_-(\cdot,\mathcal{E}_0))
\end{equation}
As $\mathcal{E}_0\not\in Q\cup\{E_l,\ l\geq1\}$, the Wronskian
$w(\Psi_+(\cdot ,\zeta),\Psi_-(\cdot ,\zeta))$ does not vanish.
\subsubsection{Solutions having standard asymptotic behavior}
\label{sec:standard-asymptotics}
Fix $E=E_0$. Let $D$ be a regular domain. Fix $\zeta_0\in D$ so that
$\mathcal{E}(\zeta_0)\not\in P\cup Q$. Let $\kappa$ be a branch of the
complex momentum continuous in $D$, and let $\Psi_\pm$ be the
canonical Bloch solutions defined on $D$, normalized at $\zeta_0$ and
indexed so that $\kappa$ be the quasi-momentum for $\Psi_+$.\\
We recall the following basic definition from~\cite{Fe-Kl:03e}
\begin{Def}
  \label{def:1}
  Fix $s\in\{+,-\}$. We say that $f$, a solution of~\eqref{G.2a}, has
  standard behavior (or standard asymptotics)
  $f\sim\exp(s\frac{i}{\varepsilon}
  \int^{\zeta}\kappa\,d\zeta)\cdot\Psi_s$ in $D$ if
  \begin{itemize}
  \item there exists $V_0$, a complex neighborhood of $E_0$, and $X>0$
    such that $f$ is defined and satisfies~\eqref{G.2a}
    and~\eqref{consistency:1} for any $(x,\zeta,E)\in [-X,X]\times
    D\times V_0$;
  \item $f$ is analytic in $\zeta\in D$ and in $E\in V_0$;
  \item for any $K$, a compact subset of $D$, there is $V\subset V_0$,
    a neighborhood of $E_0$, such that, for $(x,\zeta,E)\in
    [-X,X]\times K\times V$, $f$ has the uniform asymptotic
    \begin{equation*}
      \label{stand:as}
      f=e^{\dsize s\,\frac{i}{\varepsilon} \int^{\zeta}
        \kappa\,d\zeta}\, (\Psi_s+o\,(1)),\quad \text{as}\quad
      \varepsilon\to 0,
    \end{equation*}
  \item this asymptotic can be differentiated once in $x$ without
    loosing its uniformity properties.
\end{itemize}
\end{Def}
\noindent Let $(f_+,f_-)$ be two solutions of~\eqref{G.2a} having
standard behavior $f_\pm\sim e^{\dsize\pm\frac{i}{\varepsilon}
  \int^{\zeta} \kappa\,d\zeta}\,\Psi_\pm$ in $D$. One computes
\begin{equation*}
  w(f_+,f_-)=w(\Psi_+,\Psi_-)+o(1).
\end{equation*}
By~\eqref{Wcanonical}, for $\zeta$ in any fixed compact subset of $D$
and $\varepsilon$ sufficiently small, the solutions $(f_+,f_-)$ are
linearly independent.
\subsection{Complex momentum and Stokes lines for
  $W(\zeta)=\alpha\cos\zeta$}
\label{sec:compl-moment-stok}
Now, following~\cite{Fe-Kl:03e}, we discuss the complex momentum and
describe the Stokes lines for $V$, $W$ and $E$ considered in this
paper. In particular, from now on, we assume that
\begin{equation}
  \label{eq:31}
   W(\zeta)=\alpha\cos(\zeta)\quad\text{hence,}\quad
   \mathcal{E}(\zeta)=E-\alpha\cos(\zeta),
\end{equation} 
that $E$ belongs to $J$, a compact interval satisfying the condition
(TIBM) from section~\ref{sec:main-assumption-w}, and that all the gaps
of the periodic operator $H_0$ are open.
\subsubsection{Complex momentum}
\label{sec:complex-momentum} 
{\bf 1.} \ The branch points of the complex momentum are located on
the lines of the set $\text{arccos}\,(\R)$ which consists of the real
line and the lines $\{\re\zeta=\pi l\}$ for $l\in\Z$. The set of
branch points of $\kappa$ is $2\pi$-periodic and symmetric with
respect both to the real line and to the imaginary axis.\\
Define the half-strip $\Pi=\{\zeta\in\C;\ 0<\re\zeta<\pi,\ 
\im\zeta>0\}$. It is a regular domain. Consider the branch points
located on $\partial\Pi$, the boundary of $\Pi$. $\mathcal{E}$
bijectively maps $\partial\Pi$ onto the real line. So, for any $j\in
\N$, there is exactly one branch point solution to~\eqref{bp}; we
denote it by $\zeta_j$. Under condition (TIBM), the branch points
$\zeta_{2n}$ and $\zeta_{2n+1}$ are located on the interval $(0,\pi)$,
i.e.  $0<\zeta_{2n}<\zeta_{2n+1}<\pi$. The branch points $\zeta_1$,
$\zeta_2$, $\dots$ $\zeta_{2n-1}$ are located on the imaginary axis
and satisfy $0<\im \zeta_{2n-1}<\dots<\im\zeta_{2}<\im \zeta_1$. The
other branch points are located on the line $\{\re\zeta=\pi\}$, and
one has $0<\im \zeta_{2n+2}<\im \zeta_{2n+3}<\dots$. In
Fig.~\ref{fig:zon-ga}, we show some of these
branch points.
%
\begin{floatingfigure}{6cm}
  \centering
  \includegraphics[bbllx=71,bblly=583,bburx=247,bbury=721,width=6cm]{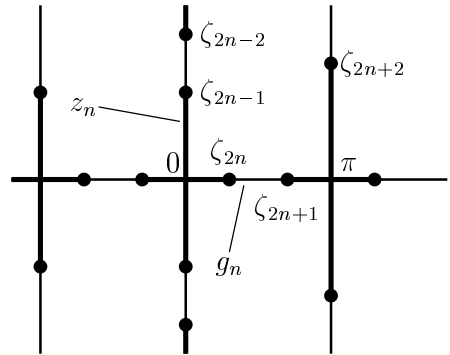}
  \caption{$(z_l)_l$ and $(g_l)_l$}\label{fig:zon-ga}
\end{floatingfigure}
%
\noindent {\bf 2.} \ $\mathcal{E}$ conformally maps the half-strip
$\Pi$ onto the upper half of the complex plane. So, on $\Pi$, we can
define a branch of the complex momentum by the formula
\begin{equation}
  \label{kappa0k0}
  \hskip-4cm
  \kappa_p(\varphi)=k_p(E-\alpha\cos\varphi),
\end{equation}
$k_p$ being the main branch of the Bloch quasi-momentum for the
periodic operator~\eqref{Ho}. We call $\kappa_p$ {\it the main branch}
of the complex momentum.
\vskip.1cm\noindent The discussion in section~\ref{SS3.2} implies the
following. First, $\kappa_p$ conformally maps $\Pi$ into the first
quadrant of the complex plane. Fix $l$, a positive integer.  The
closed segment $z_l:=[\zeta_{2l-1},\zeta_{2l}] \subset\partial \Pi$ is
bijectively mapped on the interval $[\pi(l-1),\pi l]$; on the open
segment $g_l:=(\zeta_{2l}, \zeta_{2l+1})\subset\partial\Pi$, the real
part of $\kappa$ equals to $\pi l$, and its imaginary part is
positive. Two of the intervals $(z_l)_l$ and $(g_l)_l$ are shown in
Fig.~\ref{fig:zon-ga}.
\subsubsection{Stokes lines}
\label{sec:stokes-lines-1}
Let $\zeta_0$ be a branch point of the complex momentum. 
%
\begin{floatingfigure}{7cm}
  \centering
  \includegraphics[bbllx=71,bblly=563,bburx=275,bbury=721,width=7cm]{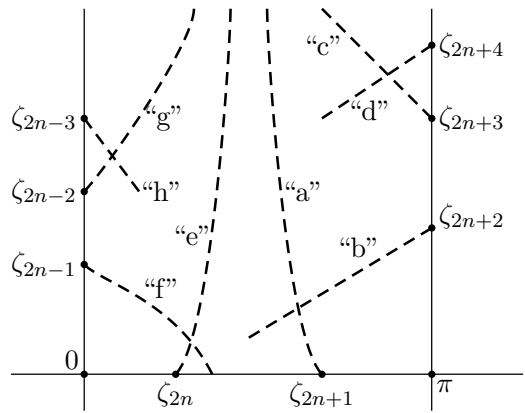}
  \caption{The Stokes lines}\label{bem:St-lines}
\end{floatingfigure}
%
\noindent A {\it Stokes line} beginning at $\zeta_0$ is a curve
$\gamma$ defined by the equation $\text{Im}\,\int_{\zeta_0}^\zeta
(\kappa\,(\xi) -\kappa\,(\zeta_0))d\xi=0$ (where $\kappa$ is a branch
of the complex momentum continuous on $\gamma$). There are three
Stokes lines beginning at each branch point of the complex momentum.
The angles between them at the branch point are all equal to $2\pi/3$.
\smallpagebreak Let us discuss the set of Stokes lines for
$W(\zeta)=\alpha\cos\zeta$. Due to the symmetry properties of
$\mathcal E$, the set of the Stokes lines is $2\pi$-periodic and
symmetric with respect to both the real and the imaginary axes.
 So, it suffices to describe the Stokes lines in $\Pi$.
Here, we follow~\cite{Fe-Kl:03e}.
\smallpagebreak In Fig.~\ref{bem:St-lines}, we have represented
Stokes lines in $\Pi$ by dashed lines.
\vskip.2cm\noindent{\it Elementary properties of Stokes lines.\/}
Recall that the ends of the intervals $(g_l)_l$ are branch points and,
reciprocally, any branch point located on $\partial \Pi$ is an
end of one of the $g_l$'s.\\
Consider the Stokes lines beginning at the ends of $g_n$. The right
end of $g_n$ is $\zeta_{2n+1}$. One of the Stokes lines beginning
at this point goes to the right of $\zeta_{2n+1}$ along $\R$; the
two other Stokes lines beginning at $\zeta_{2n+1}$ are symmetric
with respect to the real line.  Similarly, one of the Stokes lines
beginning at $\zeta_{2n}$, the left end of $g_n$, goes to the left
of $\zeta_{2n}$ along $\R$, and the two other Stokes lines
beginning at $\zeta_{2n}$ are symmetric with respect to the real
line.\\
Consider the Stokes lines beginning at the ends of $g_l$ for either
$l\ge n+1$ or $l\le n-1$. One of these Stokes lines coincides with
$g_{l}$. Let $\zeta_0$ be one of the ends of $g_{l}$. The two Stokes
lines beginning at $\zeta_0$ and different from $g_l$ are symmetric
with respect to the line $\{\re\zeta=\re\zeta_0\}$, see
Fig.~\ref{bem:St-lines}.
\vskip.2cm\noindent{\it Global properties of the Stokes lines in
  $\Pi$.}  First, we discuss the Stokes lines starting at
$\zeta_{2n+1}$,..., $\zeta_{2n+4}$ and $\zeta_{2n}$ denoted
respectively by ``a'',..., ``d'' and ``e''. They are shown in
Fig.~\ref{bem:St-lines} and described by
\begin{Le}[\cite{Fe-Kl:03e}]
  \label{BEM:le:Sl}
  The Stokes lines ``a'',..., ``d'' and ``e'' have the following
  properties:
  \begin{itemize}
  \item the Stokes lines ``a'' and ``e'' stay inside $\Pi$, are
    vertical and disjoint;
  \item the Stokes line ``c'' stays between ``a'' and the line
    $\{\re\zeta=\pi\}$ (without intersecting them) and is vertical;
  \item before leaving $\Pi$, the Stokes lines ``b'' stays vertical
    and intersects ``a'' at a point with positive imaginary part;
  \item before leaving $\Pi$, the Stokes lines ``d'' stays vertical
    and intersects ``c'' above $\zeta_{2n+3}$, the beginning of ``c''.
  \end{itemize}
\end{Le}
\noindent The term ``vertical line'' used in this lemma
means a smooth curve intersecting the lines $\{\im\zeta=C\}$
transversally. The proof of Lemma~\ref{BEM:le:Sl} can be found
in~\cite{Fe-Kl:03e}.\vskip.1cm\noindent
Now, consider the Stokes lines located in $\Pi$ and starting at
$\zeta_{2n-1}$, $\zeta_{2n-2}$ and $\zeta_{2n-3}$. We respectively
denote them by ``f'', ``g'' and ``h'', see Fig.~\ref{bem:St-lines}.
One proves
\begin{Le}
  \label{BEM:le:Sl:left}
  The Stokes lines ``f'', ``g'' and ``h'' have the following
  properties:
  \begin{itemize}
  \item the Stokes line ``g'' is vertical and stays between ``e'' and
    the line $\{\re\zeta=0\}$ without intersecting them;
  \item before leaving $\Pi$, the Stokes lines ``f'' stays vertical
    and intersects ``e'' at a point with positive imaginary part;
  \item before leaving $\Pi$, the Stokes lines ``h'' stays vertical
    and intersects ``g'' above $\zeta_{2n-2}$, the beginning of ``g''.
  \end{itemize}
\end{Le}
\noindent We omit the proof of this lemma as it is similar to the proof of
Lemma~\ref{BEM:le:Sl}.
\subsection{Two consistent solutions}
\label{two-bases}
We now introduce two solutions of~\eqref{G.2}
satisfying~\eqref{consistency:1}. For $y>0$, we define $S_y=\{|\im
\zeta|< y\}$. \\
Fix $\tilde Y>\im\zeta_{2n+4}$. The solutions we describe are analytic in the
strip $S_{\tilde Y}$.\\
We first describe the branch of the complex momentum used to write the
asymptotics of these solutions. Define the strip 
\begin{equation*}
  S^p=\{\zeta\in\C; 0<\im\zeta<\min(\im\zeta_{2n-1},\im\zeta_{2n+2})\}.
\end{equation*}
It is regular.  Analytically continue $\kappa_p$ to this strip.
Recall that the integer $n$ in the condition (TIBM) is even. Let
\begin{equation}
  \label{kappaPkappa}
  \kappa(\zeta)=\kappa_p(\zeta)-n\pi.
\end{equation}
As $n$ is even, the discussion in the section~\ref{sec:kappa} shows
that $\kappa$ is a branch of the complex momentum. It is continuous up
to the boundary of the strip $S^p$; one has
\begin{equation*}
  \label{kappa:special}
  \kappa(\zeta_{2n})=\kappa(\zeta_{2n+1})=0.
\end{equation*}
\subsubsection{The solution $f_\pi$} 
\label{sec:solution-f_pi}
\begin{figure}[htbp]
  \centering \subfigure[The domain $\mathcal{D}_\pi$]{
    \includegraphics[bbllx=71,bblly=591,bburx=205,bbury=721,width=4.5cm]{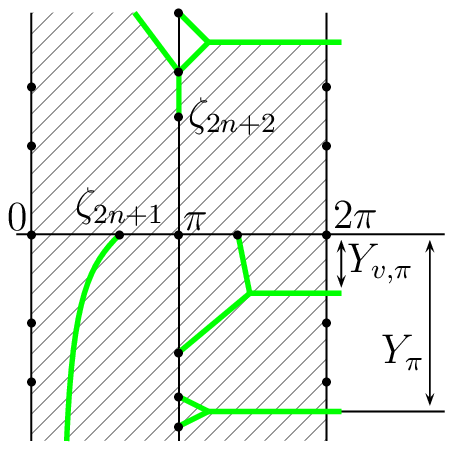}
    \label{BEM:fig:dp:1}}
  \hskip3cm \subfigure[The domain $\mathcal{D}_0$]{
    \includegraphics[bbllx=71,bblly=600,bburx=204,bbury=721,width=4.5cm]{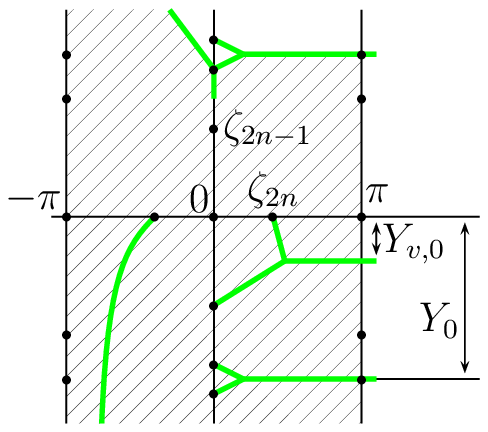}
    \label{BEM:fig:dp:2}}
  \caption{The continuation diagrams}\label{fig:prol}
\end{figure}

Consider ${\mathcal D}_\pi$, the subdomain of the domain
$D_\pi=\{|\im\zeta|< \tilde Y,\ 0<\re\zeta<2\pi\}$ shown in
Fig.~\ref{BEM:fig:dp:1}.
%
%
\noindent Its boundary consists of the lines bounding $D_\pi$ and of
the segments of Stokes lines and lines $\{\re\zeta=C\}$ beginning at
the intersection points of Stokes lines. The domain ${\mathcal D}_\pi$
is simply connected.\\
Let $\kappa_\pi$ be the analytic continuations of $\kappa$ from
$S^p$ to ${\mathcal D}_\pi$, i.e., for $\zeta\in {\mathcal
  D}_\pi\cap S^p$
\begin{equation}
  \label{kappa-pi}
  \kappa_\pi(\zeta)=\kappa(\zeta).
\end{equation}
Let $\Psi_+^{(\pi)}$ be the canonical Bloch solution analytic
${\mathcal D}_\pi$, normalized at $\pi$ and such that $\kappa_\pi$ is
its Bloch quasi-momentum. In~\cite{Fe-Kl:03e}, we have proved
\begin{Pro}[\cite{Fe-Kl:03e}]
  \label{f:pi:global}
  Fix $E=E_*\in J$. For sufficiently small $\varepsilon$, there exists
  $f_\pi$, a solution to~\eqref{G.2}, satisfying~\eqref{consistency:1}
  and analytic in the strip $S_{\tilde Y}$ that, on ${\mathcal
    D}_\pi$, has the standard behavior
  \begin{equation}
    \label{eq:5}
    f_\pi\sim \exp\left(\frac i\varepsilon\int_{\pi}^\zeta
    \kappa_\pi d\zeta\right)\,\Psi_+^{(\pi)}.
  \end{equation}
\end{Pro}
\subsubsection{The solution $f_0$}
\label{sec:solution-f_0}
Consider ${\mathcal D}_0$, the subdomain of the domain
$D_0=\{|\im\zeta|<\tilde Y,\ -\pi<\re \zeta<\pi \}$ shown in
Fig.~\ref{BEM:fig:dp:2}.
\noindent  Its boundary consists of the lines bounding
$D_0$ and of the segments of Stokes lines and lines $\{\re\zeta=C\}$
beginning at the intersection points of Stokes lines. The domain
${\mathcal D}_0$ is simply connected.\\
Let $\kappa_0$ be the analytic continuation of $-\kappa$ from
$S^p$ to ${\mathcal D}_0$ i.e., for $\zeta\in {\mathcal
  D}_0\cap S^p$,
\begin{equation}\label{kappa-0}
  \kappa_0(\zeta)=-\kappa(\zeta).
\end{equation}
Let $\Psi_+^{(0)}$ be the canonical Bloch solution analytic ${\mathcal
  D}_0$, normalized at $0$ and such that $\kappa_0$ is its Bloch
quasi-momentum. One has
\begin{Pro}
  \label{f:0:global}
  Fix $E=E_*\in J$. For sufficiently small $\varepsilon$, there exists
  $f_0$, a solution to~\eqref{G.2}, satisfying~\eqref{consistency:1}
  and analytic in the strip $S_{\tilde Y}$ that, on ${\mathcal D}_0$,
  has the standard behavior
  \begin{equation}
    \label{eq:4}
    f_0\sim \exp\left(\frac i\varepsilon\int_{0}^\zeta
    \kappa_0 d\zeta\right)\,\Psi_+^{(0)}.
  \end{equation}
\end{Pro}
\noindent The proof of Proposition~\ref{f:0:global} is similar to that of
Proposition~\ref{f:pi:global}; we omit it.
\section{Two consistent bases} 
\label{sec:two-consistent-bases}
\noindent In this section, we construct the consistent bases used in
the section~\ref{sec:transition-matrices}.\\
Fix $\nu\in\{0,\pi\}$. The solution $f_\nu^*$ is related to $f_\nu$ by
the transformation~\eqref{star}. First, we compute the asymptotics of
$f_\nu^*$.  Then, we compute the asymptotic of the Wronskian
$w(f_\nu,f_\nu^*)$.  This Wronskian is constant up to a factor of the
form $(1+o(1))$.  Finally, we correct $f$ so that
\begin{enumerate}
\item the Wronskian $w(f_\nu,f_\nu^*)$ be constant (and, thus, $f_\nu$
  and $f_\nu^*$ form a consistent basis),
\item the ``new'' solutions $f_\nu$ and $f_\nu^*$ have the ``old''
  behavior in the strip $S_{\tilde Y}$.
\end{enumerate}
The constructions described here are standard for the adiabatic
complex WKB method. The proofs of Lemmas~\ref{le:f*}
and~\ref{basis:Wronskian}, and of Theorem~\ref{thm:basis} below
essentially repeat the proofs of the analogous statements found
in~\cite{Fe-Kl:03f} and are therefore omitted.
\subsection{Asymptotics of $f_\nu^*$}
\label{sec:asymptotics-f_nu}
To discuss the asymptotic behavior of $f_\nu^*$, we need some
additional material.
\subsubsection{Preparation}
\label{sec:preparation}
Define $\mathfrak{z}_0=(-\zeta_{2n},\zeta_{2n})\subset\R$ and
$\mathfrak{z}_\pi=(\zeta_{2n+1},2\pi-\zeta_{2n+1})\subset\R$. Note
that $\mathcal E$ maps $\mathfrak{z}_0$ into the $n$-th spectral band,
and $\mathfrak{z}_\pi$ into the $(n+1)$-st spectral band.\\
Recall that the leading terms of the asymptotics of the solutions
having standard asymptotic behavior are fixed up to the choice of $q$,
the branch of $\sqrt{k'({\mathcal E}(\zeta))}$ from the definition of
the canonical Bloch solution, see~\eqref{canonicalBS}. Let $q_\nu$ be
the branch of $q$ from the definition of $\Psi_+^{(\nu)}$. Fix it so
that
\begin{equation*}
  q_\nu(\zeta)>0\quad\text{for}\quad\zeta\in{\mathfrak z}_\nu.
\end{equation*}
This choice is possible as, inside any spectral band of the periodic
operator, $k_p'>0$.
\subsubsection{The asymptotics}
\label{sec:asymptotics}
Let ${\mathcal D}_\nu^*$ be the domain symmetric to ${\mathcal D}_\nu$
with respect to the real line. Note that
$\mathfrak{z}_\nu\subset\mathcal{D}_\nu\cap \mathcal{D}_\nu^*$. One
has
\begin{Le}
  \label{le:f*}
  In $\mathcal{D}_\nu^*$, the solution $f_\nu^*$ has the
  standard behavior
  \begin{equation}
    \label{f*:as}
    f_\nu^*\sim e^{-\frac i\varepsilon\int_{\nu}^\zeta
    \kappa_{\nu,*} d\zeta}\,\Psi_-^{(\nu),*}(x,\zeta).
  \end{equation}
  Here, $\kappa_{\nu,*}$ is the branch of the complex momentum
  analytic in $\mathcal{D}_\nu^*$ that coincides with $\kappa_\nu$ on
  $\mathfrak{z}_\nu$; the function $\Psi_-^{(\nu),*}$ is the canonical
  Bloch solution analytic in $\mathcal{D}_\nu^*$ that coincides with
  $\Psi_-^{(\nu)}$ (complementary to $\Psi_+^{(\nu)}$ from the
  asymptotics of $f_\nu$) on ${\mathfrak z}_\nu$.
\end{Le}
\noindent The proof of Lemma~\ref{le:f*} mimics that of Lemma 6.1
in~\cite{Fe-Kl:03f}.\\
Note that $\kappa_{\nu,*}=\kappa_\nu^*$, and that
$\Psi_-^{(\nu),*}=(\Psi_+^{(\nu)})^*$.
\subsection{The Wronskian of $f_\nu$ and $f_\nu^*$}
\label{sec:wronskian-f-f}
The solution $f_\nu$ and $f_\nu^*$ are analytic in the strip $S_{\tilde Y}$; so
does their Wronskian. As both $f_\nu$ and $f_\nu^*$ satisfy
condition~\eqref{consistency:1}, it is $\varepsilon$-periodic in
$\zeta$. One has
\begin{Le}
  \label{basis:Wronskian}
  The Wronskian of $f_\nu$ and $f_\nu^*$ admits the asymptotic
  representation:
  \begin{equation}
    \label{basis:Wronskian:as}
    w(f_\nu,f_\nu^*)= w(\Psi_+^{(\nu)},\Psi_-^{(\nu)})|_{\zeta=\nu}+
    g_\nu,\quad \quad \zeta\in
    S_{\tilde Y}.
  \end{equation}
  Here, $g_\nu$ is a function analytic in $S_{\tilde Y}$ such that,
  for real $\zeta$ and $E$, $\re g_\nu=0$. Moreover, $g_\nu=o(1)$
  locally uniformly in any compact of $S_{\tilde Y}$ provided that $E$
  is in a sufficiently small complex neighborhood of $E_0$.
\end{Le}
\noindent The proof of Lemma~\ref{basis:Wronskian} mimics that of
Lemma 6.2 in~\cite{Fe-Kl:03f}.
\begin{Rem}
  \label{rem:4}
  Note that $w(\Psi_+^{(\nu)},\Psi_-^{(\nu)})|_{\zeta=\nu}\ne 0$ as
  $\mathcal{E}(\nu)\not\in P\cup Q$.
\end{Rem}
\noindent As $g_\nu$, the error term in~\eqref{basis:Wronskian:as}, may
depend on $\zeta$, we redefine the solution $f_\nu$ setting
\begin{equation*}
  f_\nu:=f_\nu/Q\quad\text{where}\quad
   Q=\sqrt{1+g/w(\Psi_+^{(\nu)},\Psi_-^{(\nu)})|_{\zeta=\nu}}.
\end{equation*}
In terms of this new solution $f_\nu$, we define the new $f_\nu^*$.
The solutions $(f_\nu,f_\nu^*)$ form the basis the monodromy matrix of
which we study. For these ``new'' $f_\nu$ and $f_\nu^*$, we have
\begin{Th}
  \label{thm:basis}
  The solutions $f_\nu$ and $f_\nu^*$ are analytic in $S_{\tilde Y}$,
  satisfy the condition~\eqref{consistency:1}, and
  \begin{equation}
    \label{new-basis:Wronskian}
    w(f_\nu,f_\nu^\ast)=w(\Psi_+^{(\nu)},\Psi_-^{(\nu)})|_{\zeta=\nu}.
  \end{equation}
  Moreover, $f_\nu$ has the standard behavior,~\eqref{eq:4}
  or~\eqref{eq:5}, in ${\mathcal D}_\nu$, and $f_\nu^*$ has the
  standard behavior~\eqref{f*:as} in $\mathcal{D}_\nu^*$.
\end{Th}
\noindent The proof of Theorem~\ref{thm:basis} mimics that of Theorem
6.1 from~\cite{Fe-Kl:03f}.\\
Let $\zeta\mapsto \psi_\pm(x,{\mathcal E}(\zeta))$ be the two branches
of the Bloch solution $\zeta\mapsto \psi(x,{\mathcal E}(\zeta))$ that
are analytic in $\zeta\in S^p$ and such that $\kappa$, the branch of
the complex momentum defined in the beginning of the
section~\ref{two-bases}, is the Bloch quasi-momentum for $\psi_+$.
By~\eqref{Wcanonical} and the definitions of the canonical Bloch
solutions $\Psi_\pm^\nu$, one has
\begin{equation}
  \label{new-basis:Wronskian:1}
  w(\Psi_+^{(\nu)},\Psi_-^{(\nu)})|_{\zeta=\nu}
  =s(\nu)k_p'({\mathcal E}(\nu))\, w(\psi_+,\psi_-)|_{\zeta=\nu},
  \quad{\rm where}\quad s(\nu)=\left\{\begin{array}{l} 1 \text{ if }
  \nu=\pi,\\ -1 \text{ if } \nu=0.\end{array}\right.
\end{equation} 
%


%
\section{Transition matrices}
\label{sec:transition-matrices1}
\noindent In this section, we compute the asymptotics of the
transition matrices $T_\nu$ defined by~\eqref{Tmatrices} for the bases
$(f_\nu,f^*_\nu)$ for $\nu\in\{0,\pi\}$.
\subsection{Elementary properties of the transition matrices}
\label{sec:elem-prop-trans}
One can easily express the coefficients of the transition matrices,
see~\eqref{Tmatrices}, via the Wronskians of the basis solutions;
formulas~\eqref{Tmatrices} immediately imply
\begin{Le}
  \label{wronsk}
  One has
  \begin{equation}
    \label{a,b:pi}
    a_\pi(\zeta)=\frac{w(f_\pi(\cdot,\zeta+2\pi),f_0^*(\cdot,
      \zeta))}{w(f_0(\cdot,\zeta),f_0^*(\cdot, \zeta))},\quad
    b_\pi(\zeta)=\frac{w(f_0(\cdot,\zeta),f_\pi(\cdot,
      \zeta+2\pi))}{w(f_0(\cdot,\zeta),f_0^*(\cdot, \zeta))}.
  \end{equation}
  and
  \begin{equation}
    \label{a,b:0}
    a_0(\zeta)=\frac{w(f_0(\cdot,\zeta),f_\pi^*(\cdot,
      \zeta))}{w(f_\pi(\cdot,\zeta),f_\pi^*(\cdot, \zeta))},\quad
    b_0(\zeta)=\frac{w(f_\pi(\cdot,\zeta),f_0(\cdot,
      \zeta))}{w(f_\pi(\cdot,\zeta),f_\pi^*(\cdot, \zeta))}.
  \end{equation}
\end{Le}
\noindent For $\nu\in\{0,\pi\}$, by the definition of the standard
behavior, the basis $\{f_\nu,f_\nu^*\}$ is defined and analytic for
$(\zeta,E)\in S_{\tilde Y}\times V(\tilde Y)$ where $V(\tilde Y)$ is a
neighborhood of $E_*\in J$; this neighborhood is independent of
$\varepsilon$. One has
\begin{Le}
  \label{m-m:general}
  Pick $\nu\in\{0,\pi\}$. The matrix $T_\nu$ is analytic and
  $\varepsilon$-periodic in $\zeta\in S_{\tilde Y}$ and analytic in
  $E\in V(\tilde Y)$.  Moreover, $\det T_\nu$ is independent of
  $\zeta$ and does not vanish.
\end{Le}
\demo As the solutions $f_\nu$ and $f_\nu^*$ are analytic functions of
the variables $\zeta$ and $E$, so are the Wronskians in
formulae~\eqref{a,b:pi} and~\eqref{a,b:0}. Moreover,
by~\eqref{new-basis:Wronskian}, the Wronskians in the denominators
of~\eqref{a,b:pi} and~\eqref{a,b:0} do not vanish. This implies the
analyticity of the coefficients of the transition matrices. The
periodicity in $\zeta$ follows from the fact that all the solutions
satisfy~\eqref{consistency:1}.  Finally, relations~\eqref{Tmatrices}
imply that 
\begin{equation}
  \label{eq:35}
  w(f_\pi(x,\zeta+2\pi),f_\pi^*(x, \zeta+2\pi))=\det T_\pi\,
  w(f_0(x,\zeta),f_0^*(x, \zeta)). 
\end{equation}
Now,~\eqref{new-basis:Wronskian} imply that $\det T_\pi$ is
independent of $\zeta$. Similarly one checks that $\det T_0$ is
independent of $\zeta$.  This completes the proof of
Lemma~\ref{m-m:general}.\qed
\subsection{The asymptotics of the transition matrices}
\label{sec:asympt-trans-matr}
\begin{figure}[htbp]
  \centering
    \includegraphics[bbllx=71,bblly=464,bburx=559,bbury=721,width=14cm]{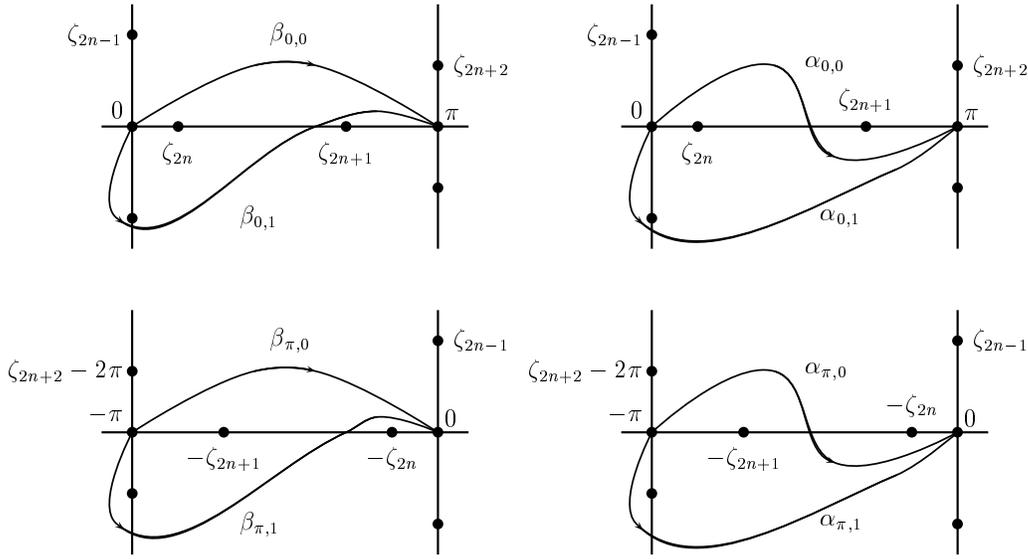}
  \caption{The integrations paths for Theorem~\ref{th:T-matrices}}
  \label{Curves-alpha-and-beta}
\end{figure}

We first introduce some notations:
\begin{enumerate}
\item For the Fourier coefficients of $a_\nu$ and $b_\nu$ we use the
  notations introduced in~\eqref{a,b:F}, and recall that
  $p(z)=e^{2\pi|\im z|}$.
\item Let $Y_\pi$, $Y_{v,\pi}$ and $Y_0$, $Y_{v,0}$ be the distances
  marked in Fig.~\ref{BEM:fig:dp:1} and Fig.~\ref{BEM:fig:dp:2}
  respectively. E.g., $Y_0$ is the imaginary part of the point of
  intersection of the Stokes lines $\overline{``g''}$ and
  $\overline{``h''}$ (see Lemma~\ref{BEM:le:Sl:left}). Note that, for
  any $\nu\in\{0,\pi\}$, one has
\begin{equation} 
  \label{YandY}
  0<Y_{v,\nu}<Y_\nu<\tilde Y.
\end{equation}
\item We use the branch $\kappa$ introduced in the beginning of the
  section~\ref{two-bases}; $\psi_\pm$ (resp.  $\Omega_\pm$)
  are the branches of $\psi(x,{\mathcal E}(\cdot))$ (resp. $\Omega$)
  such that $\kappa$ is the Bloch quasi-momentum of $\psi_+$. When 
  integrating $\kappa$ (resp. integrating $\Omega$ or continuing analytically $\psi$) 
  along a curve, we choose a branch of $\kappa$ (resp.
  $\Omega$, $\psi$) near the starting point of the curve and 
  then continue it along the curve.
\item Let $\gamma$ be a curve and $g$ be a function continuous on
  $\gamma$. We denote by $\Delta\arg q|_\gamma$ the increment of the
  argument $q(\zeta)$ along the curve $\gamma$.
\end{enumerate}
The asymptotics of the transition matrices coefficients are described
by
\begin{Pro} 
  \label{pro:a,b:as} 
  Pick $\nu\in\{0,\pi\}$. Fix $Y$ so that $Y_{v,\nu}<Y<Y_\nu$. There
  exists $V_\nu(Y)$, a complex neighborhood of $E_*$ independent of
  $\varepsilon$, such that, for $\varepsilon$ sufficiently small,
  $j\in\{0,1\}$ and $E\in V_\nu(Y)$, one has the uniform asymptotics
  \begin{gather}
    \label{Fourier-coefa}
    a_{\nu,j}= \exp\left(s\frac i\varepsilon \int_{\alpha} \kappa
      d\zeta-j\frac{2\pi(\pi-\nu) i}\varepsilon+\int_{\alpha} \Omega_s +
      i\Delta\arg
      q|_{\alpha} +o(1)\right),\quad \alpha=\alpha_{\nu,j},\\
    \label{Fourier-coefb}
    b_{\nu,j}=\exp\left(s\frac i\varepsilon \int_{\beta} \kappa
      d\zeta -j\frac{2\pi(\pi-\nu) i}\varepsilon+\int_{\beta} \Omega_s +
      i\Delta\arg q|_{\beta} +o(1)\right),\quad \beta=\beta_{\nu, j},\\
    \nonumber\text{where }\quad s=+1\ \text{if}\quad\nu=\pi,\text{ and
    } s=-1\ \text{if}\quad\nu=0.
  \end{gather}
  In~\eqref{Fourier-coefa} and~\eqref{Fourier-coefb}, one integrates
  along the curves shown in Fig.~\ref{Curves-alpha-and-beta} chosen
  such that ${\mathcal E}(\zeta)\not\in (P\cup Q)$ along them; for
  each of the integration curves, $q$ denotes a branch of
  $\zeta\mapsto\sqrt{k'({\mathcal E}(\zeta))}$ continuous on this curve.\\
  For $(\zeta,E)\in S(Y)\times V_\nu(Y)$, one has the uniform estimates
  \begin{equation}
    \label{Fst:est}
    a_{\nu,-1}(\zeta)=o(a_{\nu,0}),\quad
    b_{\nu,-1}(\zeta)=o(b_{\nu,0}),\quad
    a_{\nu,2}(\zeta)=o(p(\zeta/\varepsilon)a_{\nu,1}),\quad
    b_{\nu,2}(\zeta)=o(p(\zeta/\varepsilon)b_{\nu,1}).
  \end{equation}
\end{Pro}
%
%
\noindent In the remaining part of the present section, we first
explain how Theorem~\ref{th:T-matrices} is deduced from
Proposition~\ref{pro:a,b:as}. Then, we turn to the proof of
Proposition~\ref{pro:a,b:as}.  We begin with describing general
asymptotic formulae for the Wronskians of two solutions having
standard behavior; this material mostly stems from~\cite{Fe-Kl:03f}.
Then, using these formulae, we compute the Wronskians in the formulae
for the transition matrix coefficients (see Lemma~\ref{wronsk}) and,
thus, complete the proof of Proposition~\ref{pro:a,b:as}. Note that we
carry out the analysis only for the asymptotics and the estimates for
$a_0$ and $b_0$. The coefficients $a_\pi$ and $b_\pi$ are analyzed in
a similar way.
\subsection{The proof of Theorem~\ref{th:T-matrices}}
\label{sec:proof-theor-refth:t-1}
In section~\ref{sec:phase-integr-tunn}, we study the actions
$(S_{v,\nu})_{\nu\in\{0,\pi\}}$ and prove
\begin{Le}
  \label{le:action-Y} 
  Pick $E_*\in J$. For each $\nu\in\{0,\pi\}$, one has
  $S_{v,\nu}(E_*)=2\pi Y_{v,\nu}(E_*)$.
\end{Le}
\noindent Lemma~\ref{le:action-Y}  and the condition (T), see
section~\ref{sec:une-descr-gross}, imply that, in
Proposition~\ref{pro:a,b:as}, we can choose $Y$ so that (1) $2\pi Y>
\max_{E\in J}S_h(E)$ and (2) $Y_{v,\nu}<Y<Y_\nu$ simultaneously for
$\nu=0$ and $\nu=\pi$. We then define $V_*=V_0\cap V_\pi$. With this,
each of the basis solutions $f_0$, $f^*_0$, $f_\pi$ and $f_\pi^*$ is
defined and analytic in the domain~\eqref{analyticity:dom:1}. This and
Lemma~\ref{m-m:general} imply the first and the second point
of Theorem~\ref{th:T-matrices}.\\
In section~\ref{sec:fur}, we derive the estimates of the third point
of Theorem~\ref{th:T-matrices} from the
asymptotics~\eqref{Fourier-coefa} and~\eqref{Fourier-coefb}. \\
Finally, the last point of Theorem~\ref{th:T-matrices} is an immediate
consequence of the estimates~\eqref{Fst:est}. So,
Theorem~\ref{th:T-matrices} is proved.
\subsection{General asymptotic formulae}
\label{sec:gener-asympt-form}
We recall results from section 8 of~\cite{Fe-Kl:03f}. Consider
equation~\eqref{G.2a} assuming only that $W$ is analytic and that $E$
is fixed, say $E=E_0$. Let $h$ and $g$ be two solutions
of~\eqref{G.2a} having the standard asymptotic behavior in regular
domains $D_h$ and $D_g$:
\begin{equation}
  \label{hg:as}
  h\sim e^{\frac i\varepsilon\int_{\zeta_h}^{\zeta}\kappa_h
  d\zeta}\,\Psi_{h}(x,\zeta),\quad
  g\sim e^{\frac i\varepsilon\int_{\zeta_g}^{\zeta}\kappa_g
  d\zeta}\,\Psi_{g}(x,\zeta).
\end{equation}
Here, $\kappa_h$ (resp. $\kappa_g$) is a branch of the complex
momentum analytic in $D_h$ (resp. $D_g$), $\Psi_h$ (resp. $\Psi_g$) is
the canonical Bloch solution defined on $D_h$ (resp.  $D_g$) and
having the quasi-momentum $\kappa_h$ (resp. $\kappa_g$), and $\zeta_h$
(resp. $\zeta_g$) is the normalization point for $h$ (resp. $g$).\\
As the solutions $h$ and $g$ satisfy the consistency condition, their
Wronskian is $\varepsilon$-periodic in $\zeta$. First,
following~\cite{Fe-Kl:03f}, we describe the asymptotics of this
Wronskian and of its Fourier coefficients. Then, we develop simple
tools to compute some constants coming up in these formulae.
\subsubsection{Asymptotics of the Wronskian}
\label{sec:asympt-wronsk}
Let $d$ be a simply connected domain such that $d\subset D_h\cap
D_g$.\\
{\bf Arcs.} \ Let $\gamma$ be a curve connecting $\zeta_g$ to
$\zeta_h$ going first from $\zeta_g$ to some point in $d$ while in
$D_g$ and, then, from this point to $\zeta_h$ while in $D_h$. We call
$\gamma$ an arc {\it associated to the triple $(g,h,d)$} and
denote it by $\gamma(g,h,d)$.\\
Two  arcs associated to one and the same triple are called {\it equivalent}.
\smallpagebreak Continue $\kappa_h$ and $\kappa_g$ analytically along
$\gamma(g,h,d)$. Then, there exists $m\in\Z$ and $\sigma\in\{-1,+1\}$
such that for $\zeta$ close to $\gamma$, one has
\begin{equation}
  \label{kh,kg}
  \kappa_g(\zeta)=\sigma\kappa_h(\zeta)+2\pi m.
\end{equation}
We call $\sigma=\sigma(g,h,d)$ {\it the signature} of $\gamma$, and
$m=m(g,h,d)$ {\it the index} of $\gamma$. These two integers do not
change when we replace the arc $\gamma$ by an equivalent one.\\
{\bf Meeting domains.} \ A domain $d$ is called {\it a meeting domain}
if the functions $\im\kappa_h$ and $\im\kappa_g$ do not vanish and are
of opposite sign in $d$. One has
\begin{Le}[\cite{Fe-Kl:03f}]
  \label{M-d:sign}
  Suppose the functions $\im\kappa_h$ and $\im\kappa_g$ do not vanish
  in $d$. Then, $d$ is a meeting domain if and only if
  $\sigma(g,h,d)=-1$.
\end{Le}
\noindent{\bf Fourier coefficients.} \ Let $S(d)$ be the smallest
strip of the form $\{C_1<\im\zeta<C_2\}$ containing the domain $d$.
One has
\begin{Pro}[\cite{Fe-Kl:03f}]
  \label{pro:w:as}
  Fix $E_0$. Let $d=d(h,g)$ be a meeting domain for $h$ and $g$, and
  $m=m(g,h,d)$ be the corresponding index (at energy $E_0$). Then,
  there exists $V_0$ a neighborhood of $E_0$ such that for
  $\varepsilon$ sufficiently small, for $E\in V_0$ and $\zeta\in
  S(d)$, the Wronskian of $h$ and $g$ is given by the formulae
  \begin{equation}
    \label{w:as}
    w(h,g)=\tilde w_m\,e^{\frac{2\pi
    im}{\varepsilon}(\zeta-\zeta_h)}(1+o(1)), 
  \end{equation}
  where
  \begin{equation}
    \label{wm}
    \tilde w_m=\left.\left(q_g/q_h\right)\right|_{\zeta=\zeta_h}\,
    \exp\left(\frac i\varepsilon\int_{\gamma(g,h,d)}\kappa_g
      d\zeta+\int_{\zeta_g}^{\zeta_h}\Omega_g\right)\,
    w(\Psi_+,\Psi_-)_{|\zeta=\zeta_g}.
  \end{equation}
  In these formulae:
  \begin{itemize}
  \item $\tilde w_m$ is independent of $\zeta$;
  \item we choose the arc $\gamma(g,h,d)$ so that, along it,
    $\mathcal{E} (\zeta)\not\in P\cup Q$;
  \item $\zeta\mapsto q_g(\zeta)=\sqrt{k'(\mathcal{E}(\zeta))}$ and
    $\zeta\mapsto \Omega_g(\zeta)=\Omega_g(\mathcal{E}(\zeta))$ are
    the analytic continuations of the function and the 1-form from the
    definition of $\Psi_{g}$ along $\gamma(g,h,d)$.
  \item $\Psi_+=\Psi_h$, and $\Psi_-$ is the canonical Bloch solution
    ``complementary'' to $\Psi_+$.
  \end{itemize}
  Fix $K$, a compact subset of $S(d)$. Then, there exists $V^K_0$ a
  neighborhood of $E_0$ in $V_0$ such that the
  asymptotics~\eqref{w:as} is uniform in $K\times V^K_0$.
\end{Pro}
\noindent The factor $\tilde w_m$ is the leading term of the
asymptotics of the $m$-th Fourier coefficient of $w(h,g)$.
\subsubsection{Closed curves and the index $m$}
\label{index-m}
In practice, it is not too difficult to compute the index $m$.
However, as one needs to control several Fourier coefficients of each
Wronskian, the computations become lengthy. Fortunately, there is an
effective way to compare the indexes of two (non-equivalent) arcs.  To
this end, we define the index of a closed curve.
\smallpagebreak {\it Closed curves.\/} Let $c$ be an oriented closed
curve containing no branch points of the complex momentum. 
Pick $\zeta_0\in c$. In $V_0$, a regular neighborhood
of $\zeta_0$, fix $\kappa$, an analytic branch of the complex
momentum. We call the  triple $(c,\zeta_0,\kappa)$ a {\it loop}.\\
We shall consider $c$ as disjoint at $\zeta_0$ and speak about its
beginning and its end. Continue $\kappa$ analytically along $c$. This
yields a new branch of the complex momentum in $V_0$. Denote it by
$\kappa|_c$. Hence, there exists $\sigma\in\{-1,+1\}$ and $m\in\Z$
such that, for $\zeta\in V_0$
\begin{equation}
  \label{m:cl-cv}
  \kappa|_c(\zeta)=\sigma \kappa(\zeta)+2\pi m.
\end{equation}
The numbers $\sigma=\sigma(c,\zeta_0,\kappa)$ and
$m=m(c,\zeta_0,\kappa)$ are called {\it the signature} and {\it the
  index } of the loop $(c,\zeta_0,\kappa)$.\\
Consider two loops $(c_1,\zeta_1,\kappa_1)$ and
$(c_2,\zeta_2,\kappa_2)$. Assume that one can continuously deform
$c_1$ into $c_2$ without intersecting any branching point. Assume
moreover that, in result of the same deformation, $\zeta_1$ becomes
$\zeta_2$. This deformation defines an analytic continuation of
$\kappa_1$ to a neighborhood of $\zeta_2$. If this analytic
continuation coincides with $\kappa_2$, we say that the loops are {\it
  equivalent}. The indexes $m$ and $\sigma$ calculated for equivalent
loops coincide.
\smallpagebreak Let us explain how to compute the indexes $m$ and
$\sigma$. Let $G$ be the pre-image with respect to $\mathcal E$ of
the set of the spectral gaps of the periodic operator~\eqref{Ho}. Note
that
\begin{itemize}
\item on any connected component of $G$, the value of the real part of
  the complex momentum is constant and belongs to $\{\pi l;\ 
  l\in\Z\}$;
\item locally, outside $\{\zeta;\ W'(\zeta)=0\}$, all the connected
  components of $G$ are analytic curves.
\end{itemize}
Now, we can formulate
\begin{Le}
  \label{G}
  Assume that $c$ does not start at a point of $G$. Assume moreover
  that $c$ intersects $G$ exactly $N$ times ($N<\infty$) and that, at
  the intersection points, $W'\ne 0$.  Let $r_1$, $r_2$, \dots, $r_N$
  be the values that $\re\kappa$ takes consecutively at these
  intersection points as $\zeta$ moves along $c$ (from the beginning
  to the end). Then,
  \begin{equation}\label{eq:m:cl-cv}
   \sigma(c,\zeta_0,\kappa)=(-1)^N,\quad\text{ and }\quad
    m(c,\zeta_0,\kappa)=
    \frac1{\pi}\,\left(r_N-r_{N-1}+r_{N-2}-\dots+(-1)^{N-1}r_1\right).
  \end{equation}
\end{Le}
\noindent The proof of Lemma~\ref{G} mimics the proof of Lemma 8.2
in~\cite{Fe-Kl:03f} which describes the index of a $2\pi$-periodic
curve.
\smallpagebreak{\it Comparing the indexes of arcs.\/} Let $d$ and
$\tilde d$ be two (distinct) meeting domains for the solutions $h$ and
$g$, and let $\gamma$ and $\tilde \gamma$ be the corresponding arcs.
One can write
\begin{equation}
  \label{arc:cl-cur}
  \tilde\gamma=c+\gamma,
\end{equation}
where $c$ is a closed regular curve; its orientation is induced by
those of $\gamma$ and $\tilde\gamma$.\\
As $\sigma(g,h,\tilde d)=\sigma(g,h,d)=-1$, one has
$\sigma(c,\zeta_g,\kappa_g)=1$. As an immediate consequence of the
definitions, we also get
\begin{equation}
  \label{m:rel}
  m(g,\tilde h,\tilde d)= m(c,\zeta_g,\kappa_g)+m(g,h,d).
\end{equation}
This formula and Lemma~\ref{G} give an effective way to compute the
indexes of arcs.

\subsection{The asymptotics of the coefficient $b_0$}
\label{sec:asympt-coeff-t_0}
The coefficient $b_0$ of the matrix $T_0$ is given in~\eqref{a,b:0}.
As $w(f_\pi,f_\pi^*)$ is given by formula~\eqref{new-basis:Wronskian},
we have only to compute $w(f_\pi(\cdot,\zeta),f_0(\cdot ,\zeta))$. One
applies the constructions of section~\ref{sec:gener-asympt-form} with
\begin{gather}
  h(x,\zeta)=f_\pi(x,\zeta),\quad g(x,\zeta)=f_0(x,\zeta),\quad
  D_h=\mathcal{D}_\pi,\quad D_g=\mathcal{D}_0;\nonumber\\
  \label{zeta0:h,g}
  \zeta_h=\pi,\quad \zeta_g=0;\\
  \label{kappa:h,g}
  \kappa_h(\zeta)=\kappa(\zeta)\text{ for }\zeta\sim
  \pi,\quad\text{and}\quad \kappa_g(\zeta)=-\kappa(\zeta)\text{ for
  }\zeta\sim 0.
\end{gather}
In~\eqref{kappa:h,g}, $\kappa$ is the branch of the complex
momentum defined in~\eqref{kappaPkappa}.\\
Let $Y_0$ and $Y_{v,0}$ be the distances marked in
Fig.~\ref{BEM:fig:dp:2}. They satisfy~\eqref{YandY}.
\subsubsection{The asymptotics in the strip $\{-Y_{v,0}<\im\zeta<Y_0\}$}
\label{sec:asympt-above-real:b}
Let us describe $d_0$, the meeting domain, and
$\gamma(f_0,f_\pi,d_0)$, the arc used to compute $w(f_\pi,f_0)$ in the
strip
\begin{equation*}
  S_0=\{\zeta\in\C;\ -Y_{v,0}<\im\zeta<Y_0\}.
\end{equation*}
\smallpagebreak {\it The meeting domain $d_0$.} It is the subdomain of
the strip $S_0$ between the lines $\gamma_1$ and $\gamma_2$ defined by
\begin{itemize}
\item the line $\gamma_1$ consists of the following lines: the Stokes
  line $\overline{\text{``e''}}$ symmetric to the Stokes line ``e''
  with respect to the real line, the segment $[0,\zeta_{2n}]$ of the
  real line, the segment $[0,\zeta_{2n-2}]$ of the imaginary axis and
  the Stokes line ``g'' (see Fig.~\ref{bem:St-lines});
\item the line $\gamma_2$ consists of the following lines: the Stokes
  line $\overline{\text{``a''}}$ symmetric to the Stokes line ``a''
  with respect to the real line, the segment $[\zeta_{2n+1},\pi]$ of
  the real line, the segment $[\pi,\zeta_{2n+3}]$ of the line
  $\re\zeta=\pi$ and the Stokes line ``c'' (see
  Fig.~\ref{bem:St-lines}).
\end{itemize}
The Stokes lines mentioned here are described by
Lemmas~\ref{BEM:le:Sl} and~\ref{BEM:le:Sl:left}. In particular, these
lemmas imply that $\gamma_1\cap\gamma_2=\emptyset$.\\
Note that $d_0$ does not intersect $Z$, the pre-image of the set of
the bands of the periodic operator~\eqref{Ho} with respect to the
mapping $\mathcal E$. So, in $d_0$, one has $\im\kappa\ne 0$.
\smallpagebreak {\it The arc $\gamma(g,h,d_0)$.\/} It is the curve
$\beta_{0,0}$ shown in Fig.~\ref{Curves-alpha-and-beta}; it stays
in $d_0$ and connects $\zeta_g=0$ to $\zeta_h=\pi$.\\
{\it Index $m$.\/} In view of~\eqref{kappa:h,g}, in $d_0$, one has
$\kappa_h=-\kappa_g$. This implies that $m(g,h,d_0)=0$.\\
{\it The result.\/} \ Proposition~\ref{pro:w:as},
formulae~\eqref{a,b:0} and~\eqref{new-basis:Wronskian} imply that, for
$\zeta\in S_0$,
\begin{equation}
  \label{F:b:0}
  b_0=\tilde b_0 (1+o(1)),\quad \tilde b_0=
  \exp\left(-\frac i\varepsilon \int_{\beta}
  \kappa d\zeta+\int_{\beta} \Omega_- + i\Delta\arg
  q|_{\beta}\right)\text{ where }\beta=\beta_{0,0},
\end{equation}
and, as $q$, one can take any branch of the function $\zeta\mapsto
\sqrt{k'({\mathcal E}(\zeta))}$ continuous on $\beta$.\\
When deriving the formula for $\tilde b_0$, we have used the facts that 
\begin{itemize}
\item $\Omega_g$ is the branch of $\Omega_-$ corresponding to the
  branch $\kappa$ chosen above;
\item $(q_g/q_h)(\zeta_h)=e^{i\Delta \arg q_g|_\beta}$ as, at
  $\zeta_h$, $q_h$ is real and $|q_g/q_h|=1$;
\item the quantity $\Delta\arg q_g|_{\beta}$ does not depend on the
  branch of $q_g$ as long as it is continuous on $\beta$.
\end{itemize}
\subsubsection{The asymptotics in the strip $\{-Y_0<\im\zeta<-Y_{v,0}\}$}
\label{sec:asympt-strip-1:b}
Let us describe $d_1$, the meeting domain, and
$\gamma(f_\pi,f_0,d_1)$, the arc used to compute $w(f_\pi,f_0)$ in
this strip
\begin{equation*}
  S_1=\{\zeta\in\C;\ -Y_0<\im\zeta<-Y_{v,0}\}.
\end{equation*}
{\it The meeting domain $d_1$.\/} Let $d_1$ be the subdomain of the
strip $S_1$ located between the Stokes line $\overline{\text{``a''}}$
(symmetric to ``a'' with respect to the real line) and $\gamma_3$, the
curve which consists of the following lines:
\begin{itemize}
\item the Stokes line $\overline{\text{``f''}}$ symmetric to the
  Stokes line ``f'' with respect to the real line, the segment
  $[\overline{\zeta_{2n-1}},\overline{\zeta_{2n-2}}]$ of the imaginary
  axis, and the Stokes line $\overline{\text{``g''}}$ symmetric to the
  Stokes line ``g'' with respect to the real line.
\end{itemize}
The domain $d_1$ is a meeting domain in view of
\begin{Le}
  \label{le:1}
  In $d_1$, one has $\im\kappa_\pi=-\im\kappa_0>0$.
\end{Le}
\demo The sign of $\im\kappa$ remains the same in any regular domain
$D$ such that $D\cap Z=\emptyset$. Moreover, the sign of $\im\kappa$
flips as $\zeta$ intersects (transversally) a connected component of
$Z$ at a point where $W'$ does not vanish.\\
By~\eqref{kappaPkappa} and~\eqref{kappa-pi}, one has
$\im\kappa_\pi=\im\kappa=\im\kappa_p>0$ in ${\mathcal D}_\pi\cap \Pi$.
As one goes from $\Pi$ to $d_1$ in ${\mathcal D}_\pi$ without
intersecting $Z$, we get $\im\kappa_\pi(\zeta)>0$ for $\zeta\in d_1$.
Similarly, by~\eqref{kappaPkappa} and~\eqref{kappa-0}, one has $\im
\kappa_0=-\im\kappa_p<0$ in ${\mathcal D}_0\cap\Pi$. Furthermore, to
go from $\Pi$ to $d_1$ staying in ${\mathcal D}_0$, one has to
intersect two connected components of $Z$, namely, the segment
$[-\zeta_{2n},\zeta_{2n}]$ of the real line and the segment
$[-\zeta_{2n-1},\zeta_{2n-1}]$ of the imaginary axis. Hence,
$\im\kappa_0(\zeta)<0$ for $\zeta\in d_1$. This completes the proof of
Lemma~\ref{le:1}. \qed
\smallpagebreak {\it The arc $\gamma(g,h,d_1)$.\/} It is the curve
$\beta_{0,1}$ shown in Fig.~\ref{Curves-alpha-and-beta}; it connects
$\zeta_g=0$ to $\zeta_h=\pi$.
\smallpagebreak {\it Index $m$.\/} One has
\begin{equation*}
  \gamma(g,h,d_1)=c_0+\gamma(g,h,d_0),
\end{equation*}
where $c_0$ is the closed curve shown in Fig.~\ref{Closed-curves}.
By~\eqref{m:rel}, we get
\begin{equation*}
  m(g,h,d_1)= m(c_0,0,\kappa_g)+m(g,h,d_0)=m(c_0,0,\kappa_g).
\end{equation*}
%
\begin{floatingfigure}[r]{7cm}
  \centering
  \vskip-.2cm
  \includegraphics[bbllx=71,bblly=577,bburx=304,bbury=721,width=7cm]{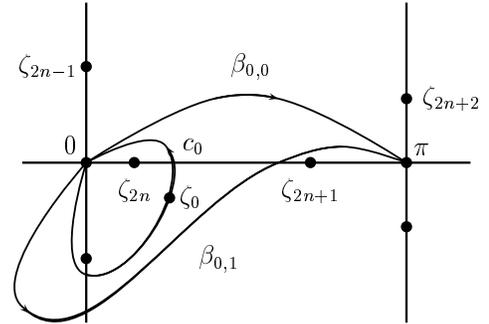}
  \caption{The curve $c_0$}
  \label{Closed-curves}
\end{floatingfigure}
%
\noindent So, the index $m(g,h,d_1)$ is equal to the index of the loop
$(c_0,0,\kappa_g)$. Recall that the indexes of equivalent loops
coincide. To compute the index, we pick a point $\zeta_0\in c_0$ as
shown in Fig.~\ref{Closed-curves} and we replace the loop
$(c_0,0,\kappa_g)$ by the equivalent loop defined by the same curve
$c_0$ and the point $\zeta_0$. The branch of the complex momentum
fixed for this new loop is the analytic continuation of the old branch
along $c_0$ from $0$ to $\zeta_0$ in the clockwise
direction. For this new branch, we keep the old notation $\kappa_g$.\\
In view of Lemma~\ref{G}, it is sufficient to compute
$\kappa_g$ at the intersections of $c_0$ and $G$. The set $G$ is
$2\pi$-periodic and symmetric with respect to the real line and to the
imaginary axis. The connected components of $G$ located in the
$\{0\le\im\zeta,\ 0\le\re\zeta\le\pi\}$ are described in
section~\ref{sec:complex-momentum}, part 2.\\
In Fig.~\ref{Closed-curves}, the curve $c_0$ intersects two connected
components of $G$, the segment
$[\overline{\zeta_{2n-1}},\overline{\zeta_{2n-2}}]$ of the imaginary
axis and the segment $[\zeta_{2n},\zeta_{2n+1}]$ of the real line. So,
Lemma~\ref{G} implies that
\begin{equation}
  \label{m:b:1}
  m(c_0,0,\kappa_g)=m(c_0,\zeta_0,\kappa_g)=
  \frac1\pi\left(\re\kappa_g(\overline{\zeta_{2n-1}})-
  \re\kappa_g(\zeta_{2n})\right),
\end{equation}
as $\re\kappa$ stays constant on any connected component of $G$.  As
$\kappa_g$ is defined by the formulae~\eqref{kappa:h,g}
and~\eqref{kappaPkappa}, one has
\begin{equation}
  \label{eq:30}
  \kappa_g(\zeta_{2n})=-\kappa(\zeta_{2n})=-(\kappa_p(\zeta_{2n})-\pi
  n)=-(\pi n-\pi n)=0.
\end{equation}
Along the interval $[-\zeta_{2n},\zeta_{2n}]$, one has
$\kappa_g(\zeta)=-\kappa(\zeta)\in\R$; hence,
\begin{equation*}
  \kappa_g(\overline{\zeta_{2n-1}})=-\overline{\kappa(\zeta_{2n-1})}=
  -\overline{(\kappa_p(\zeta_{2n-1})-\pi n)}=-(\pi (n-1)-\pi n)=\pi.
\end{equation*}
Substituting this and~\eqref{eq:30} into~\eqref{m:b:1}, we finally get
\begin{equation*}
   m(g,h,d_1)=m(c_0,0,\kappa_g)=1.
\end{equation*}
{\it The result.\/} \ Proposition~\ref{pro:w:as},
formulae~\eqref{a,b:0} and~\eqref{new-basis:Wronskian} imply that, for
$\zeta\in S_1$,
\begin{equation}
  \label{F:b:1}
  b_0=\tilde b_{1} e^{\frac{2\pi i\zeta}\varepsilon}(1+o(1)),\quad \tilde b_{1}=
  \exp\left(-\frac i\varepsilon \int_{\beta}
  \kappa d\zeta-\frac{2\pi^2i}\varepsilon+\int_{\beta} \Omega_- +
  i\Delta\arg q|_{\beta}\right)\text{ where }\beta=\beta_{0, 1}.
\end{equation}
{\it Completing the analysis.\/} \ The coefficient $b_0$ being
$\varepsilon$-periodic, we write its Fourier series
\begin{equation}
  \label{Fourier}
  b_0(\zeta)= \sum_{l=-\infty}^{\infty} b_{0,l}\,
  e^{2\pi\,l\,\zeta/\varepsilon}\quad\text{where}\quad
  b_{0,l}=\frac1\varepsilon \int_{\tilde\zeta}^{\tilde\zeta+2\pi}
  b_0(\zeta)e^{-2\pi\,l\,\zeta/\varepsilon} d\zeta\text{ for }l\in\Z,
\end{equation}
As $b_0$ is analytic in the strip $\{|\im\zeta|< Y_0\}$, $\tilde\zeta$
can be taken arbitrarily in the strip $\{|\im\zeta|< Y_0\}$. The
asymptotics and the estimates for $b_0$ in
Proposition~\ref{pro:a,b:as} are obtained by analyzing its Fourier
coefficients. To estimate the Fourier coefficients with non-positive
index, one uses~\eqref{F:b:0} and~\eqref{Fourier} with $\tilde\zeta\in
S_0$. To study the Fourier coefficients with positive index, one
uses~\eqref{F:b:1} and~\eqref{Fourier} with $\tilde\zeta\in S_1$. We
omit the elementary details and note only that $\tilde b_0$
in~\eqref{F:b:0} is the leading term of the asymptotics of $b_{0,0}$,
and that $\tilde b_{1}$ in~\eqref{F:b:1} is the leading term for
$b_{0,1}$.
\subsection{The asymptotics of the coefficient $a_0$}
\label{a:pi}
By~\eqref{a,b:0}, it suffices to compute the Wronskian
$w(f_0(\cdot ,\zeta),f_\pi^*(\cdot , \zeta))$. The computations of the
coefficient $a_0$ follow the same scheme as the ones of $b_0$.
So, we only outline them. Now,
\begin{gather}
  h=f_\pi^*,\quad g=f_0;\quad\quad
  D_h=\mathcal{D}_\pi^*,\quad D_g=\mathcal{D}_0;\\
  \zeta_h=\pi,\quad \zeta_g=0;\\
  \label{a:kappa:h,g}
  \kappa_h(\zeta)=-\overline\kappa(\bar\zeta)\text{ for }
  \zeta\sim\pi,\quad\text{and}\quad
  \kappa_g(\zeta)=-\kappa(\zeta)\text{ for } \zeta\sim 0.
\end{gather}
Recall that the complex momentum is real on
$[\zeta_{2n+1},2\pi-\zeta_{2n+1}]$. This imply that
\begin{equation}
  \label{a:kappas}
  \kappa_h(\zeta)=-\kappa(\zeta)\text{ for }\zeta\sim\pi.
\end{equation}
\subsubsection{The asymptotics in the strip $S_0$}
\label{sec:asympt-above-real:a}
In this case, the meeting domain $\tilde d_0$ is the subdomain of the
strip $S_0$ located between the lines the lines $\gamma_1$ and
$\overline{\gamma_2}$ symmetric to $\gamma_2$ with respect to the real
line (see section~\ref{sec:asympt-above-real:b}). These two
lines do not intersect.\\
The arc $\gamma(g,h,\tilde d_0)$ is the curve $\alpha_{0,0}$
shown in Fig.~\ref{Curves-alpha-and-beta}. One has $m(g,h,\tilde
d_0)=0$.\\
The asymptotics of $a_0$ for $\zeta\in S_0$ is described by
\begin{equation}
  \label{F:a:0}
  a_0=\tilde a_0 (1+o(1)),\quad \tilde a_0=
  \exp\left(-\frac i\varepsilon \int_{\alpha}
  \kappa d\zeta+\int_{\alpha} \Omega_- +i \Delta\arg
  q|_{\alpha}\right)\text{ where }\alpha=\alpha_{0, 0}.
\end{equation}
\subsubsection{The asymptotics in the strip $S_1$}
\label{sec:asympt-strip-1:a}
Now, the meeting domain $\tilde d_1$ is the subdomain of the strip
$S_1$ located between the line $\gamma_3$ (see
section~\ref{sec:asympt-strip-1:b}) and the line
$\overline{\gamma_2}$. \\
The arc $\gamma(g,h,\tilde d_1)$ is the curve $\alpha_{0,1}$
shown in Fig.~\ref{Curves-alpha-and-beta}. One has
\begin{equation*}
  \gamma(g,h,\tilde d_1)=c_0+\gamma(g,h,\tilde d_0),
\end{equation*}
where $c_0$ is the closed curve shown in Fig.~\ref{Closed-curves}. The
computation done for $b_0$ in $S_1$ yields
\begin{equation*}
  m(g,h,\tilde d_1)=m(c_0,0,\kappa_g)=1.
\end{equation*}
In result, for $\zeta\in S_1$, we get the asymptotic formula
\begin{equation}
  \label{F:a:1}
  a_0=\tilde a_{1}e^{\frac{2\pi i\zeta}\varepsilon} (1+o(1)),\quad
  \tilde a_{1}= \exp\left(-\frac i\varepsilon \int_{\alpha}
  \kappa d\zeta-\frac{2\pi^2i}\varepsilon+\int_{\alpha} \Omega_- +
  i\Delta\arg q|_{\alpha}\right)\text{ where }\alpha=\alpha_{0, 1}.
\end{equation}
The asymptotics~\eqref{F:a:0} and~\eqref{F:a:1} imply the formulae and
the estimates for $a_0$ in Proposition~\ref{pro:a,b:as}.
\section{Phase integrals, tunneling coefficients and the iso-energy surface}
\label{sec:phase-integr-tunn}
In this section, we first check the statements found in
section~\ref{sec:acti-integr-tunn-1}. We also prove
Lemma~\ref{le:action-Y} giving a geometric interpretation of the
vertical tunneling coefficients.\\
Then, we analyze the geometry of the iso-energy curves $\Gamma$ and
$\Gamma_\R$ (see~\eqref{isoen} and~\eqref{isoenr}) and justify the
interpretation of the phase integrals and tunneling coefficients in
terms of these curves.
\subsection{The complex momentum on the integration contours}
\label{sec:clos-curv-compl}
The phase integrals and the tunneling coefficients were defined as
contour integrals of the complex momentum along the curves shown in
Fig.~\ref{bp-tibm} and~\ref{TIBMfig:2}.
We have claimed that, on each of these curves, one can fix a
continuous branch of the complex momentum, which we justify in
\begin{Le}
  \label{contours}
  Let $\gamma$ be one of the curves  $\tilde\gamma_{0}$,
  $\tilde\gamma_{\pi}$, $\tilde\gamma_{h,0}$, $\tilde\gamma_{h,\pi}$,
  $\tilde\gamma_{v,0}$ and $\tilde\gamma_{v,\pi}$.
  Any branch of the complex momentum, analytic in a neighborhood of a
  point of $\gamma$, can be analytically continued to a single valued
  function on $\gamma$.
\end{Le}
\demo The curve $\gamma$ goes exactly around two branch points of the
complex momentum. They are of square root type (see
section~\ref{sec:kappa}). So, it suffices to check that, at the branch
points, the values of the complex momentum coincide. For the curve
$\tilde\gamma_{h,\pi}$, this follows from the facts that ${\mathcal
  E}$ (defined in~\eqref{eq:31}) bijectively maps the interval
$[\zeta_{2n},\zeta_{2n+1}]$ onto the $n$-th spectral gap of the
periodic operator, and that the values of a branch of the Bloch
quasi-momentum coincide at the ends of a gap. For
$\tilde\gamma_{\pi}$, this holds as ${\mathcal E}$ maps the interval
$(\zeta_{2n+1},2\pi-\zeta_{2n+1})$ into the $n$-th spectral band so
that both ends are mapped on $E_{2n+1}$. For $\tilde\gamma_{v,\pi}$,
it holds as ${\mathcal E}$ maps the segment
$(\zeta_{2n+2},2\pi-\zeta_{2n+2})$ into the $(n+1)$-st spectral band
so that both its ends are mapped on $E_{2n+2}$. The analysis of the
other curves is done in the same way. \qed
\subsection{Independence of  the tunneling coefficients and phase integrals
on the branch of the complex momentum in their definitions}
\label{t,Phi:kappa}
The independence  follows from the observations:
\begin{itemize}
\item only the signs of the integrals defining the phase integrals and
  the tunneling coefficients depend on the choice of the branches of
  the complex momentum being integrated;
\item the branches of the complex momentum being chosen, each of the
  phase integrals and each of the tunneling action is real and
  non-zero.
\end{itemize}
Let us check the first observation. Let $\gamma$ be one of the curves
$\tilde\gamma_{0}$, $\tilde\gamma_{\pi}$, $\tilde\gamma_{h,0}$,
$\tilde\gamma_{h,\pi}$, $\tilde\gamma_{v,0}$ and
$\tilde\gamma_{v,\pi}$.  Let $\kappa$ be a branch of the complex
momentum continuous on $\gamma$.  The formula~\eqref{allbr} describes
all the other branches continuous on $\gamma$. As $\gamma$ is closed,
this shows that only the sign of the integral $\oint_\gamma \kappa
d\zeta$ depends on the choice of the branch $\kappa$.\\
Recall that $\kappa_p$ is analytic in the strip $S^p$ (see
section~\ref{two-bases}). To prove the second observation, we fix a
branch of the complex momentum on each of the integration contours.
For $\gamma_{\nu}$ and $\gamma_{h,\nu}$, we fix this branch so that
$\kappa=\kappa_p-\pi n$ on the parts of the contours in $\C_+$; for
$\gamma_{v,\nu}$, we choose $\kappa=\kappa_p-\pi n$ on the parts of
the contours in $\C_+\cap\{\nu<\re\zeta\}$.  We orient the contours
$\tilde\gamma_{\pi}$, $\tilde\gamma_{h,\pi}$ and
$\tilde\gamma_{v,\pi}$ clockwise, and we orient the contours
$\tilde\gamma_{0}$, $\tilde\gamma_{h,0}$ and $\tilde\gamma_{v,0}$
anticlockwise. Then, the second observation follows from
\begin{Le} 
  \label{le:2}
  For $E\in J$, for the above definitions of the integration contours
  and of the branches of the complex momentum defined on them, each of
  the functions $\Phi_\nu$, $S_{h,\nu}$ and $S_{v,\nu}$ is positive.
\end{Le}
\demo Begin with $\Phi_\pi$. As $\zeta_{2n+1}$ is a square root branch
point of $\kappa$, and, as $\kappa(\zeta_{2n+1})=0$, we get
\begin{equation*}
  \Phi_\pi(E)=\int_{\zeta_{2n+1}}^{2\pi-\zeta_{2n+1}}
  \kappa(\zeta+i0)\,d\zeta,
\end{equation*}
where one integrates along $\R$. As ${\mathcal E}(\zeta)$ is even, one proves that 
\begin{equation}\label{Phi-segment}
  \Phi_\pi(E)=2\int_{\zeta_{2n+1}}^{\pi}
  \kappa(\zeta+i0)\,d\zeta=2\int_{\zeta_{2n+1}}^{\pi}
  (\kappa_p(\zeta)-\pi n)\,d\zeta.
\end{equation}
Inside the integration interval, one has $\im\kappa_p=0$, and $\pi
n<\re\kappa_p<\pi(n+1)$. This implies the positivity of $\Phi_\pi$.\\
Arguing as above, for $S_{h,\pi}$, we get
\begin{equation}
  \label{S-h-segment}
  S_{h,\pi}(E)=-i\int_{\zeta_{2n}}^{\zeta_{2n+1}}
  (\kappa_p(\zeta)-\pi n)\,d\zeta,
\end{equation}
where one integrates along $\R$. Inside the integration interval, one
has  $\re \kappa_p=\pi n$ and $\im\kappa_p>0$ so that $S_{h,\pi}>0$.\\
For  $S_{v,\pi}$, one obtains 
\begin{equation}\label{S-v-pi:simple}
  S_{v,\pi}(E)=-2i\int_{\zeta_{2n+2}}^{\pi}
  (\kappa_p(\zeta)-\pi(n+1))\,d\zeta,
\end{equation}
where one integrates along $\pi+i\R$. Inside the integration interval,
$\pi n<\re \kappa_p<\pi(n+1)$ and $\im\kappa_p=0$ which implies
$S_{v,\pi}>0$.\\
Arguing similarly, one proves the positivity of $\Phi_0$, $S_{v,0}$ and
$S_{h,0}$. We omit further details.
\qed
\subsection{Proof of the inequalities~\eqref{eq:21}}
\label{sec:proof-ineq-eqref}
One has
\begin{equation*}
  \Phi_\pi(E)=2\int_{\zeta_{2n+1}}^{\pi} (\kappa_p(\zeta)-\pi
  n)\,d\zeta\quad\text{and}\quad 
  \Phi_0(E)=-2\int_0^{\zeta_{2n}}  (\kappa_p(\zeta)-\pi n)\,d\zeta.
\end{equation*}
The first equality was established when proving Lemma~\ref{le:2}. The
second is proved similarly. In view of~\eqref{kappa0k0}, we get
\begin{equation*}
  \Phi_\pi'(E)=2\int_{\zeta_{2n+1}}^{\pi} k_p'(E-\alpha
  \cos\zeta)\,d\zeta\quad\text{and}\quad 
  \Phi_0'(E)=-2\int_0^{\zeta_{2n}} k_p'(E-\alpha \cos\zeta)\,d\zeta,
\end{equation*}
where $k_p$ is the main branch of the Bloch quasi-momentum described
in section~\ref{SS3.2}. As, inside any spectral band of the periodic
operator $H_0$, the derivative $k'_p$ is positive, this
proves~\eqref{eq:21}.
\subsection{Proof of~\eqref{parity-h}}
\label{sec:t-h-0-pi}
We can choose the oriented contours $\tilde \gamma_{h,0}$ and $\tilde
\gamma_{h,\pi}$ so that one be the symmetric of the other with respect
to the origin. As ${\mathcal E}(\zeta)$ is even, for $\zeta\in
\gamma_{h,\pi}$, one has $\kappa(-\zeta)=\kappa(\zeta)$. These two
remarks imply relations~\eqref{parity-h}.
\subsection{Proof of Lemma~\ref{le:action-Y}}
\label{sec:proof-lemma-refl}
We shall prove the statement of Lemma~\ref{le:action-Y} for $\nu=\pi$.
For $\nu=0$ the argument is similar. As
$S_{v,\pi}(E_*)\in\R$,~\eqref{S-v-pi:simple} implies that
\begin{equation}
  \label{S-v-pi:simple-1}
  S_{v,\pi}(E_*)= \re S_{v,\pi}(E_*)=2\im
  \int_{\zeta_{2n+2}}^{\pi}(\kappa_p(\zeta)-\pi(n+1))\,d\zeta.
\end{equation}
Let us deform the integration contour in the right hand side so 
that it go successively
\begin{itemize}
\item from $\zeta_{2n+2}$ along the Stokes line ``b'' to $\zeta_{ba}$,
  the point of intersection of the Stokes lines ``b'' and ``a'' (see
  Fig.~\ref{bem:St-lines}),
\item from $\zeta_{ba}$ along the Stokes line ``a'' to $\zeta_{2n+1}$,
\item from $\zeta_{2n+1}$ to $\pi$ along the interval
  $[\zeta_{2n+1},\pi]$ which also is a Stokes line.
\end{itemize}
As $\kappa_p(\zeta_{2n+1})=\pi n$ and $\kappa_p(\zeta_{2n+2})=\pi
(n+1)$, the definitions of the Stokes lines then imply that
\begin{align*} 
  S_{v,\pi}(E_*)&=2\im \int_{\zeta_{2n+2},{\rm \ along \ 
      ``b''}}^{\zeta_{ba}}(\kappa_p(\zeta)-\pi(n+1))\,d\zeta+ 2\im
  \int_{\zeta_{ba},{\rm \ along \ 
      ``a''}}^{\zeta_{2n+1}}(\kappa_p(\zeta)-\pi(n+1))\,d\zeta \\ &+
  2\im \int_{\zeta_{2n+1},{\rm \ along \ 
    }\R}^{\pi}(\kappa_p(\zeta)-\pi(n+1))\,d\zeta=0+2\pi
  \im\zeta_{ba}+0 =2\pi\im\zeta_{ba}.
\end{align*}
As the set of the Stokes lines is symmetric with respect to both the
real line and the line $\pi+i\R$, the definition of $Y_{v,\pi}$
implies that $\im\zeta_{ba}=Y_{v,\pi}(E_*)$.  This and the result of
the last computation imply that $S_{v,\pi}(E_*)=2\pi Y_{v,\pi}(E_*)$.
The proof of Lemma~\ref{le:action-Y} is complete.
\subsection{The iso-energy curve}
\label{sec:iso-energy-curve-1}
The iso-energy curve $\Gamma$ is defined by~\eqref{isoen}. A point
$(\zeta,\kappa)\in\C^2$ belongs to $\Gamma$ if and only if $\kappa$ is
one of the values of the complex momentum at the point $\zeta$.\\
We now discuss the iso-energy curve under the assumptions (H), (O) and
(TIBM).
\subsubsection{The real branches}
\label{sec:real-branches}
Consider the real iso-energy curve $\Gamma_\R$ defined
by~\eqref{isoenr}.  Its connected components are {\it the real
  branches} of the iso-energy curve.  One has
\begin{Le}
  \label{le:real-branches}
  The real iso-energy curve is $2\pi$-periodic in both the $\kappa$-
  and $\zeta$-directions; it is symmetric with respect to each of the
  lines $\{\kappa=\pi n\}$ and $\{\zeta=\pi m\}$ for $m,n\in\Z$.\\
  Any periodicity cell contains exactly two real branches of $\Gamma$.
  Each of them is homeomorphic to a  circle.\\
  There exists $\gamma_0$ and $\gamma_\pi$, two disjoint connected
  components of $\Gamma_\R$ such that the convex hull of $\gamma_0$
  contains the point $(0,\pi n)$, and the convex hull of $\gamma_\pi$
  contains the point $(\pi,\pi n)$.\\
  The curves $\gamma_0$ and $\gamma_\pi$ are disjoint and are inside
  the strip $\{\pi(n-1)<\kappa<\pi(n+1)\}$.\\
  Any other real branch of $\Gamma$ can be obtained either from
  $\gamma_0$ or $\gamma_\pi$ by $2\pi$-translations in $\kappa$-
  or/and in $\zeta$-directions.
\end{Le} 
\demo The analysis of $\Gamma_\R$ is quite standard. A detailed
example can be found in~\cite{Fe-Kl:03f}. So, we only outline the
proof of Lemma~\ref{le:real-branches}. The periodicity and the
symmetries of $\Gamma_\R$ in $\zeta$ follows from the symmetry and
periodicity of the cosine and from formula~\eqref{allbr}.\\
Describe two real branches of $\Gamma$. Recall that one has
$\kappa_p([\zeta_{2n+1},\pi])\subset [\pi n,\pi(n+1)[$,
$\kappa_p([0,\zeta_{2n}])\subset]\pi(n-1),\pi n]$ and
$\kappa_p([\zeta_{2n},\zeta_{2n+1}])\subset\pi n+i\R_+$.  On the first
two intervals, $\kappa_p$ is monotonously increasing; on the last
interval, the imaginary part of $\kappa_p$ has only one maximum; this
maximum is non degenerate.  The graphs of $\kappa_p$ on each of the
intervals $[0,\zeta_{2n}]$ and $[\zeta_{2n+1},\pi]$ belong to
$\Gamma_\R$.  The real branch $\gamma_0$ is obtained from the graph on
$[0,\zeta_{2n}]$ by the reflections with respect to the lines
$\{\kappa=\pi n\}$ and $\{\zeta=0\}$.  The real branch $\gamma_\pi$ is
obtained from the graph on $[\zeta_{2n+1},\pi]$ by the reflections
with respect to the lines $\{\kappa=\pi n\}$ and $\{\zeta=\pi\}$.\\
We omit further elementary details of the proof. \qed
\subsubsection{Complex loops} 
\label{sec:complex-loops}
We prove
\begin{Le}
  \label{le:loops}
  The closed curve $\tilde \gamma_0$ (resp. $\tilde \gamma_\pi$,
  $\tilde\gamma_{h,0}$, $\tilde\gamma_{h,\pi}$, $\tilde\gamma_{v,0}$
  and $\tilde\gamma_{v,\pi}$) (see figures~\ref{bp-tibm}
  and~\ref{TIBMfig:2}) is the projection on the $\zeta$-plane of a
  loop $\gamma_0$ (resp. $\gamma_\pi$, $\gamma_{h,0}$,
  $\gamma_{h,\pi}$, $\gamma_{v,0}$ and $\gamma_{v,\pi}$) that is
  located on $\Gamma$. These loops satisfy:
  \begin{itemize}
  \item the loop $\gamma_{h,\pi}$ connects the real branches
    $\gamma_\pi$ and $\gamma_0$;
  \item the loop $\gamma_{h,0}$ connects the real branches $\gamma_0$
    and $\gamma_\pi-(2\pi,0)$;
  \item the loop $\gamma_{v,\pi}$ connects the real branches
    $\gamma_\pi$ and $\gamma_\pi+(0,2\pi)$;
  \item the loop $\gamma_{v,0}$ connects the real branches $\gamma_0$
    and $\gamma_0+(0,2\pi)$.
  \end{itemize}
\end{Le}
\noindent In Fig.~\ref{TIBMfig:actions}, we sketched the loops
described in Lemma~\ref{le:loops}.\\
{\it Proof of Lemma~\ref{le:loops}.} By Lemma~\ref{contours}, the
complex momentum can be analytically continued along each of the above
closed curves on $\C$. This implies that each of them is the
projection to $\C$ of a loop on $\Gamma$. Fix $\nu\in\{0,\pi\}$. For
$d\in\{h,v\}$, the loops discussed in the lemma satisfy:
\begin{equation}
  \label{loops:def}
      \gamma_{\nu}=\{(\zeta,\tilde\kappa_p(\zeta));\ \zeta\in
      \tilde\gamma_{\nu}\},\quad {\rm and}\quad
      \gamma_{d,\nu}=\{(\zeta,\tilde\kappa_p(\zeta));\ \zeta\in
      \tilde\gamma_{d,\nu}\}.
\end{equation}
Here, for $\gamma_{v,\nu}$, $\tilde\kappa_p$ denotes the branch of the
complex momentum that coincides with $\kappa_p$ on the parts of the
contours in $\C_+\cap\{\nu<\re\zeta\}$; for $\gamma_{\nu}$ and
$\gamma_{h,\nu}$, it is the branch that coincides with $\kappa_p$ on
the parts of the contours in $\C_+$. Therefore, we note that the curve
$\tilde\gamma_{h,\pi}$ intersects $\tilde\gamma_0$ and
$\tilde\gamma_\pi$. At the intersection point of
$\tilde\gamma_{h,\pi}$ and $\gamma_\pi$ (resp. $\gamma_0$), the
branches of $\tilde\kappa_p$ fixed on these curves coincide. This
implies that $\gamma_{h,\pi}$ connects the real branches $\gamma_\pi$
and $\gamma_0$. \\
The analysis of the other loops is done in the same way; we omit
further details.  \qed
\subsubsection{Interpretation of the phase integrals  and the
tunneling coefficients in terms of the iso-energy curve}
\label{sec:defin-phase-integr}
Let $E$ be real.  Pick $\nu\in\{0,\pi\}$ and $d\in\{v,h\}$.
Formula~\eqref{loops:def} shows that, up to the sign, $\Phi_\nu$ and $
S_{d,\nu}(E)$ coincide with $\frac12 \oint_{\gamma_\nu} \kappa d\zeta$
and $ -\frac i2\oint_{\gamma_{d,\nu}}\kappa d\zeta$.  So, choosing the
orientations of $\gamma_\nu$ and $\gamma_{d,\nu}$ in a suitable way,
we get $\Phi_\nu=\frac12 \oint_{\gamma_\nu} \kappa d\zeta$ and $
S_{d,\nu}(E)= -\frac i2\oint_{\gamma_{d,\nu}}\kappa d\zeta$.
\section{Properties of the Fourier coefficients}
\label{sec:fur}
We now prove the estimates and the asymptotics of the Fourier
coefficients found in Theorem~\ref{th:T-matrices} which will complete
the proof of this result.
\subsection{Computing the semi-classical factors}
\label{sec:comp-semicl-fact}
Proposition~\ref{pro:a,b:as} shows that the leading terms of the first
Fourier coefficients of $a_\nu$ and $b_\nu$ contain factors of the
form $e^{\frac i\varepsilon\int_\gamma \kappa d\zeta}$. They are
computed in
\begin{Le}\label{le:semi-cl-fact}
For $E\in J$, one has 
\begin{gather}
  \label{0-0}
  \exp\left(-\frac
    i\varepsilon\int_{\alpha_{0,0}}\kappa\,d\zeta\right)= e^{i\frac
    {\Phi_0+\Phi_\pi}{2\varepsilon}}\,t_{h,\pi}^{-1},\quad
  \exp\left(-\frac
    i\varepsilon\int_{\beta_{0,0}}\kappa\,d\zeta\right)=
  e^{i\frac {\Phi_0-\Phi_\pi}{2\varepsilon}}\,t_{h,\pi}^{-1},\\
  \label{0-1}
  \exp\left(-\frac
    i\varepsilon\int_{\alpha_{0,1}}\kappa\,d\zeta\right)= e^{-i\frac
    {\Phi_0-\Phi_\pi-4\pi^2}{2\varepsilon}}\,t_{v,0}\,t_{h,\pi}^{-1},\quad
  \exp\left(-\frac
    i\varepsilon\int_{\beta_{0,1}}\kappa\,d\zeta\right)= e^{-i\frac
    {\Phi_0+\Phi_\pi-4\pi^2}{2\varepsilon}}\,t_{v,0}\,t_{h,\pi}^{-1},\\
  \label{pi-0}
  \exp\left(\frac
    i\varepsilon\int_{\alpha_{\pi,0}}\kappa\,d\zeta\right)= e^{i\frac
    {\Phi_0+\Phi_\pi}{2\varepsilon}}\,t_{h,0}^{-1},\quad
  \exp\left(\frac
    i\varepsilon\int_{\beta_{\pi,0}}\kappa\,d\zeta\right)=
  e^{-\frac i{2\varepsilon}\,(\Phi_0-\Phi_\pi)}\,t_{h,0}^{-1},\\
  \label{pi-1}
  \exp\left(\frac
    i\varepsilon\int_{\alpha_{\pi,1}}\kappa\,d\zeta\right)= e^{-i\frac
    {\Phi_\pi-\Phi_0-4\pi^2}{2\varepsilon}}\,t_{v,\pi}\,t_{h,0}^{-1},\quad
  \exp\left(\frac
    i\varepsilon\int_{\beta_{\pi,1}}\kappa\,d\zeta\right)=e^{-i\frac
    {\Phi_\pi+\Phi_0-4\pi^2}{2\varepsilon}}\,t_{v,\pi}\,t_{h,0}^{-1}.
\end{gather}
Here, the integration contours are the curves shown in
Fig.~\ref{Curves-alpha-and-beta}; \ in each of the integrals, $\kappa$
is the branch of the complex momentum obtained from the one introduced
in the beginning of the section~\ref{two-bases} by analytic
continuation along the integration contour from its beginning to its
end.
\end{Le}
\demo All the formulae~\eqref{0-0}~--~\eqref{pi-1} are proved
similarly. Check the first formula in~\eqref{0-0}. Therefore, we
deform the curve $\alpha_{0,0}$ so that it go along the real line
going around the branch points $\zeta_{2n}$ and $\zeta_{2n+1}$ along
infinitesimally small circles. We get
\begin{equation*}
  -\int_{\alpha_{0,0}}\kappa\,d\zeta=I_1+I_2+I_3
\end{equation*}
where
\begin{equation}
  \label{0-0:check}
  I_1=-\int_{0}^{\zeta_{2n}}\kappa(\zeta+i0)\,d\zeta,\quad
  I_2=-\int_{\zeta_{2n}}^{\zeta_{2n+1}}\kappa(\zeta+i0)\,d\zeta
  \quad\text{and}\quad 
  I_3=-\int_{\zeta_{2n+1}}^\pi\tilde \kappa(\zeta-i0)\,d\zeta. 
\end{equation}
Here, in $I_1$ and $I_2$, we integrate the branch of the complex
momentum $\kappa$ introduced in the beginning of the
section~\ref{two-bases}, and, in $I_3$, $\tilde\kappa$ is the branch
obtained from $\kappa$ by analytic continuation from the interval
$(\zeta_{2n+1},\pi)+i0$ to the interval $(\zeta_{2n+1},\pi)-i0$ around
the branch point $\zeta_{2n+1}$ in the anti-clockwise direction.\\
Consider $I_3$. As $\zeta_{2n+1}$ is a square root branch point of
$\kappa$ and as $\kappa(\zeta_{2n+1})=0$, we have
$\tilde\kappa(\zeta-i0)=-\kappa(\zeta+i0)$ for $\zeta\in
(\zeta_{2n+1},\pi)\subset \R$. So, $\D I_3=
\int_{\zeta_{2n+1}}^{\pi}\kappa(\zeta+i0)\,d\zeta=\int_{\zeta_{2n+1}}^{\pi}
(\kappa_p(\zeta)-\pi n)\,d\zeta$. Comparing this with the right hand
side of~\eqref{Phi-segment}, we get $I_3=\frac12\Phi_\pi$.  Similarly,
one proves that $I_1=\frac12\Phi_0$. In view of~\eqref{S-h-segment},
one has $I_2=-iS_{h,\pi}$.
Combining the obtained expressions for $I_1$, $I_2$ and $I_3$, we get
\begin{equation*}
  \exp\left(-\frac i\varepsilon\int_{\alpha_{0,0}}\kappa\,d\zeta\right)=
  \exp\left(\frac{i}{\varepsilon}(I_1+I_2+I_3)\right)=
  \exp\left(\frac i{2\varepsilon}\,(\Phi_0+\Phi_\pi) +\frac1\varepsilon\,S_{h,\pi}\right).
\end{equation*}
This and the definition of $t_{h,\pi}$ implies the first formula
from~\eqref{0-0}. The second formula is proved similarly.\\
Describe the computation of the integrals in~\eqref{0-1}.  Let
$\int_\gamma\kappa d\zeta$ be one of them. First, using a symmetry
argument, we rewrite the integral in terms of the branch $\kappa_p$.
As $\kappa$ is real analytic in a neighborhood of $0$, one notes that
$\int_\gamma\kappa d\zeta=\overline{\int_{\overline{\gamma}}\kappa
  d\zeta}$, where $\overline{\gamma}$ is the oriented contour
symmetric to $\gamma$ with respect to the real line. One expresses the
integral $\int_{\overline{\gamma}}\kappa d\zeta$ in terms of the
tunneling actions and phase integrals using arguments similar the ones
presented above, and, then one computes $\int_\gamma\kappa d\zeta$
using the fact that the phase integrals and the actions are
real for real $E$. We omit further details.\\
Describe the computation of the integrals in~\eqref{pi-0}
and~\eqref{pi-1}.  Let $\int_\gamma\kappa d\zeta$ be one of them.
Again, using a symmetry argument, we rewrite the integral in terms of
the branch $\kappa_p$. As the function $\zeta\to \kappa(i\zeta)$ is
real analytic in a neighborhood of $0$, one notes that
$\int_\gamma\kappa d\zeta=-\overline{\int_{-\overline{\gamma}}\kappa
  d\zeta}$, where $-\overline{\gamma}$ is the oriented contour
symmetric to $\gamma$ with respect to the imaginary axis. Then, one
computes the integral $\int_{-\overline{\gamma}}\kappa d\zeta$ as the
integrals in~\eqref{0-0} and~\eqref{0-1}.  We omit further details.
This completes the proof of Lemma~\ref{le:semi-cl-fact}.\qed
\subsection{Proof of~\eqref{abs-Fourier:1}~--~\eqref{phases-Fourier:2}}
\label{sec:prop-four-coeff}
Being valid for $E\in J$, formulae~\eqref{0-0}~--~\eqref{pi-1} remain
valid in some neighborhood of $J$ independent of $\varepsilon$ (as
equalities between analytic functions).  The
formulae~\eqref{abs-Fourier:1} and~\eqref{phases-Fourier:2} follow
from the asymptotics~\eqref{Fourier-coefa} and~\eqref{Fourier-coefb},
and from formulae~\eqref{0-0}~--~\eqref{pi-1}. To illustrate this, let
us prove the formulae for $a_{0,0}$. Let $V_0$ be the neighborhood of
$E_*$ from Proposition~\ref{pro:a,b:as}. Using~\eqref{Fourier-coefa}
and~\eqref{0-0}, for $E\in V_0$, we get
\begin{equation}
  \label{a00:rep}
  a_{0,0}=t_{h,\pi}^{-1}\, \exp\left(\frac
    i{2\varepsilon}(\Phi_\pi+\Phi_0)+
    \int_{\alpha_{0,0}} \Omega_- +i\Delta\text{Arg}q_{|\alpha_{0,0}}
    +o(1)\right)=
    t_{h,0}^{-1}\, \exp\left(\frac
    i{2\varepsilon}(\Phi_\pi+\Phi_0)+O(1)\right),
\end{equation}
where we have used~\eqref{eq:15} and the fact that $\Omega_-$ and $q$
are independent of $\varepsilon$.  As $E\mapsto t_{h,\pi}(E)$,
$E\mapsto \Phi_0(E)$ and $E\mapsto \Phi_\pi(E)$ are real
analytic,~\eqref{a00:rep} implies the representations concerning
$a_{00}$ from~\eqref{abs-Fourier:1} and~\eqref{phases-Fourier:1}.
\subsection{Proof of~\eqref{z-nu-prime}}\label{sec:z-nu}
\label{sec:proof-eqrefz-nu}
Pick $\nu\in\{0,\pi\}$. Let $V_*$ be the neighborhood of $E_*$ from
Theorem~\ref{th:T-matrices}. By means
of~\eqref{z-nu:definition}~\eqref{phases-Fourier:1}
and~\eqref{phases-Fourier:2}, for $E\in V_*$, we get
$z_\nu=O(1/\varepsilon)$. The Cauchy estimates then imply that
$z_\nu'=O(1/\varepsilon)$ in any fixed compact of $V_*$.  So, at
expense of reducing somewhat $V_*$, we have proved~\eqref{z-nu-prime}.
%

%
\section{Combinations of Fourier coefficients}
\label{sec:four-coeff-comb}
\noindent Here, we study the asymptotics of the quantities
$\theta$, $T_h$, $T_{v,0}$, $T_{v,\pi}$, $\check \Phi_0$,
$\check\Phi_\pi$ and $z_0$, $z_\pi$.  We always use the branches
$\kappa$, $\psi_\pm$ and $\Omega_\pm$ described in the beginning of
section~\ref{sec:asympt-trans-matr}. Also, we systematically use the
notations and constructions from section~\ref{S3}.\\
Let ${\mathcal E}_0$ be a point in $ \mathcal{S}$.  Assume that it is
not a branch point, and that $\pi({\mathcal E}_0)\in\R$.  Consider
$U$, a neighborhood of $\mathcal{E}_0$ where $\pi^{-1}$ is analytic.
On $U$, we define the mapping $*:{\mathcal E}\mapsto
\pi^{-1}(\overline{\pi({\mathcal E})})$.
For $\gamma$, an oriented curve in ${\mathcal S}$ containing no branch
points and beginning at ${\mathcal E}_0$, we continue the map $*$
along $\gamma$ and, thus, define the oriented curve $\gamma^*$.
\subsection{The constant $\theta$ and the coefficients $T_h$,
  $T_{v,0}$, $T_{v,\pi}$}
\label{sec:constant-theta}
They are defined in~\eqref{theta} and~\eqref{T:definitions}. The
asymptotics~\eqref{theta:as} and~\eqref{T:as} are obtained in the same
way; so, we justify only the asymptotic for $\theta$.\\
The proof that, for sufficiently small $\varepsilon$, in the case of
Theorem~\ref{th:M-matrices}, one has~\eqref{theta:as} with the
constant $\theta_n$ defined in~\eqref{Lambda-omega}, consists of three
steps.
\subsubsection{Asymptotics of $\babs{\frac{a_{0,0}}{a_{\pi,0}}}$}
\label{theta:babs}
\begin{figure}
  \includegraphics[bbllx=71,bblly=638,bburx=474,bbury=721,width=14cm]{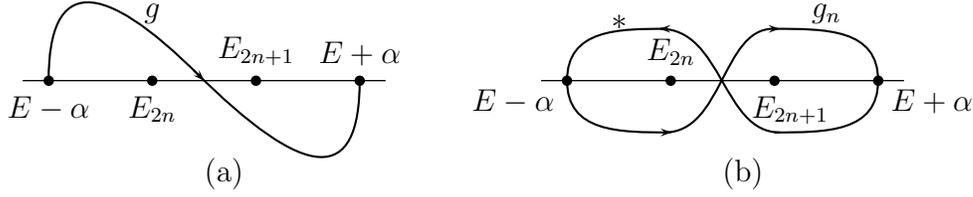}
  \caption{The curves $g$ and $\tilde g_n$}
  \label{g's}
\end{figure}

Let $g$ be a curve on $ \mathcal{S}$ that goes around the branch
points as shown in Fig.~\ref{g's}, part a, and that, for
$\pi({\mathcal E})>0$, is on the sheet of $ \mathcal{S}$ where
$k_p(\pi({\mathcal E}))$ is the Bloch quasi-momentum of $\psi(x,
{\mathcal E})$. We check that
%
%
\begin{equation}
  \label{theta:1}
  \babs{\frac{a_{0,0}}{a_{\pi,0}}}=
  \exp\left(\int_{g}\Omega(\hat {\mathcal
  E})+\int_{g^*}\Omega({\mathcal E})+o(1)\right).
\end{equation}
The representations~\eqref{Fourier-coefa} and the first formulae
from~\eqref{0-0} and~\eqref{pi-0} imply that
\begin{equation}
  \label{theta:11}\dsize
  \begin{split}
    \babs{\frac{a_{0,0}}{a_{\pi,0}}}&= \babs{\frac{t_{h,0}}{t_{h,\pi}}
      \exp\left(\int_{\alpha_{0,0}}\Omega_-(\zeta)-
        \int_{\alpha_{\pi,0}}\Omega_+(\zeta)+o(1)\right)}\\&=
    \babs{\exp\left(\int_{\alpha_{0,0}}\Omega_-(\zeta)-
        \int_{\alpha_{\pi,0}}\Omega_+(\zeta)+o(1)\right)}
  \end{split}
\end{equation}
as $t_{h,\pi}=t_{h,0}$, see~\eqref{parity-h}. \\
Recall that the curves $(\alpha_{\nu,0})_{\nu\in\{0,\pi\}}$ are shown
in Fig.~\ref{Curves-alpha-and-beta}.  We can and do assume that
$-\alpha_{\pi,0}$ is the symmetric to $\alpha_{0,0}$ with respect to
the origin.\\
As there are only two different branches of $\zeta\mapsto
\Omega(\zeta)$, and as the branch points of $\Omega$ coincide with
those of $\kappa$, the analytic continuation of $\Omega_+$ along
$\alpha_{\pi,0}$, near $0$, the end of $\alpha_{\pi,0}$, coincides
with $\Omega_-$. Therefore, \eqref{theta:11} can be rewritten in the
form
\begin{equation}
  \label{theta:13}\dsize
  \babs{\frac{a_{0,0}}{a_{\pi,0}}}=
  \babs{\exp\left(\int_{\alpha_{0,0}}\Omega_-+
      \int_{-\alpha_{\pi,0}}\Omega_-+o(1)\right)}
\end{equation}
Now, we make the change of variables $\zeta\mapsto {\mathcal
  E}(\zeta)$.  It maps each of the curves $\alpha_{0,0}$ and
$-\alpha_{\pi,0}$ on $g$, and we get
\begin{equation*}\dsize
  \exp\left(\int_{\alpha_{0,0}}\Omega_-(\zeta)+
    \int_{-\alpha_{\pi,0}}\Omega_-(\zeta)\right)=
  \exp\left(2\int_{g}\Omega(\hat {\mathcal E})\right),
\end{equation*}
where we have used that, for $\zeta$ near $0$, the branches $\zeta\to
\Omega_\pm(\zeta)$ correspond to the Bloch solutions
$\zeta\mapsto\psi_\pm(x,\mathcal{E}(\zeta))$ with the quasi-momenta
$\zeta\mapsto\pm k_p(\mathcal{E}(\zeta))$.
In section~\ref{sec:Omega}, we have formulated 
general properties of $\Omega$.  The fifth property implies that
\begin{equation}
  \label{eq:32}
  \overline{\int_{g}\Omega(\hat {\mathcal E})}=\int_{g^*}
  \Omega({\mathcal E}). 
\end{equation}
So,
\begin{equation*}
  \dsize\babs{\exp\left(\int_{\alpha_{0,0}}\Omega_-(\zeta)+
      \int_{-\alpha_{\pi,0}}\Omega_-(\zeta)\right)}=\exp\left(\int_{g}\Omega(\hat
    {\mathcal E})+\int_{g^*}\Omega({\mathcal E})\right).
\end{equation*}
This and~\eqref{theta:13} imply~\eqref{theta:1}.
\subsubsection{Computation of $\det T_\pi$}
\label{sec:computation-det-t_pi}
Here, we prove that
\begin{equation}
  \label{theta:2}
  \det T_\pi=-\exp\left(-\int_{g}\Omega({\mathcal E})+\Omega(\hat
  {\mathcal E})\right).
\end{equation}
Relations~\eqref{eq:35},~\eqref{new-basis:Wronskian},
and~\eqref{new-basis:Wronskian:1} imply that
\begin{equation}
  \label{theta:21}
  \dsize\det T_\pi= 
  \frac{\left.w(f_\pi,f^*_\pi)\right|_{\zeta+2\pi}}
  {\left.w(f_0,f^*_0)\right|_{\zeta}}=
  -\frac{k_p'(E+\alpha)\,w(\psi_+(\,\cdot\,,E+\alpha),
    \psi_-(\,\cdot\,,E+\alpha))}
  {k'_p(E-\alpha)\,w(\psi_+(\,\cdot\,,E-\alpha),
    \psi_-(\,\cdot\,,E-\alpha))}.
\end{equation} 
Furthermore, it follows directly from the definition of $\Omega$ that
$\Omega({\mathcal E})+\Omega(\hat {\mathcal E})= -d\log \int_0^1
\psi(x,{\mathcal E})\psi(x,\hat {\mathcal E})\, dx$.  Note that
$\psi(x,{\mathcal E})\psi(x,\hat {\mathcal E})$ remains the same when
we interchange ${\mathcal E}$ and $\hat {\mathcal E}$.  Therefore, it
depends only on $E=\pi({\mathcal E})$ and is single valued on the complex
plane. So, we get
\begin{equation}\label{theta:22}
\dsize
  \exp\left(\int_g \Omega({\mathcal E})+\Omega(\hat {\mathcal E})\right)=
  \frac{\left.\int_0^1 \psi_+(x,e)\psi_-(x, e)\, dx\right|_{e=E-\alpha}}
       {\left.\int_0^1 \psi_+(x,e)\psi_-(x, e)\, dx\right|_{e=E+\alpha}}.
\end{equation}
On any simply connected domain of $\C$ containing no branch points of
$\psi$, one has (see, for example,~\cite{MR2002h:81069})
\begin{equation*}
  \int_0^1 \psi_+(x,E)\psi_-(x,E)\, dx=
  -ik'(E)w(\psi_+(\cdot,E),\psi_-(\cdot, E)), 
\end{equation*}
where $\psi_\pm$ are two different branches of $\psi$ and $k$ is the
Bloch quasi-momentum of $\psi_+$.  This formula,~\eqref{theta:22}
and~\eqref{theta:21} imply~\eqref{theta:2}.
\subsubsection{Completing the proof of~\eqref{theta:as}}
\label{sec:compl-proof-eqrefth}
Let $\tilde g_n\subset \mathcal S$ be the curve shown in
Fig.~\ref{g's}, part b; its part marked by ``$*$'' is on the part of
$\mathcal S$ where $k_p(\pi(\mathcal{E}))$ is the Bloch quasi-momentum
of $\psi(x,\mathcal{E})$.  Relations~\eqref{theta:1},~\eqref{theta:2}
and~\eqref{theta} imply that $\theta=\exp\left(\oint_{\tilde g_n}
  \Omega({\mathcal E})+o(1)\right)$. \\
Now, let us compare $\oint_{\tilde g_n} \Omega({\mathcal E})$ with
$\oint_{g_n} \Omega({\mathcal E})$ where $g_n$ is the curve
in~\eqref{Lambda-omega}. Note that, on ${\mathcal S}$, modulo
contractible curves, one has $g_n=\tilde g_n$.
When deforming on ${\mathcal S}$ the curve $\tilde g_n$ to $g_n$, one
may intersect poles of $\Omega$. The poles and the residues of
$\Omega$ are described in section~\ref{sec:Omega}. This description
implies that the above two integrals coincide modulo $2\pi i$. So, we
have $\theta=\exp\left(\oint_{g_n} \Omega({\mathcal E})+o(1)\right)$.
This completes the proof of~\eqref{theta:as}.
\subsection{The phases $\{\check \Phi_\nu\}_{\nu=0,\pi}$ and
  $\{z_\nu\}_{\nu=0,\pi}$}
\label{sec:vrai-phases:vrai-as}
These are defined in~\eqref{vrai-phases} and~\eqref{z-nu:definition}.
The asymptotics of all the phases (see~\eqref{check-Phi:as}
and~\eqref{z-nu:as}) are obtained in the same
way; we justify only the asymptotic for $\check \Phi_\pi$.\\
So, we prove here that, for sufficiently small $\varepsilon$, in the
case of Theorem~\ref{th:M-matrices}, $\check \Phi_\pi$ admits the
asymptotics~\eqref{check-Phi:as}.
\smallpagebreak The asymptotics~\eqref{Fourier-coefa}
and~\eqref{Fourier-coefb} and formulae~\eqref{0-0} and~\eqref{pi-0}
imply that
\begin{equation}
  \label{check:1}
  \frac1\varepsilon
  \check\Phi_\pi=\frac1\varepsilon\Phi_\pi+\frac1{4i}
  \left(S-\overline{S}\right)+\frac12s+o(1),
\end{equation}
where 
\begin{gather}
  {\dsize S=\int_{\alpha_{\pi,0}}\Omega_+ +
    \int_{\alpha_{0,0}}\Omega_-
    + \int_{\beta_{\pi,0}} \Omega_+   - \int_{\beta_{0,0}}\Omega_-,}\\
  s=\left.\Delta\arg q\right|_{\alpha_{\pi,0}} + \left.\Delta\arg
    q\right|_{\alpha_{0,0}} + \left.\Delta\arg
    q\right|_{\beta_{\pi,0}} - \left.\Delta\arg
    q\right|_{\beta_{0,0}},
\end{gather}
where $(\alpha_{\nu,0},\beta_{\nu,0})_{\varepsilon\in\{0,\pi\}}$ are
sketched in Fig.~\ref{Curves-alpha-and-beta}. We can and, below, we
assume that, as the oriented curve $\alpha_{0,0}$ (resp.
$\beta_{0,0}$) is symmetric to the oriented curve $-\alpha_{\pi,0}$
(resp. $-\overline{\beta_{\pi,0}}$) with respect to zero. \\
First, show that $S-\overline{S}=0$. Arguing as when
deducing~\eqref{theta:13} from~\eqref{theta:11}, we get
\begin{equation*}\dsize
S=-\int_{-\alpha_{\pi,0}}\Omega_-  +  \int_{\alpha_{0,0}}\Omega_- 
   -\int_{-\beta_{\pi,0}} \Omega_+   - \int_{\beta_{0,0}}\Omega_-.
\end{equation*}
Now, we make the change of variables $\zeta\mapsto {\mathcal
  E}(\zeta)$.  As ${\mathcal E}(\alpha_{0,0})={\mathcal
  E}(-\alpha_{\pi,0})$, and ${\mathcal E}(-\beta_{\pi,0})={\mathcal
  E}(\overline{\beta_{0,0}})$, we get
\begin{equation}
  \label{eq:36}
  \dsize
  S=-\int_{({\mathcal E}(\beta_{0,0}))^*} \Omega({\mathcal E}) -
  \int_{{\mathcal E}(\beta_{0,0})} \Omega(\hat{\mathcal E}).
\end{equation}
As when proving~\eqref{eq:32}, we see that the terms in~\eqref{eq:36}
differ only by complex conjugation. So, $S$ is real and
$S-\overline{S}=0$.
\smallpagebreak Finally, we show that that $s=0$. This will complete
the proof of the asymptotics
of  $\check\Phi_\pi$.\\
When computing the increments of the argument of
$q(\zeta)=\sqrt{k'({\mathcal {\mathcal E}}(\zeta)}$, we choose the
(continuous) branch of this function which is positive on the interval
$(-\zeta_{2n},\zeta_{2n})$. Then, in a neighborhood of zero,
$q^*(\zeta)=q(\zeta)$ and $q(-\zeta)=q(\zeta)$.  Therefore, and due to
our ``symmetric'' choice of the curves
$(\alpha_{\nu,0})_{\nu\in\{0,\pi\}}$ and
$(\beta_{\nu,0})_{\nu\in\{0,\pi\}}$, we get
\begin{equation*}
  \left. \Delta\arg q\right|_{\alpha_{\pi,0}}=-\left. \Delta\arg
    q\right|_{\alpha_{0,0}}\quad\text{and}\quad 
  \left. \Delta\arg q\right|_{\beta_{\pi,0}}=-\left. \Delta\arg
    q\right|_{\overline{\beta_{0,0}}}=
  \left. \Delta\arg q\right|_{\beta_{0,0}}.
\end{equation*}
This and the definition of $s$ implies that $s=0$.  
\qed
%
%

%
\def\cprime{$'$} \def\cydot{\leavevmode\raise.4ex\hbox{.}}

%
%
\end{document}